\documentclass[twocolumn, tighten, trackchanges, resetfootnote]{aastex7}

\usepackage{amsmath}
\usepackage{rotating}

\shorttitle{Bayesian Inference of Gravity with 3D Modeling of Wide Binaries}
\shortauthors{Chae}

\begin{document}

\title{Bayesian Inference of Gravity through Realistic 3D Modeling of Wide Binary Orbits: \\ General Algorithm and a Pilot Study with HARPS Radial Velocities}

\author[orcid=0000-0002-6016-2736, gname=Kyu-Hyun, sname=Chae]{Kyu-Hyun Chae}
\email{chae@sejong.ac.kr}\thanks{corresponding author: chae@sejong.ac.kr \\ kyuhyunchae@gmail.com}
\affiliation{Department of Physics and Astronomy, Sejong University, 209 Neungdong-ro Gwangjin-gu, Seoul 05006, Republic of Korea}

\begin{abstract}
When 3D relative displacement $\mathbf{r}$ and velocity $\mathbf{v}$ between the pair in a gravitationally-bound system are precisely measured, the six measured quantities at one phase can allow elliptical orbit solutions at a given gravitational parameter $G$. Due to degeneracies between orbital-geometric parameters and $G$, individual Bayesian inferences and their statistical consolidation are needed to infer $G$ as recently suggested by a Bayesian 3D modeling algorithm. Here I present a fully general Bayesian algorithm suitable for wide binaries with two (almost) exact sky-projected relative positions (as in the Gaia data release 3) and the other four sufficiently precise quantities. Wide binaries meeting the requirements of the general algorithm to allow for its full potential are rare at present, largely because the measurement uncertainty of the line-of-sight (radial) separation is usually larger than the true separation. As a pilot study, the algorithm is applied to 32 Gaia binaries for which precise HARPS radial velocities are available. The value of $\Gamma \equiv \log_{10}\sqrt{G/G_{\rm N}}$ (where $G_{\rm N}$ is Newton's constant) is $-0.002_{-0.018}^{+0.012}$ supporting Newton for a combination of 24 binaries with Newtonian acceleration $g_{\rm N}>10^{-9}$m\,s$^{-2}$, while it is $\Gamma=0.134_{-0.036}^{+0.056}$ ($0.143_{-0.041}^{+0.068}$) for 8 (6) binaries with $g_{\rm N}<10^{-9}$ ($<10^{-9.5}$) m\,s$^{-2}$ representing $\ga 3.5\sigma$ discrepancy with Newton. However, one system (Stars HD189739 and HD189760) dominates the signal. Without it, the tension with Newton is significantly lessened with $\Gamma=0.063_{-0.041}^{+0.065}$. Thus, to verify the tentative signal, many such systems need to be discovered and their kinematic nature such as any possibility of hidden tertiary stars needs to be thoroughly addressed. The pilot study demonstrates the potential of the algorithm in measuring and testing gravity at low acceleration with future samples of wide binaries.
\end{abstract}


\section{Introduction}\label{sec:intro}

Gravitationally-bound binary systems have played pivotal roles in astronomy and physics. The orbital motion of planets in the Sun-planet binary systems provided the early experimental foundation of Newtonian gravitational dynamics. The gravitational anomaly in Mercury's orbital motion in a relatively strong gravity regime provided an early experimental success of Einstein's general relativity. Binary (and higher-multiplicity) stars have been playing fundamental roles in determining stellar masses and radii and constraining star formation and evolution models (e.g., \citealt{Duchene:2013,ElBadry:2024}). 

A currently interesting use of binary stars in studies of fundamental physics is to probe gravity in the low-acceleration ($\la 10^{-9}$m\,s$^{-2}$) regime with wide binaries. The dark matter (DM) or missing gravity phenomena in galaxies indicate a critical acceleration scale $a_0\approx 1.2\times 10^{-10}$m\,s$^{-2}$, known as Milgrom's or modified Newtonian Dynamics \citep[MOND; ][]{Milgrom:1983} constant (see the reviews by \citealt{Sanders:2002,Famaey:2012,BanikZhao:2022}). As Milgrom proposed, the presence of $a_0$ in galaxies may indicate the breakdown of standard Newton-Einstein gravity in the low-acceleration regime, opening possibilities of gravity theories breaking the strong equivalence principle \citep{Milgrom:1983}. This can be directly tested by wide binaries in the solar neighborhood as wide binary internal dynamics cannot be affected by the dark matter density inferred for the Milky Way (\citealt{Hernandez:2012}; see also, e.g., \citealt{Acedo:2020}) while the \emph{phantom} dark matter density around a small-scale star is much higher than that of the Milky Way (see, e.g., \citealt{Pf-A:2025}). 

There is a fundamental limitation in probing gravity with individual wide binaries because those of interest have long periods of $\ga 10^5$~yr and thus their orbits cannot be traced for significant segments. When the relative displacement $\mathbf{r}$ and the relative velocity $\mathbf{v}$ between the pair are measured with sufficient precision, the six components at one phase can be used to reconstruct the orbit assuming Newtonian gravitational dynamics (see, e.g., \cite{Kervella:2017} and \cite{Akeson:2021} for reconstructed orbits of Proxima around $\alpha$ Cen AB). However, the same data may be equally well fitted by nonstandard theories of gravity with varied orbits. In other words, globally different orbits for different theories of gravity may fit the instantaneous displacement and velocity equally well. Thus, such degeneracies must be dealt with in using wide binaries for testing gravity through 3D orbit modeling. In the specific case of Proxima, its exceptional proximity to the Sun may allow a future monitoring campaign to break the degeneracy between standard and nonstandard gravities \citep{BanikZhao:2018,BanikKroupa:2019}.

Wide binaries with precise $\mathbf{r}$ and $\mathbf{v}$ are rare because it is challenging to measure the relative radial (line-of-sight) displacement and the relative radial velocity (RV) with sufficiently good precision. In particular, the measurement uncertainties of the relative distances are usually (much) larger than the radial separations, whereas precise RVs are currently being obtained for increasingly larger samples. Nevertheless, the potential importance of 3D velocities $\mathbf{v}$ with precise RVs was recognized by earlier investigators of wide binaries \citep{Scarpa:2017,PittordisSutherland:2018}.

Recently, two studies considered 3D orbit modeling to test gravity based on samples of limited quantity and/or quality. For 32 binaries selected from the Gaia data release 3 \citep[DR3; ][]{Gaia:2023} database, \cite{Saglia:2025} derived very precise RVs from the High Accuracy Radial Velocity Planet Searcher (HARPS) spectrograph database. Their sample covers a broad range of Newtonian gravitational acceleration $10^{-11}\la g_{\rm N}\la 10^{-6}$~m\,s$^{-2}$ including about 8 wide binaries with $g_{\rm N}< 10^{-9}$~m\,s$^{-2}$ as will be shown below. Notably, their sample includes only a few binaries with uncertainty of distance difference ($\Delta d$) satisfying $\sigma_{\Delta d}< s/\sqrt{2}$ where $s$ is the sky-projected 2D separation. Thus, individual orbit inferences and gravity tests are limited by overall imprecise radial separations (note that tangential separations derived from Gaia astrometry have negligibly small errors). Moreover, even if $\sigma_{\Delta d}$ is sufficiently small, degeneracies between orbits and gravity theories need to be dealt with. \cite{Saglia:2025} presented only non-unique Newtonian orbit solutions for all their binaries except for one system (\#24) with stars HD189739 and HD189760 which they interpret as an unbound system. They have not derived uncertainties or probability distributions of orbit parameters, but they noted that some orbit phases seemed unnatural.

\cite{Chae:2025} introduced a Bayesian methodology to infer for each binary a posterior probability distribution of (nuisance) orbit parameters along with the gravitational boost (or anomaly) parameter
\begin{equation}
    \Gamma \equiv \log_{10} \sqrt{\gamma_g} \equiv  \log_{10} \sqrt{G/G_{\rm N}},
    \label{eq:Gamma}
\end{equation}
where $G$ is the generalized gravitational parameter and $G_{\rm N}$ is Newton's constant. The individual PDFs of $\Gamma$ are then statistically consolidated \citep{Hill:2011} for binaries having similar internal gravitational accelerations to derive $\Gamma$ as a function of acceleration. The methodology was applied to a sample of 312 wide binaries with relative RV uncertainties $<380$m\,s$^{-2}$ selected from the Gaia DR3. One caveat of \cite{Chae:2025} was that the orbit configuration suppressed one of the two parameters representing the direction of the orbit axis with respect to the observer's view, in part because the observational constraints were not sufficient. For this reason, the \cite{Chae:2025} methodology will be referred to as `simple Bayesian' throughout.

The purpose of this work is two-fold. One is to present a fully general version of the \cite{Chae:2025} methodology that can be used for wide binaries with sufficiently precise measurements of all components of $\mathbf{r}$ and $\mathbf{v}$. At present, such wide binaries are extremely rare, but it is in principle more correct to use a fully general model even for insufficient data (although the parameters may be less well constrained). The other is to to apply the methodology to the \cite{Saglia:2025} sample with extremely precise HARPS RVs, as a pilot study expecting (much) larger samples of similar qualities in the future. This pilot study can also complement \cite{Saglia:2025} results by deriving Bayesian probability distributions of not only orbit and orientation parameters but also $\Gamma$. However, note that except for a few binaries of the \cite{Saglia:2025} sample, $\Gamma$ and the orbit parameters are not well constrained individually. The Python codes implementing the general Bayesian algorithm \citep{Chae:Zenodo2025} and the Bayesian outputs for the \cite{Saglia:2025} sample are available on Zenodo under an open-source Creative Commons Attribution license: 10.5281/zenodo.17113129.

\section{Methodology} \label{sec:method}

Following \cite{Chae:2025} I work only with elliptical orbits, meaning that Newtonian or pseudo-Newtonian gravity with a generalized gravitational parameter is considered. In this approach, any nonstandard gravity is approximated to be pseudo-Newtonian in wide binary internal dynamics. This assumption is completely valid in testing Newtonian gravity and quantifying a degree of deviation from it if there arises a discrepancy with Newton. However, this assumption is approximate in testing a specific nonstandard gravity model because its predicted numerical orbits may not be closed ellipses. 

\begin{figure}[!htb]
    \centering
    \includegraphics[width=1.05\linewidth]{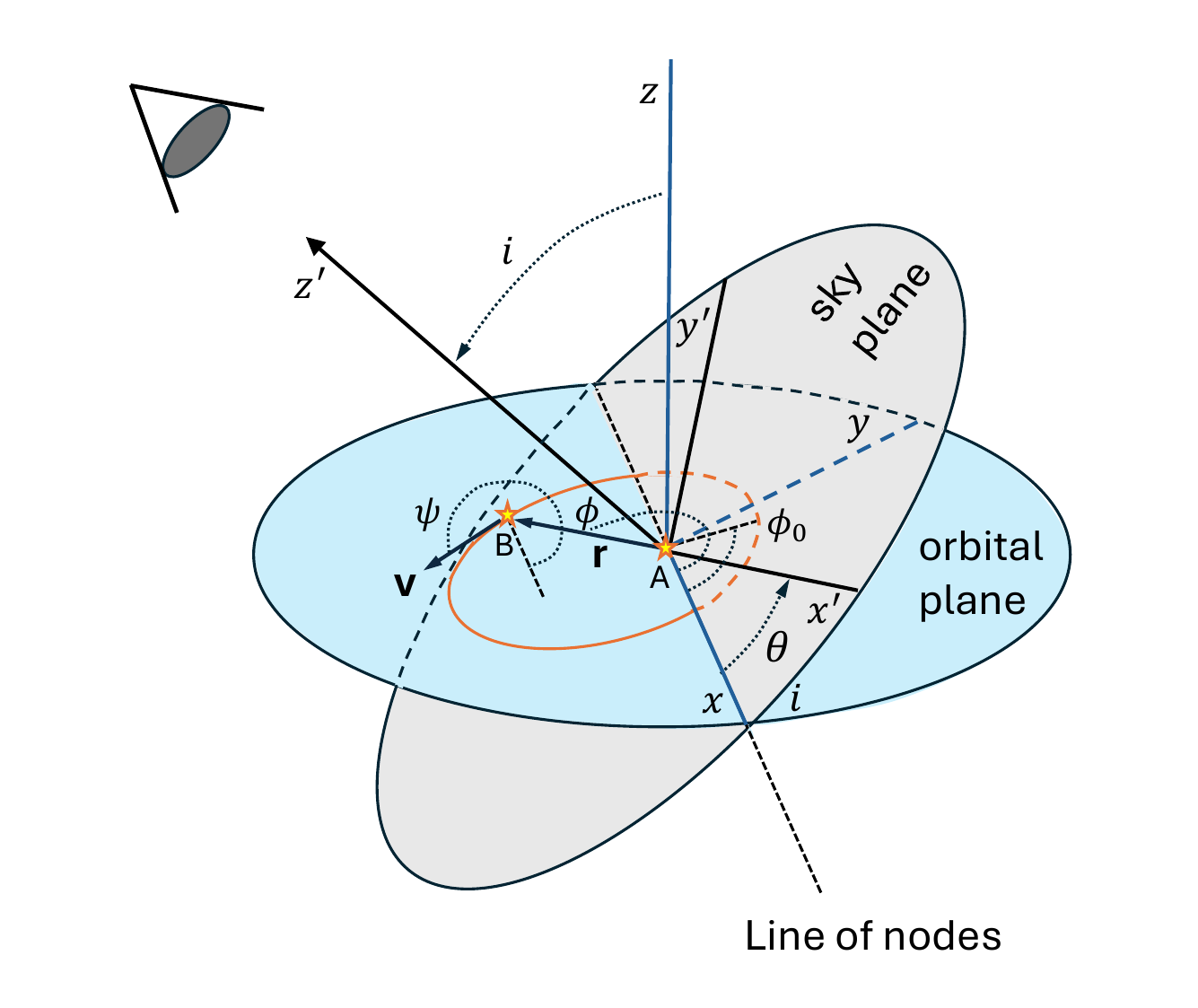}
    \caption{
    A general 3D geometry of observing an elliptical orbit of the relative motion between the stars of a binary system whose plane makes an inclination angle of $i$ with the sky plane. Two angles $i$ and $\theta$ can represent any arbitrary orientation of the observer's direction (the $z^\prime$-axis) with respect to the $z$-axis of the orbital plane.  
    }
    \label{fig:Euler2angles}
\end{figure}

Figure~\ref{fig:Euler2angles} shows a general 3D configuration of the binary orbit with respect to the observer. The six components of the relative displacement and the relative velocity given by $\mathbf{r}=(x^\prime,y^\prime,z^\prime)$ and $\mathbf{v}=(v_{x^\prime}, v_{y^\prime}, v_{z^\prime})$ in the observer's $x^\prime y^\prime z^\prime$ frame are determined by four orbit parameters, i.e., $a$ (semi-major axis), $e$ (eccentricity), $\phi_0$ (argument of the periastron), and $\phi$ (phase), and two orientation parameters $i$ (inclination) and $\theta$ (rotation about the $z^\prime$ axis which is related to the argument of the ascending node $\Omega$ by $\Omega =\pi -\theta$ in the figure). When $x^\prime$ (separation along the R.A. direction) and $y^\prime$ (separation along the decl.\ direction) are measured with negligibly small errors as in Gaia DR3, they can be used to reduce the number of free parameters by two with mathematical relations. I will work with four free parameters $e$, $\phi_0$, $\phi$, and $i$ given the four observational constraints $v_{x^\prime}$, $v_{y^\prime}$, $v_{z^\prime}$, and $z^\prime$ along with their uncertainties. Parameters $a$ and $\theta$ follow from the other parameters and the mathematical relations for $x^\prime$ and $y^\prime$ as will be shown below. 

For a binary of stars $A$ (brighter) and $B$ with an observational total mass $M_{\rm tot} \equiv M_A + M_B$ and sky-projected separation $s\equiv \sqrt{(x^\prime)^2+(y^\prime)^2}$, we have the magnitude of the 3D displacement
\begin{equation}
    r = \frac{a(1-e^2)}{1+e\cos(\phi-\phi_0)} = \frac{s}{\sqrt{\cos^2\phi + \cos^2 i \sin^2 \phi}},
    \label{eq:r}
\end{equation}
and the magnitude of the 3D velocity
\begin{eqnarray}
    v = v(r) & = & \sqrt{\gamma_g \frac{G_{\rm N}f_M M_{\rm tot}}{r} \left(2-\frac{r}{a}\right)} \nonumber \\ 
    & = & 941.9\text{ m s}^{-1} \sqrt{\gamma_gf_M\frac{M_{\rm tot}/M_\odot}{r/\text{kau}} \left(2-\frac{r}{a}\right) },
    \label{eq:v}
\end{eqnarray}
where $f_M$ is a factor to take into account the uncertainty of the observational mass. Note that variables $r$ and $a$ appearing in Equations~(\ref{eq:r}) and (\ref{eq:v}) depend on the other variables in Equation~(\ref{eq:r}), i.e., the directly measurable quantity $s$ and the independent variables $\phi$, $\phi_0$, $i$, and $e$. 

With (exact) values of $s$ and $b\equiv y^\prime /x^\prime$ given from observations, we have 
\begin{equation}
    \tan\psi = - \frac{\cos\phi + e\cos\phi_0}{\sin\phi + e\sin\phi_0}
    \label{eq:psi}
\end{equation}
and
\begin{equation}
    \tan\theta = \frac{\tan\phi \cos i - b}{1 + b\tan\phi \cos i},
    \label{eq:theta}
\end{equation}
from which follows $\phi=\tan^{-1}(b/\cos i)$ (Equation~(14) of \cite{Chae:2025}) when $\tan\theta = 0$. Then, we have the following prediction on the four remaining quantities $z^\prime$, $v_{x^\prime}$, $v_{y^\prime}$, and $v_{z^\prime}$ expressed as functions of $e$, $i$, $\phi_0$, and $\phi$ only:
\begin{equation}
   z^\prime =  - s \sin\phi \sin i /\sqrt{\cos^2\phi + \cos^2 i \sin^2 \phi}  
\label{eq:zprime}
\end{equation}
and
\begin{equation}
    \left[\begin{array}{c} v_{x^\prime} \\ v_{y^\prime}  \\ v_{z^\prime}   \end{array} \right] = k v  |\cos\psi|\left[\begin{array}{c}
        \cos\theta + \tan\psi  \cos i \sin\theta  \\
       -\sin\theta + \tan\psi \cos i \cos\theta  \\
       -\tan\psi \sin i   \\
    \end{array} \right],
    \label{eq:relvel_model}  
\end{equation}
where $k(=\pm 1)$ indicates the sign of the quantity $dx/d\phi$ or ($\sin\phi + e\sin\phi_0$). Equations~(\ref{eq:zprime}) and (\ref{eq:relvel_model}) follow from Equations~(A5) and (A6) of \cite{Chae:2025}.

For the binary of stars $A$ and $B$, $x^\prime$ and $y^\prime$ come from the precise R.A. ($\alpha_I$ with $I=A,B$ hereafter) and decl.\ ($\delta_I$) values given in units of degrees provided by the Gaia DR3 database:
\begin{equation}
\left. \begin{array}{cl}
   x^\prime = & -3600 d_M \cos((\delta_A+\delta_B)\pi/360) \Delta\alpha \text{ au} \\ 
   y^\prime = & 3600 d_M \Delta \delta \text{ au} 
\end{array} \right\},
\label{eq:relpos_sky}
\end{equation}
where $\Delta\alpha\equiv \alpha_B-\alpha_A)$, $\Delta\delta \equiv \delta_B-\delta_A$, and the minus sign on the right-hand side of the first equation follows from the fact that R.A. increases in the $-x^\prime$ direction of Figure~\ref{fig:Euler2angles}. In Equation~(\ref{eq:relpos_sky}), $d_M$ refers to the distance in units of pc to the binary from the Sun given by the error-weighted mean of the two measured distances, as would be the case in the plane geometry when the separation is negligibly small compared with the distances. 

The observational quantities to match the four theoretical quantities given by Equations~(\ref{eq:zprime}) and (\ref{eq:relvel_model}) are as follows:
\begin{equation}
   z^\prime_{\rm obs} = -(d_B - d_A)648000/\pi \text{ au}, 
\label{eq:zprime_obs}
\end{equation}
and
\begin{equation}
\left. \begin{array}{rl}
  v_{x^\prime,{\rm obs}} = & -4.7404 d_M (\mu^\star_{\alpha,B}-\mu^\star_{\alpha,A})\text{ m s}^{-1} \\ 
  v_{y^\prime,{\rm obs}} = & 4.7404 d_M (\mu_{\delta,B}-\mu_{\delta,A})\text{ m s}^{-1} \\
  v_{z^\prime,{\rm obs}} = & -1000(\text{RV}_B - \text{RV}_A) \text{ m s}^{-1}
\end{array} \right\},
\label{eq:relvel_obs}
\end{equation}
where $\mu^\star_{\alpha,I}$ and $\mu_{\delta,I}$ are PMs in the R.A. and decl.\ directions given in units of of mas~yr$^{-1}$, and RV$_i$ are the measured RVs given in units of km~s$^{-1}$. In estimating uncertainties of $v_{x^\prime,{\rm obs}}$ and $v_{y^\prime,{\rm obs}}$, the system distance $d_M$ is fixed as the velocities refer to internal relative velocities within the system.

For the general case of modeling, the free parameters are $\mathbf{\Theta}=\{e,i,\phi_0,\Delta\phi(\equiv\phi-\phi_0),\log_{10}f_M,\Gamma \}$. In the Bayesian approach, the posterior probability distribution of the parameters $p(\mathbf{\Theta})$ is defined by
\begin{equation}
    \ln p(\mathbf{\Theta}) = \ln\mathcal{L} + \sum_l \ln f_\text{pr}(\Theta_l),
\label{eq:postprob}    
\end{equation}
where $\mathcal{L}$ is the likelihood function, and $f_\text{pr}$($\Theta_l$) ($l=1,\cdots,6$) is the prior probability density for parameter $\Theta_l$, as described below. The likelihood function is defined by
\begin{equation}
    \ln\mathcal{L} = -\frac{1}{2}\sum_j \left[ \left(\frac{X_{j,\rm{mod}}(\mathbf{\Theta})-X_{j,\rm{obs}}}{\sigma_j} \right)^2 +\ln(2\pi\sigma_j^2) \right],
\label{eq:likelihood}    
\end{equation}
where $\{X_{j,\rm{mod}}(\mathbf{\Theta})\}$ ($j=1,2,3,4$) are the model predictions given by Equations~(\ref{eq:zprime}) and (\ref{eq:relvel_model}) while $X_{j,\rm{obs}}$ and $\sigma_j$ are the observed quantities (Equations~(\ref{eq:zprime_obs}) and (\ref{eq:relvel_obs})) and their uncertainties.

Physical ranges of the explicit angular variables are: $0\le i \le \pi$, $0\le\phi_0 <2\pi$, and $0\le \Delta\phi <2\pi$. The implicit variable $\theta$ has the physical range $-\pi\le\theta <\pi$. However, because the prior Equation~(\ref{eq:theta}) returns only the range $-\pi/2<\theta <\pi/2$, it is more convenient to consider $-\pi\le i \le \pi$ along with $-\pi/2<\theta <\pi/2$ using the mathematical property that $\mathbf{r}$ and $\mathbf{v}$ are invariant under the transformations: $i\rightarrow -i$, $\phi_0\rightarrow\phi_0\pm \pi$, $\phi\rightarrow\phi\pm \pi$ (keeping $\Delta\phi$ invariant), and $\theta\rightarrow\theta\pm \pi$. In doing so, values of $\theta$ outside the range $-\pi/2<\theta <\pi/2$ can be represented by the range $-\pi/2<i<0$, and, if desired, transformations can be used to obtain values in the physical ranges of $i$ and $\theta$.

Priors on the model parameters are considered depending on how gravity is probed/tested with the Bayesian methodology. Two approaches are considered. In the first approach, as in \cite{Saglia:2025}, gravity is fixed for each wide binary and the model parameters are constrained with or without priors, and the distributions of the inferred parameters are tested against the expectations for the assumed gravity. This approach will be considered in this work not only for Newtonian gravity, but also for pseudo-Newtonian gravities (where $G$ is allowed to vary from system to system depending on the internal Newtonian acceleration) inspired by existing MOND gravity models. In the second approach, as in \cite{Chae:2025}, logical or empirical priors are imposed on all the free parameters except for $\Gamma$, and PDFs of $\Gamma$ are derived. For the purpose of inferring/testing gravity, this will be the main approach of this work.

The prior on a parameter can be either a logical requirement or an empirical/reasonable constraint on the parameter from observational or numerical simulation studies. From the randomness of observed wide binaries, there are logical requirements on $i$ (inclination), $\phi_0$ (periastron), and orbital phase $\Delta\phi$ (known as the true anomaly) as follows: $f_\text{pr}$($i$) = $|\sin(i)|$, $f_\text{pr}$($\phi_0$) = uniform, and 
\begin{equation}
    f_\text{pr}(\Delta\phi) = \frac{(1-e^2)^{3/2}}{2\pi} \frac{1}{[ 1+ e \cos(\Delta\phi) ]^2}. 
\label{eq:PrDelphi}   
\end{equation}
 The prior on $\Delta\phi$ given by Equation~(\ref{eq:PrDelphi}) is required to ensure uniform probability in time (rather than $\Delta\phi$) along the orbit within a period for physically realizable Newtonian and pseudo-Newtonian elliptical orbits.\footnote{MONDian gravity models may not be accurately described by pseudo-Newtonian orbits with a generalized $G$. In such cases, the latter should be understood as working approximations.} This prior is of particular importance in testing gravity regardless of the approach. If the approach is to infer a consolidated PDF of $\Gamma$ from a random population of systems, the prior must be imposed for all the systems so that Equation~(\ref{eq:PrDelphi}) is satisfied by the population. In the approach of fixed gravity with no priors, the inferred values of $\Delta\phi$ can be compared with Equation~(\ref{eq:PrDelphi}).

 Further priors are considered for mass parameter $f_M$ and eccentricity $e$. Because the total mass of each system was determined observationally, I consider a lognormal probability distribution of $f_M$ with mean $=0$ and s.d.\ $=5\%$. This prior on $f_M$ is always imposed regardless of the other priors. Because a very precise 3D velocity has some constraining power on $e$, a prior on $e$ is not a significant factor (unlike \citealt{Chae:2025}). I consider two cases: a generic distribution $f_\text{pr}(e) = 2e$ (the thermal probability distribution) and the flat/no prior within the range $0<e<1$. Finally, when $\Gamma$ is constrained for each binary, it is assumed to be in the range $-1<\Gamma<1$, which turns out to include most PDFs sufficiently well.

Markov chain Monte Carlo (MCMC) procedure is performed to derive the posterior distribution $p(\mathbf{\Theta})$ with the public Python package {\tt emcee} \citep{emcee}. The numerical procedure is as follows. First, approximate global best-fit values of $e$, $i$, $\phi_0$, $\Delta\phi$, and $\Gamma$ are determined by a grid search of the parameter space to minimize $\chi^2=\Sigma_j((X_{j,\rm{mod}}(\mathbf{\Theta})-X_{j,\rm{obs}})/{\sigma_j})^2$ (see Equation~(\ref{eq:likelihood}) for the definition of symbols). These values along with $\log_{10}f_M=0$ are the starting values of random walks. The number of random walkers is {\tt nwalkers}$=200$. With six parameters ({\tt ndim}$=6$) that is one more than the simple Bayesian approach of \cite{Chae:2025}, I consider longer chains of iterations than \cite{Chae:2025}. The particular choice to derive the nominal results is the following: {\tt niter}$=400000$, {\tt discard}$=200000$ (to well exclude the burn-in phase), and {\tt thin}$=20$, so that a total of {\tt nwalkers} $\times$ ({\tt niter}$-${\tt discard}) $/$ {\tt thin} $= 2\times 10^6$ models are obtained to derive posterior probability density functions (PDFs) of the parameters.

\section{A pilot study with HARPS radial velocities: Results and Discussion} \label{sec:result}

For each binary from the \cite{Saglia:2025} sample, Bayesian results are produced based on the two approaches described in Section~\ref{sec:method}: (1) fixed gravity with or without priors on the other parameters (except for the mass parameter $f_M$ whose prior is always imposed); (2) free $\Gamma$ with priors on the other parameters. In the first approach, two cases are considered. In one case, gravity is Newtonian for all 32 binaries. In the other case, hereafter to be referred to as the ``MOND-type toy model'', $G$ is fixed for each binary at a value depending on its internal Newtonian acceleration (inferred from the Newtonian model) $g_{\rm N}$ as follows: $G=G_{\rm N}$ for $g_{\rm N}>10^{-9}$~m\,s$^{-2}$ and $G=1.4G_{\rm N}$ for $g_{\rm N}<10^{-9}$~m\,s$^{-2}$. This toy model is clearly not a whole theoretical prediction of MOND or any other modified gravity. It is introduced here to serve as a control nonstandard model (or a working approximation) that is intended to mimic a transition from the verified Newtonian regime of $g_{\rm N}\ga 10^{-8}$~m\,s$^{-2}$ \citep{Chae:2025} to the low-acceleration MOND regime of $g_{\rm N} \la a_0$. Moreover, as will be shown in this work, a subsample of 23 binaries with $g_{\rm N}>10^{-9}$~m\,s$^{-2}$ agrees well with Newton while the other subsample does not. Thus, this toy model is self-consistent within this work.

The relative RV between the two stars $v_{z^\prime}=-(\text{RV}_B-\text{RV}_A)$ used here is obtained by correcting RVs taken from Table~B.2 of \cite{Saglia:2025} for gravitational redshift and convective flow. The velocity components are corrected for perspective effects \citep{Shaya:2011,Yoon:2025} because some binaries are close to the Sun (although they are in general negligibly small). The nominal uncertainty of $v_{z^\prime}$ is enlarged by quadratically adding $40$~m\,s$^{-1}$ to it following the recommendation of \cite{Saglia:2025}. 

While $x^\prime_{\rm obs}$ and $y^\prime_{\rm obs}$ are measured with negligibly small errors, $z^\prime_{\rm obs}=-(d_B-d_A)$ is poorly measured in general.\footnote{See Section~2.3 of \cite{Banik:2019} for a statistical inference of the radial separation that may be applicable for statistical studies with a large sample.} Thus, even if all the velocity components are very precise, we do not expect that the model parameters are well constrained, especially for the generalized gravity model. Considering that the expected radial separation is of the order of $s/\sqrt{2}$, the required precision is $\sigma_{z^\prime}<s/\sqrt{2}$ for $S/N>1$. Only three systems meet this requirement: \#{5}, \#{12}, and \#{21} from Table~B.1 of \cite{Saglia:2025}. However, note that even for these systems, the precision of $z^\prime_{\rm obs}$ may not be sufficiently good. Thus, unfortunately, in general we do not expect narrow posterior ranges of the parameters. 

\begin{figure}[!htb]
    \centering
    \includegraphics[width=1.\linewidth]{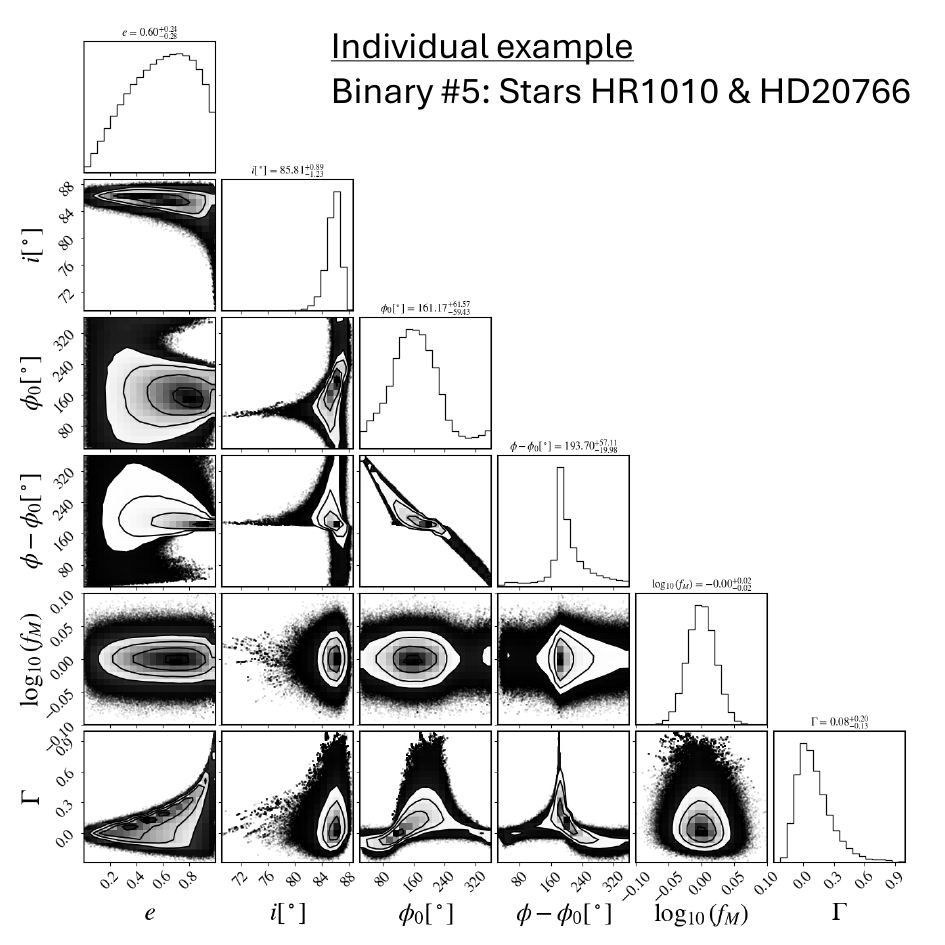}
    \caption{
    This figure shows an example of the posterior PDFs of the parameters with $\Gamma$ free (generalized gravity). This binary is one of the few that have relatively well-measured radial separations. The fitted parameters in general have broad ranges. See Tables~\ref{tab:prmt_newton} and \ref{tab:prmt_general} for the parameter values of all the binaries in Newtonian and generalized gravity. 
    }
    \label{fig:PDFexample}
\end{figure}

Figure~\ref{fig:PDFexample} shows an example of the PDFs of the model parameters in the generalized gravity case for binary \#5, which has a relatively precise value of $z^\prime_{\rm obs}$. In this case, inclination $i$ is well determined. However, parameters $e$, $\phi_0$, and $\Delta\phi$ are relatively poorly constrained covering broad ranges. Parameter $\Gamma$ is not well determined either, with the uncertainty being larger than $0.07$ (the difference between Newton and a simple MOND prediction). Thus, even with the HARPS RVs, Newton and MOND are, in general, not distinguished well by an individual system. 

\subsection{Distribution of the Bayesian Inferred Parameters} \label{sec:distribution}

\begin{figure*}[!htb]
    \centering
    \includegraphics[width=1.\linewidth]{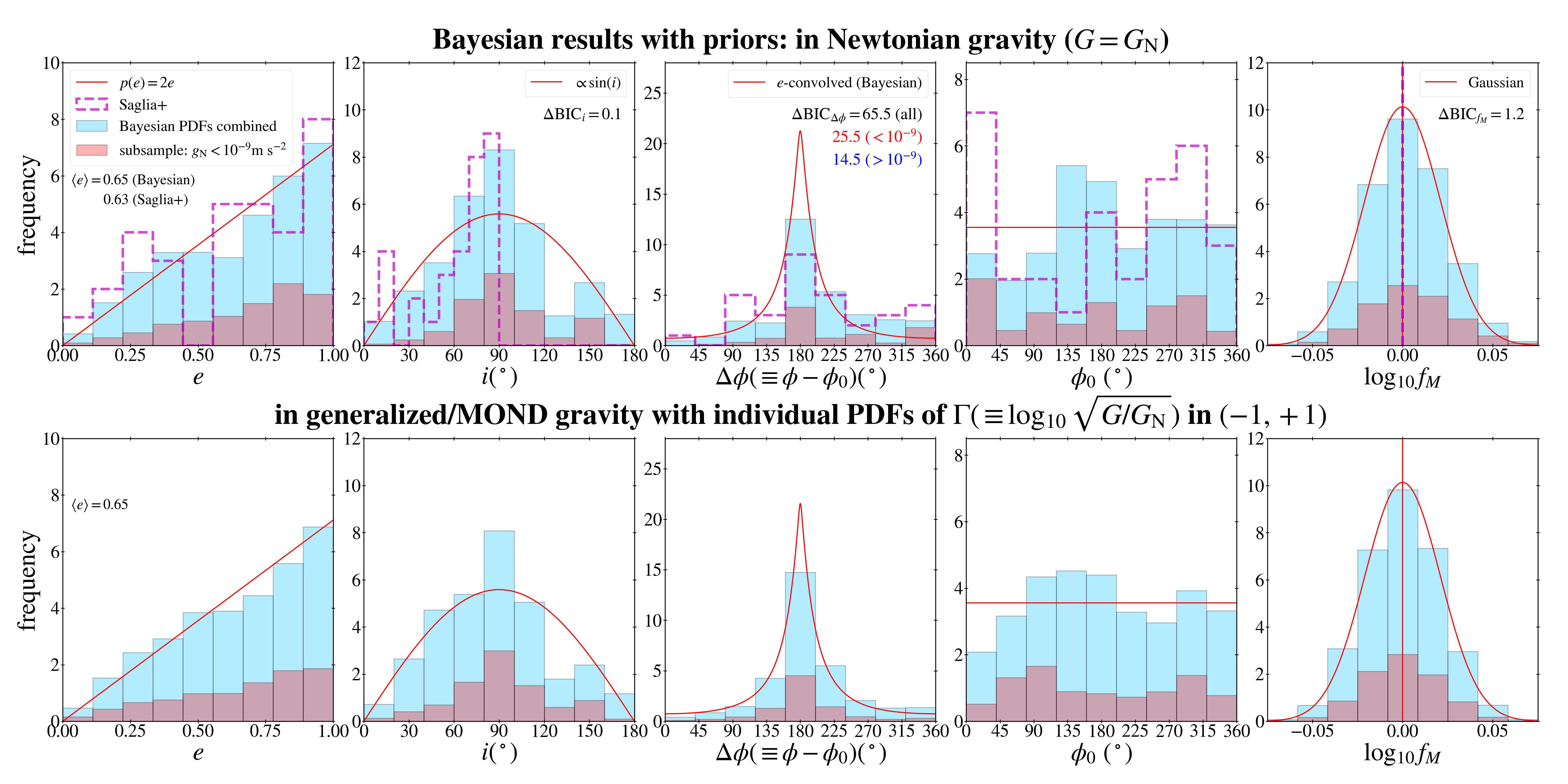}
    \caption{
    The upper and lower rows show distributions of the explicit orbit and inclination parameters as well as the mass parameter $f_M$ in Newtonian and generalized gravities. Each histogram represents the combination of all individual PDFs of the 32 binaries. The distribution of the phase $\Delta\phi(=\phi-\phi_0)$ in Newtonian gravity is in tension with the expectation in contrast to that in the generalized gravity, as the large value of $\Delta\rm{BIC}_{\Delta\phi}=\rm{BIC}_{\Delta\phi,{\rm Newton}}-\rm{BIC}_{\Delta\phi,{\rm general}}=65.3$ demonstrates. This tension is largely contributed by the subsample of 9 binaries with $g_{\rm N}<10^{-9}$~m~s$^{-2}$ (indicated by red color), which alone gives $\Delta\rm{BIC}=25.5$. The posterior distribution of $\log_{10}f_M$ in Newtonian gravity indicates a mild asymmetry due to a few systems (see the text). }
    \label{fig:distribution}
\end{figure*}

In this subsection the inferred parameters with or without priors are presented and their statistical properties are investigated in the context of testing gravity models. The results in Newtonian ($\Gamma=0$) and generalized (allowing $\Gamma$ to have its own PDF for each binary) gravities with priors are considered first, and the results in fixed gravity models without priors will follow. The two rows in Figure~\ref{fig:distribution} show the posterior distributions of the parameters (except for $\Gamma$ in the generalized gravity model) for the sample of 32 binaries. The upper and lower rows are respectively the results for the Newtonian and generalized gravity models. For each panel the colored histogram represents the combined distribution of all individual PDFs for the 32 binaries. It is thus the distribution of $32\times (2\times 10^6)$ values from the Bayesian outputs normalized so that the sum of the bar heights equals 32. 

For the Newtonian model, the values obtained by \cite{Saglia:2025} are also exhibited. The \cite{Saglia:2025} values are qualitatively in agreement with the Bayesian results. However, the \cite{Saglia:2025} ``best-fit'' values have larger scatters (with respect to the prior expectations) than the combined Bayesian distribution because the latter takes into account individual distributions themselves. The Bayesian distributions of $e$ and $\phi_0$ overall follow their prior distributions well in both gravity models. However, it does not appear that $\Delta\phi$ follows well the prior expectation that is obtained by convolving Equation~(\ref{eq:PrDelphi}) with the $e$-values shown in the leftmost column. This is due to some surplus near one of the tails that was already noted by \cite{Saglia:2025}. In the generalized gravity model, the Bayesian distribution of $\Delta\phi$ follows the prior distribution well. To estimate the statistical significance of the difference in the distribution of $\Delta\phi$ between the two gravity models, a likelihood-based comparison is considered in the following. A goodness-of-fit test of each distribution may not be so meaningful here because the Bayesian results were produced \emph{with} the priors. Such a test will be considered below for the case without priors.

The Bayesian results on $\Delta\phi$ can give the measurements $k=1,\cdots,n$ with the sorted quantities $\Delta\phi_k$ ($k=1,\cdots,n$) in the sample ($n=32$). The expectation value for the measurement $k$ is $n F(\Delta\phi_k)$ where 
\begin{equation}
    F(\Delta\phi_k)\equiv \int_0^{\Delta\phi_k} f_{e\text{-conv}}(z)dz
    \label{eq:cdf}
\end{equation}
 is the cumulative distribution function (CDF) of the $e$-convolved density function $f_{e\text{-conv}}(\Delta\phi)$ shown in the third column of Figure~\ref{fig:distribution}. The expected variance for each data point is $\sigma_k^2 = n F(\Delta\phi_k) (1 - F(\Delta\phi_k))$ from binomial statistics. The likelihood function $\mathcal{L}_{\Delta\phi}$ is then defined by 
\begin{equation}
   -2 \ln\mathcal{L}_{\Delta\phi} = \sum_{k=1}^n \frac{(k-n F(\Delta\phi_k))^2}{\sigma_k^2} + \sum_{k=1}^n\ln(2\pi\sigma_k^2)
   \label{eq:Ldelphi}
\end{equation}
The Bayesian information criterion (BIC) is given by ${\rm BIC}_{\Delta\phi}=-2 \ln\mathcal{L}_{\Delta\phi} +N_{\rm prmt}\ln n$ where the second factor on the right-hand side is irrelevant here because $f_{e\text{-conv}}(\Delta\phi)$ is fixed by the given distribution of $e$ values.

Usually, BIC is calculated using the maximum likelihood estimates (MLE) of the parameters. Following this convention, the sorted quantities $\Delta\phi_k$ may be obtained from the individual MLE values as described in Appendix~\ref{sec:MLEtables}. However, the MLE values are not likely to be optimal in the present case because the observational constraints are not sufficient (in particular, the radial separation between the two stars cannot be used for the MLE estimate: see Appendix~\ref{sec:MLEtables}). Thus, the sorted quantities $\Delta\phi_k$ are obtained from the combined distribution of all Bayesian PDFs of $\Delta\phi$, so that all posterior probabilities are taken into account. Specifically, they are derived as follows. From $2\times 10^6$ values of $\Delta\phi$ for each of the $n=32$ wide binaries we have a total of $64\times 10^6$ values for the sample. These values are sorted in the ascending order to define the combined PDF. When the numerical PDF is split into $n$ bins so that each bin has $2\times 10^6$ values, each bin occupies one occurrence in the ascending order and thus $\Delta\phi_k$ is defined by the maximum value in the $k$-th bin,\footnote{In the last bin, the maximum value can get arbitrarily close to the upper limit of the entire range of $\Delta\phi$ with $1 - F(\Delta\phi_n) \rightarrow 0$. Because there can arise a computer numerical error when $\sigma_k^2 = n F(\Delta\phi_k) (1 - F(\Delta\phi_k))$ is too close to zero, I use the 99.9th percentile value in each bin rather than the maximum value. This prescription has no impact on the likelihood but can bias the value of $A_n^2$ (Equation~(\ref{eq:ADstatistic})) just by $\approx -0.1$.} as one can verify with mock data generated with a PDF (the rightmost bottom panel provides such an example). BIC values are calculated for both the Newtonian and the generalized gravity results, and their difference $\Delta\rm{BIC}_{\Delta\phi}={\rm BIC}_{\Delta\phi,{\rm Newton}}-{\rm BIC}_{\Delta\phi,{\rm general}}$ is given in the top panel. The value of $\Delta{\rm BIC}_{\Delta\phi}=65.5$ indicates a very strong preference for the generalized gravity model based on the conventional criterion $\Delta\rm{BIC}>10$ \citep{Jeffreys:1939,Kass:1995}. If MLE values of $\Delta\phi_k$ are used, a somewhat larger value of $\Delta{\rm BIC}_{\Delta\phi}=83.9$ is obtained. If the \cite{Saglia:2025} best-fit values are used, we have a somewhat smaller value of $\Delta{\rm BIC}_{\Delta\phi}=39.1$ which is still very significant. Throughout I will use only the above standard estimates of $\Delta\phi_k$ based on the PDFs unless otherwise indicated.

Unlike $\Delta\phi$, similar procedures for $i$ and $\log_{10}f_M$ return, respectively, $\Delta{\rm BIC}_i=0.1$ and $\Delta{\rm BIC}_{f_M}=1.2$ meaning that two models are indistinguishable by these parameters as the histograms in the second and last columns of Figure~\ref{fig:distribution} indicate. However, it is worth noting that the posterior distribution of $\log_{10}f_M$ in the Newtonian model is mildly tilted to the positive direction due to a few systems (particularly one system) as will be described below in tables and in the next subsection. Because the prior on $\log_{10}f_M$ is concerned only with the observationally determined mass of each system independent of dynamics, the posterior distribution is expected to follow the prior distribution well, as is the case in the generalized gravity model. However, this is not the case (at least for one system) in the Newtonian model. 

To investigate the origin of the above large value of $\Delta\rm{BIC}_{\Delta\phi}$ in the context of testing gravity, two subsamples with $g_{\rm N}<10^{-9}$~m\,s$^{-2}$ (9 binaries) or $g_{\rm N}>10^{-9}$~m\,s$^{-2}$ (23 binaries) are considered. The subsamples have $\Delta\rm{BIC}_{\Delta\phi}=25.5$ ($<10^{-9}$) and $14.5$ ($>10^{-9}$). Although the low-acceleration subsample is much smaller than the higher-acceleration subsample, its $\Delta\rm{BIC}_{\Delta\phi}$ value is significantly larger. In detail, the low-acceleration subsample alone strongly disfavors Newton with $>10$. However, the higher-acceleration subsample also has a nominal value somewhat larger than 10. This may be interpreted as an indication that there may be some moderate problems with standard gravity near $10^{-9}$ (e.g., $10^{-8.5}<g_{\rm N}<10^{-9}$: see Figure~1 of \citealt{Chae:2024a}), and/or the flexibility of the generalized model with free $\Gamma$ accounts for the uncertainty of the individual mass compared with a fixed gravity model. I also note that the values of $\Delta\rm{BIC}_{\Delta\phi}$ for the two subsamples do not add up to the value for the total sample, meaning that an increase in sample size enhances statistical significance more steeply than linearly. This is in line with the generic Kolmogorov-Smirnov (K-S) statistics for CDF data.

\begin{figure*}[!htb]
    \centering
    \includegraphics[width=1.\linewidth]{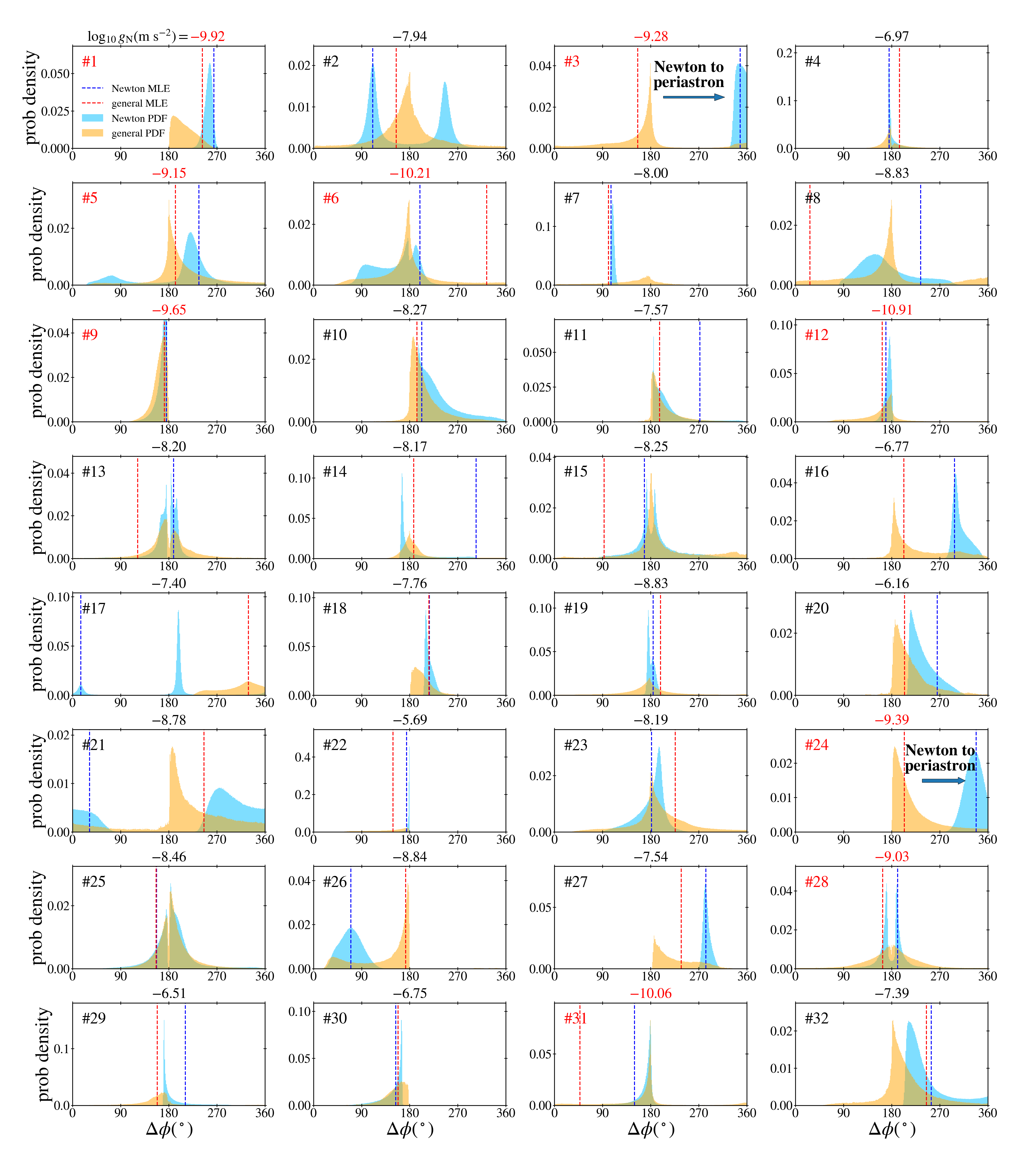}
    \caption{
    This figure exhibits the individual PDFs of $\Delta\phi$ for all 32 binaries from the Bayesian results in the Newtonian and generalized gravity models whose combined PDFs of $\Delta\phi$ are shown in the middle column of Figure~\ref{fig:distribution}. MLE values (see Appendix~\ref{sec:MLEtables}) are also indicated by dashed lines. The low-acceleration ($g_{\rm N}<10^{-9}$~m\,s$^{-2}$) wide binaries are marked by red numbers. Two ({\#}3 and {\#}24) from them have the PDFs of $\Delta\phi$ very close the periastron only in the Newtonian case.
    }
    \label{fig:distribution_delphi_individual}
\end{figure*}

Figure~\ref{fig:distribution_delphi_individual} exhibits the individual PDFs of $\Delta\phi$ to provide a deeper look at the difference in the inferred values of $\Delta\phi$ between the Newtonian and generalized gravity models. As expected, most PDFs stay away from, or are not localized at, the periastron of $\Delta\phi=0^\circ$ or $360^\circ$. There are a few exceptions. For two systems, Binary~\#3 and Binary~\#24, the Newtonian PDFs are well localized around the periastron while the generalized-gravity PDFs are not. These two binaries in the low-acceleration regime are largely responsible for the $\Delta\rm{BIC}_{\Delta\phi}$ values for both the low-acceleration subsample and the entire sample. I note that the two binaries do not allow degenerate Newtonian PDFs away from the periastron. In contrast, for several systems, Newtonian PDFs are doubly peaked (in particular, Binary~\#2, \#5, \#17, and \#21). For example, Binary~\#17 has the major peak near the apastron and a minor peak near the periastron. Although not shown in the figure, Binary~\#30 turns out to be a peculiar case with two degenerate solutions. It has a second solution narrowly localized around the periastron whose maximum likelihood is only slightly lower than that of the solution shown in the figure. However, unlike other cases with well-separated double peaks, the MCMC procedure returns only one solution depending on the starting value, and thus the shown solution is chosen manually considering that the second solution is unlikely (see Table~\ref{tab:prmt_newton} below). Binary~\#3 or Binary~\#24 does not show this kind of peculiarity. Moreover, as will be shown below, for Binary~\#24 the Newtonian solution does not provide an acceptable fit to the observed velocities and mass, on top of the concern that $\Delta\phi$ is near the periastron. 

\begin{figure*}[!htb]
    \centering
    \includegraphics[width=1.\linewidth]{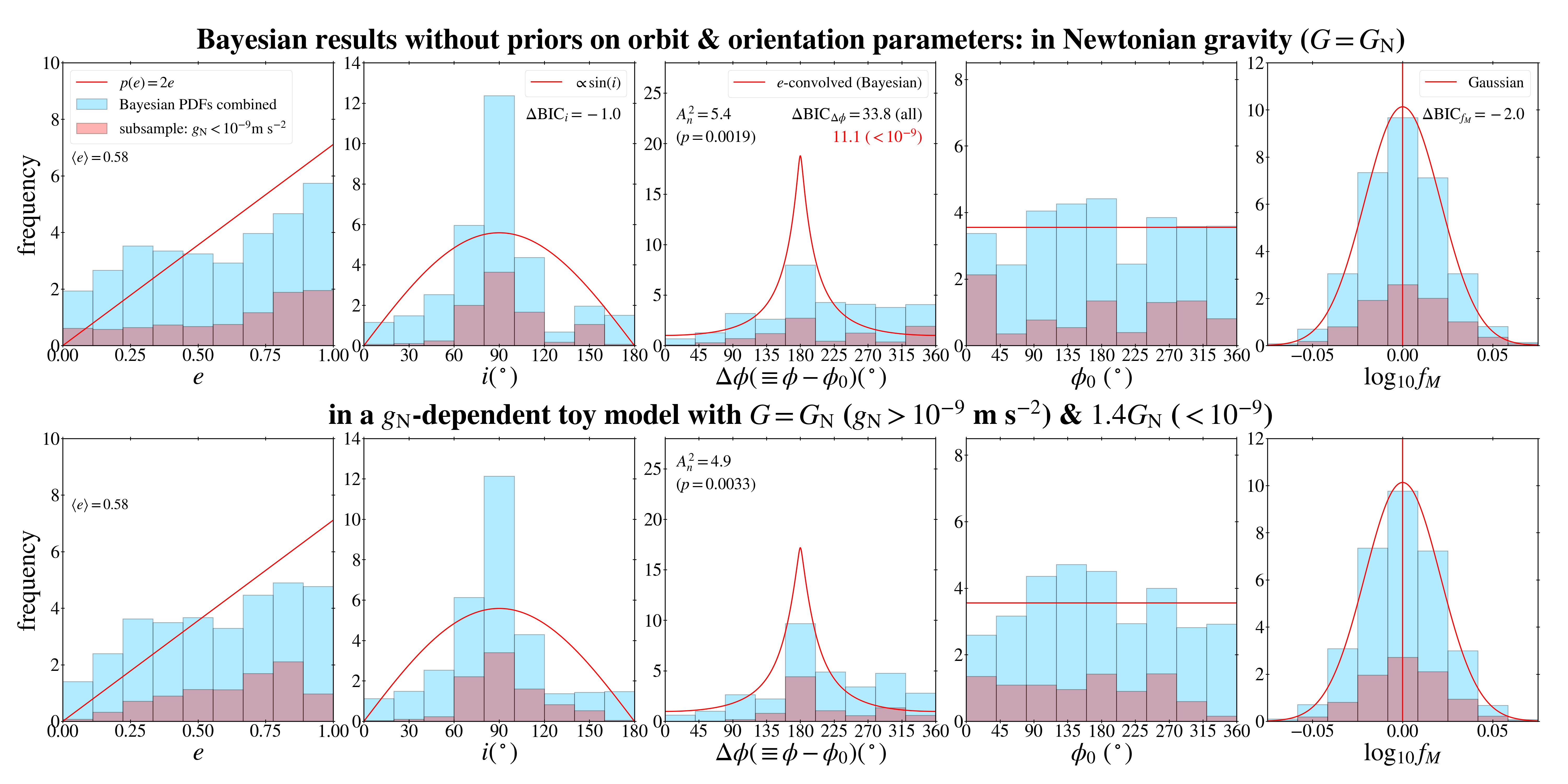}
    \caption{
    Similar to Figure~\ref{fig:distribution}, but for the Bayesian results with flat/no priors on $e$, $i$, and $\Delta\phi$. Here the generalized (i.e., allowing each wide binary to have its own PDF of $\Gamma$) gravity model of Figure~\ref{fig:distribution} is replaced by the MOND-type toy model where $G$ is fixed for each binary, but it has a value of $G_{\rm N}$ or $1.4G_{\rm N}$ depending on $g_{\rm N}$ (see the text for details). Thus, in this case the parameters $e$, $i$, and $\Delta\phi$ are determined by the data of $\mathbf{r}$ and $\mathbf{v}$ alone.
    }
    \label{fig:distribution_nopriors}
\end{figure*}

The above analyses are based on the distributions shown in Figure~\ref{fig:distribution} that were derived with the priors imposed on the parameters. It is of interest to consider Bayesian results without priors on $e$, $i$, and $\Delta\phi$ to check the roles of the priors and perform goodness-of-fit tests of the distributions of $\Delta\phi$. For this, the generalized gravity model that allows $\Gamma$ to be free for each system is replaced by the MOND-type toy model defined above in which $G$ is fixed for each binary. This is necessary because the degeneracy between $\Gamma$ and the orbit and orientation parameters requires that either $\Gamma$ is fixed or the priors are imposed on the parameters.  

Figure~\ref{fig:distribution_nopriors} shows the distributions of the parameters in the Newtonian and MOND-type toy models without priors on $e$, $i$, and $\Delta\phi$. The eccentricity distributions are somewhat flatter than the thermal distribution, but the frequency increases with $e$. The inclination distributions show some overabundance near the edge-on orientation. However, there is no distinction between the two models, as the value of $\Delta{\rm BIC}_i=-1.0$ indicates. For the distributions of $\Delta\phi$, $\Delta{\rm BIC}_{\Delta\phi}=33.8$ is obtained for the total sample and $11.1$ for the subsample with $g_{\rm N}<10^{-9}$~m\,s$^{-2}$. Because the modeling results for the 23 binaries with $g_{\rm N}>10^{-9}$~m\,s$^{-2}$ are identical between the Newtonian and MOND-type hybrid models (thus $\Delta{\rm BIC}_{\Delta\phi}=0$), only the result for $g_{\rm N}<10^{-9}$~m\,s$^{-2}$ is truly meaningful here. The result $\Delta{\rm BIC}_{\Delta\phi}> 10$ for $g_{\rm N}<10^{-9}$~m\,s$^{-2}$ indicates that Newtonian gravity is strongly disfavored in the low-acceleration regime even compared with the toy model with boosted gravity $G=1.4 G_{\rm N}$.

To further test Newtonian gravity in an absolute sense, the Anderson-Darling \citep[A-D; ][]{AndersonDarling:1954} test is considered here since it is more sensitive to the data at the tails than the K-S test. The A-D statistic is
\begin{equation}
    A_n^2 = -n -\frac{1}{n}\sum_{k=1}^n(2k-1)[\ln F(\Delta\phi_k)+\ln(1-F(\Delta\phi_{n-k+1})],
    \label{eq:ADstatistic}
\end{equation}
where $F(\Delta\phi)$ is the CDF given by Equation~(\ref{eq:cdf}). The Newtonian model has $A_n^2=5.4$ with a $p$-value \citep{Marsaglia:2004} of $1.9\times 10^{-3}$ (or $\approx 3.1\sigma$ significance) which means a significant discrepancy with the $e$-convolved distribution $f_{e\text{-conv}}(\Delta\phi)$. The MOND-type toy model has $A_n^2=4.9$ with a $p$-value of $3.3\times 10^{-3}$ ($\approx 2.9\sigma$ significance). Based on a contemporary criterion of $p<5\times 10^{-3}$ \citep{Wasserstein:2016,Wasserstein:2019} or $3\sigma$ for significance, the Newtonian model is significantly discrepant with the prior expectation while the significance is relatively weak for the MOND-type toy model although it appears not to be a good model either. Thus, in line with the model comparison with $\Delta{\rm BIC}_{\Delta\phi}$, Newtonian gravity has a significant discrepancy with the data in an absolute sense, although the discrepancy is not a $>5\sigma$ level given the small sample size. This pilot analysis of the Bayesian inferred parameters demonstrates that a much larger sample can more decisively test gravity in the future.  

\begin{figure*}[!htb]
    \centering
    \includegraphics[width=1.\linewidth]{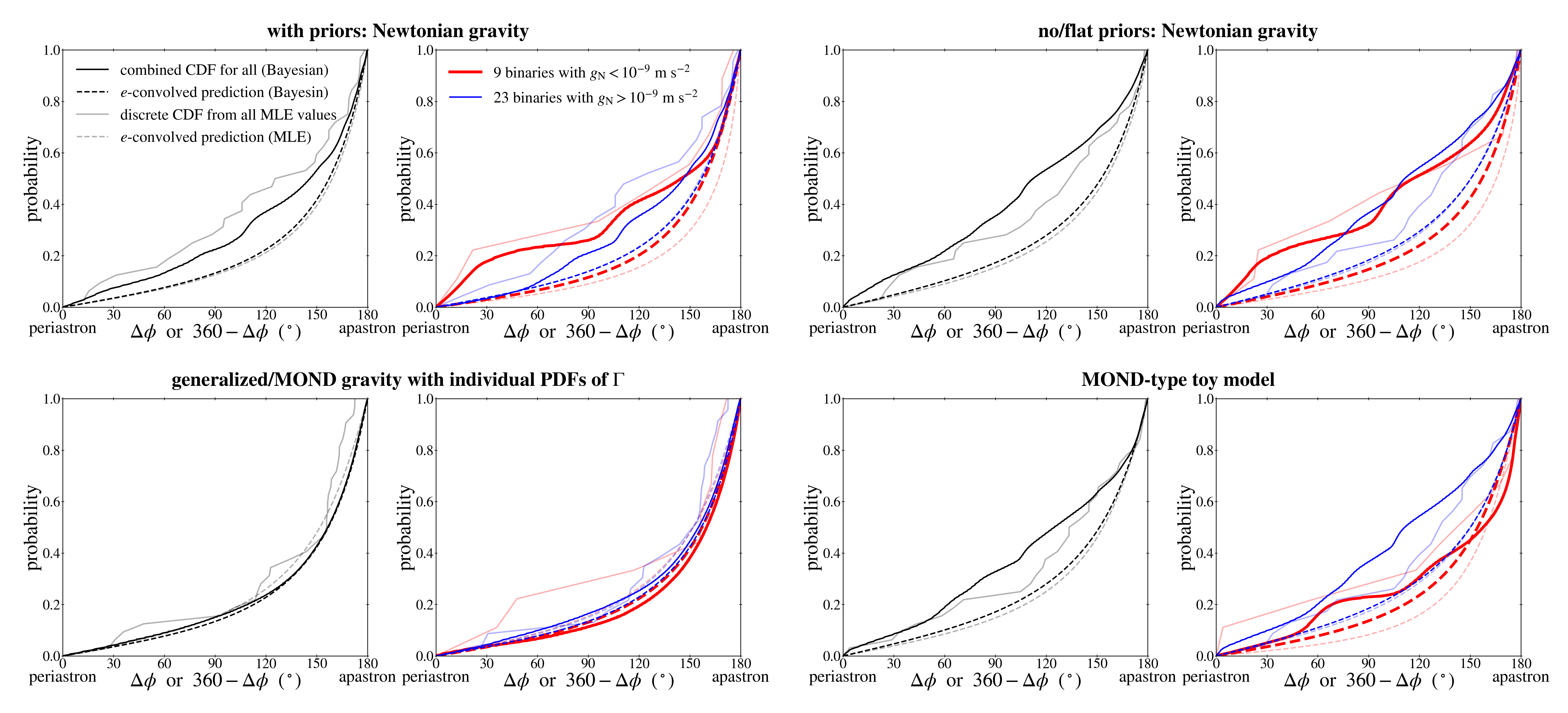}
    \caption{
    CDFs of $\Delta\phi$ are displayed for various modeling results. Considering that the expected distribution of $\Delta\phi$ is symmetrical about $180^\circ$, $360^\circ-\Delta\phi$ is shown for $\Delta\phi>180^\circ$. The panels in the left/right side show the CDFs from the modeling results with/without priors on $e$, $i$, and $\Delta\phi$. Solid curves represent the inferred CDFs while the dashed curves are the $e$-convolved expectations based on Equation~(\ref{eq:PrDelphi}). Red and blue curves show the CDFs for the low-acceleration and higher-acceleration subsamples. Only the CDFs shown in the bottom left panels (the case of generalized gravity with priors) have overall good agreement between the solid and dashed curves.
    }
    \label{fig:cdf_delphi}
\end{figure*}

The difference in $\Delta\phi$ between Newtonian gravity and the generalized or MOND-type gravity is further illustrated in Figure~\ref{fig:cdf_delphi}. The inferred CDFs of $\Delta\phi$ are compared with their $e$-convolved predictions based on Equation~(\ref{eq:PrDelphi}). The upper left panels show clearly that the Newton-inferred (with priors imposed on $e$, $i$, and $\Delta\phi$) CDF for the total sample deviates systematically from the $e$-convolved prediction, and this deviation is largely contributed by the CDF (highlighted by the thick red curve) for the low-acceleration subsample. In contrast, the lower left panels show that generalized gravity-inferred CDF matches reasonably well its $e$-convolved prediction. The panels in the right side show the results in Newtonian and MOND-type hybrid models with no/flat priors on $e$, $i$, and $\Delta\phi$. Compared with the corresponding results with priors shown in the left side, the deviations are larger overall. Nevertheless, there is a clear distinction for the low-acceleration sample between Newtonian gravity and the toy model with $G=1.4G_{\rm N}$, which is responsible for $\Delta{\rm BIC}_{\Delta\phi} > 10$ shown in Figure~\ref{fig:distribution_nopriors}.

\begin{figure}[!htb]
    \centering
    \includegraphics[width=1.\linewidth]{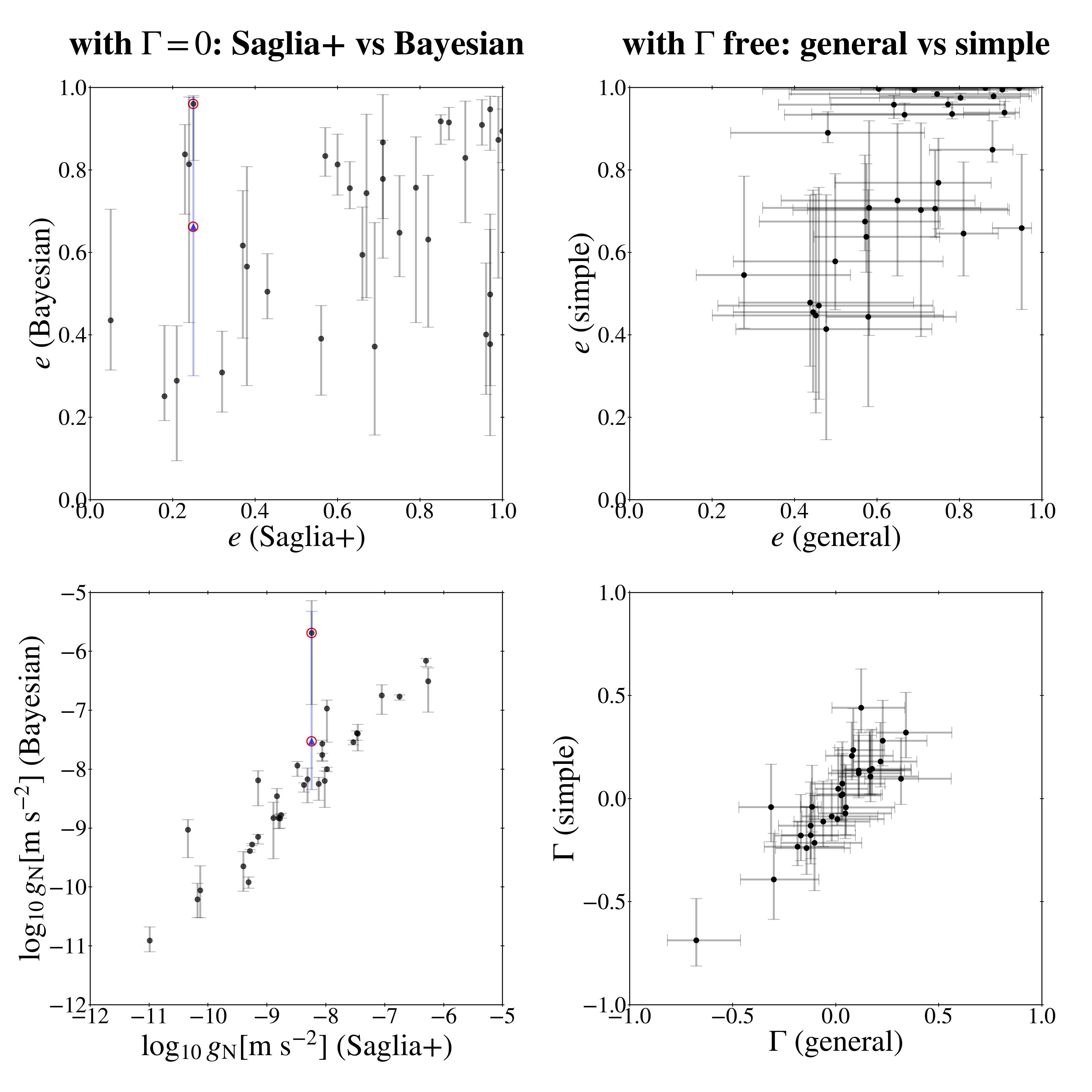}
    \caption{
    The left panels show eccentricity ($e$) and Newtonian gravitational acceleration ($g_{\rm N}$) derived in Newtonian gravity. The Bayesian results with priors are compared with the \cite{Saglia:2025} numerical solutions. The one marked with red circle (binary {\#22}: Stars 36OphB and GJ663A) exhibits a dramatic discrepancy between the two. The discrepancy is significantly reduced for the Bayesian result with no/flat priors as indicated by the blue triangle. However, the latter model overall has much less likely parameter values (see the text). The right panels show $e$ and $\Gamma$ (Equation~(\ref{eq:Gamma})) in generalized gravity. The results from the general Bayesian methodology are compared with those from the simple Bayesian methodology of \cite{Chae:2025}.
    }
    \label{fig:compare}
\end{figure}

All fitted parameters and some derived parameters for Newtonian and generalized gravities are listed in Tables~\ref{tab:prmt_newton} and \ref{tab:prmt_general}. In not only generalized gravity but also Newtonian gravity, the fitted parameters in general have broad ranges due to the measurement uncertainties. Different estimates of eccentricity and gravity parameters are compared in Figure~\ref{fig:compare}. The \cite{Saglia:2025} values and the Bayesian results of $e$ and $\log_{10}g_{\rm N}$ are compared in the left panels of Figure~\ref{fig:compare}. Here the Bayesian results refer to those obtained with priors on $e$, $i$, and $\Delta\phi$. There are reasonable agreements between the two calculations, except for one system {\#}22, which is marked by a red circle in the left panels: Stars 36OphB and GJ663A. Without system {\#}22, the Pearson correlation coefficient and the $p$-value between the \cite{Saglia:2025} values and the Bayesian medians are $0.43$ and $1.5\times 10^{-2}$ for $e$, and $0.94$ and $2.7\times 10^{-15}$ for $\log_{10}g_{\rm N}$. The binary system {\#}22 has a small projected separation of $s=0.030$kau ($0.00014$pc), perhaps similar to the Sun-Halley system. \cite{Saglia:2025} obtain $r\approx a\approx 1.2$~kau that is about 40 times $s$, an unusually large value. Consequently, the predicted Newtonian gravity has a low value of $\log_{10} g_{\rm N} (\text{m\,s}^{-2})=-8.24$. On the other hand, the Bayesian output with priors predicts $r=0.07_{-0.03}^{+0.20}$kau, $a=0.03_{-0.02}^{+0.11}$, and $\log_{10} g_{\rm N} (\text{m\,s}^{-2})=-5.7_{-1.2}^{+0.6}$. The dramatic difference stems from the fact that \cite{Saglia:2025} predict a low eccentricity of $e=0.25$ while this work predicts $e=0.96_{-0.14}^{+0.02}$ similar to that of Halley's comet. However, the Bayesian output with no/flat priors on $e$, $i$, and $\Delta\phi$ predicts $e=0.66_{-0.36}^{+0.31}$, $r=0.54_{-0.50}^{+0.85}$kau, $a=0.34_{-0.32}^{+1.12}$, and $\log_{10} g_{\rm N} (\text{m\,s}^{-2})=-7.5_{-0.8}^{+2.2}$. Because these ranges of the parameters are so broad, they are consistent with both the Bayesian results with priors and the \cite{Saglia:2025} values. However, the median values of $a$ and $r$ are perhaps too large compared with $s=0.030$kau. 

\begin{sidewaystable*}
\caption{\small{Summary of Fitted and Derived Parameters in Newtonian Gravity ($\Gamma=0$) for the sample of \cite{Saglia:2025}}}\label{tab:prmt_newton}
\scriptsize
\begin{center}
  \begin{tabular}{llccccccccccccccccc}
  \hline
 \# & Star$^a$ A/B & $r$ & $a$ &  $e$ & $i$ & $\phi_0$ & $\phi-\phi_0$ & $\log_{10} f_M$ & $\log_{10} g_{\rm N}$ & $v_{\rm obs}$ & $v_{\rm mod}$ & $v_{\rm escN}$   \\
 & & (kau) & (kau) &   & $(^\circ)$ & $(^\circ)$ & $(^\circ)$ &  & $(\text{m s}^{-2})$ & $(\text{km s}^{-1})$  & $(\text{km s}^{-1})$ & $(\text{km s}^{-1})$ &   \\
 \hline
1 & HD4552/BD+120090 & $10.47_{-0.98}^{+1.26}$ & $21.49_{-8.09}^{+19.49}$ & $0.78_{-0.10}^{+0.09}$ & $66.2_{-2.6}^{+2.3}$ & $28_{-7}^{+9}$ & $255_{-8}^{+6}$  & $0.005_{-0.021}^{+0.021}$ & $-9.92_{-0.10}^{+0.09}$  & $0.569\pm 0.015$  & $0.53_{-0.03}^{+0.03}$ & $0.61_{-0.03}^{+0.03}$  \\ 
2 & HIP6130/GJ56.3B$^c$ & $0.84_{-0.06}^{+0.20}$ & $0.86_{-0.14}^{+0.52}$ & $0.38_{-0.22}^{+0.28}$ & $109.8_{-0.9}^{+0.8}$ & $226_{-150}^{+66}$ & $142_{-36}^{+107}$  & $0.000_{-0.021}^{+0.021}$ & $-7.94_{-0.18}^{+0.07}$  & $1.210\pm 0.010$  & $1.21_{-0.04}^{+0.04}$ & $1.69_{-0.16}^{+0.08}$  \\ 
3 & HD11584/CD-50524$^b$ & $5.01_{-0.06}^{+0.12}$ & $12.22_{-2.62}^{+4.91}$ & $0.59_{-0.11}^{+0.12}$ & $145.0_{-2.4}^{+2.5}$ & $175_{-3}^{+2}$ & $347_{-8}^{+8}$  & $0.004_{-0.020}^{+0.020}$ & $-9.28_{-0.02}^{+0.02}$  & $0.799\pm 0.008$  & $0.79_{-0.02}^{+0.02}$ & $0.89_{-0.02}^{+0.02}$  \\ 
4 & HD13357A/HD13357B & $0.31_{-0.05}^{+0.28}$ & $0.17_{-0.03}^{+0.20}$ & $0.83_{-0.16}^{+0.14}$ & $88.4_{-2.1}^{+0.9}$ & $161_{-57}^{+46}$ & $177_{-2}^{+15}$  & $-0.001_{-0.021}^{+0.021}$ & $-6.97_{-0.57}^{+0.14}$  & $1.026\pm 0.011$  & $1.03_{-0.03}^{+0.03}$ & $3.14_{-0.87}^{+0.27}$  \\ 
5 & HR1010/HD20766$^b$ & $3.90_{-0.16}^{+0.59}$ & $3.78_{-0.56}^{+1.22}$ & $0.37_{-0.21}^{+0.30}$ & $85.8_{-1.2}^{+0.9}$ & $132_{-24}^{+164}$ & $220_{-119}^{+26}$  & $0.000_{-0.021}^{+0.021}$ & $-9.15_{-0.12}^{+0.04}$  & $0.635\pm 0.025$  & $0.62_{-0.04}^{+0.04}$ & $0.91_{-0.06}^{+0.03}$  \\ 
6 & HD20782/HD20781$^b$ & $13.37_{-3.56}^{+5.66}$ & $10.98_{-3.95}^{+9.67}$ & $0.43_{-0.12}^{+0.27}$ & $89.8_{-0.5}^{+0.5}$ & $255_{-70}^{+58}$ & $161_{-57}^{+31}$  & $-0.000_{-0.021}^{+0.021}$ & $-10.21_{-0.31}^{+0.27}$  & $0.321\pm 0.030$  & $0.30_{-0.04}^{+0.04}$ & $0.49_{-0.08}^{+0.08}$  \\ 
7 & HD21209A/HD21209B & $0.82_{-0.03}^{+0.04}$ & $3.58_{-1.29}^{+3.07}$ & $0.92_{-0.04}^{+0.04}$ & $72.5_{-0.6}^{+0.7}$ & $330_{-3}^{+2}$ & $109_{-2}^{+3}$  & $0.014_{-0.019}^{+0.020}$ & $-8.00_{-0.04}^{+0.03}$  & $1.518\pm 0.009$  & $1.48_{-0.03}^{+0.03}$ & $1.57_{-0.04}^{+0.04}$  \\ 
8 & HIP17414/HIP17405 & $1.99_{-0.14}^{+0.47}$ & $1.70_{-0.21}^{+0.79}$ & $0.31_{-0.10}^{+0.10}$ & $37.2_{-5.7}^{+10.1}$ & $214_{-115}^{+85}$ & $157_{-35}^{+62}$  & $-0.000_{-0.021}^{+0.021}$ & $-8.83_{-0.18}^{+0.06}$  & $0.604\pm 0.003$  & $0.60_{-0.02}^{+0.02}$ & $0.94_{-0.09}^{+0.04}$  \\ 
9 & HD24062/HD24085 & $7.69_{-1.91}^{+4.73}$ & $4.82_{-1.45}^{+4.39}$ & $0.95_{-0.10}^{+0.03}$ & $68.9_{-21.5}^{+12.1}$ & $304_{-12}^{+4}$ & $171_{-11}^{+4}$  & $-0.001_{-0.021}^{+0.021}$ & $-9.65_{-0.42}^{+0.25}$  & $0.324\pm 0.010$  & $0.32_{-0.02}^{+0.03}$ & $0.72_{-0.15}^{+0.11}$  \\ 
10 & HD26923/HD26913$^b$ & $1.49_{-0.06}^{+0.23}$ & $1.60_{-0.18}^{+0.61}$ & $0.74_{-0.25}^{+0.19}$ & $72.2_{-16.7}^{+6.9}$ & $319_{-21}^{+7}$ & $223_{-21}^{+49}$  & $0.001_{-0.021}^{+0.021}$ & $-8.27_{-0.12}^{+0.04}$  & $1.133\pm 0.010$  & $1.13_{-0.02}^{+0.02}$ & $1.55_{-0.10}^{+0.06}$  \\ 
11 & HD29167/HD29167B & $0.66_{-0.04}^{+0.28}$ & $0.49_{-0.04}^{+0.36}$ & $0.81_{-0.38}^{+0.16}$ & $78.1_{-24.5}^{+6.3}$ & $172_{-25}^{+8}$ & $202_{-14}^{+31}$  & $-0.001_{-0.021}^{+0.021}$ & $-7.57_{-0.30}^{+0.06}$  & $1.310\pm 0.012$  & $1.31_{-0.02}^{+0.02}$ & $2.31_{-0.35}^{+0.11}$  \\ 
12 & HIP33705/HIP33691 & $31.14_{-7.11}^{+7.44}$ & $17.75_{-4.50}^{+5.28}$ & $0.83_{-0.05}^{+0.07}$ & $96.2_{-1.9}^{+2.3}$ & $254_{-10}^{+13}$ & $174_{-7}^{+4}$  & $-0.001_{-0.021}^{+0.021}$ & $-10.91_{-0.19}^{+0.23}$  & $0.117\pm 0.004$  & $0.11_{-0.03}^{+0.03}$ & $0.34_{-0.04}^{+0.05}$  \\ 
13 & HIP47839/HIP47836$^b$ & $1.35_{-0.24}^{+0.91}$ & $0.87_{-0.20}^{+0.92}$ & $0.87_{-0.34}^{+0.10}$ & $85.9_{-22.2}^{+26.7}$ & $148_{-117}^{+88}$ & $177_{-16}^{+17}$  & $-0.001_{-0.021}^{+0.021}$ & $-8.20_{-0.45}^{+0.17}$  & $0.700\pm 0.098$  & $0.71_{-0.03}^{+0.09}$ & $1.58_{-0.36}^{+0.16}$  \\ 
14 & HD96273/AG+071510 & $1.31_{-0.26}^{+0.78}$ & $0.90_{-0.23}^{+0.93}$ & $0.50_{-0.22}^{+0.20}$ & $98.3_{-3.1}^{+3.5}$ & $149_{-57}^{+23}$ & $169_{-4}^{+51}$  & $-0.000_{-0.021}^{+0.021}$ & $-8.17_{-0.40}^{+0.19}$  & $0.842\pm 0.013$  & $0.84_{-0.03}^{+0.03}$ & $1.62_{-0.34}^{+0.19}$  \\ 
15 & HD104760/HD104759 & $1.42_{-0.18}^{+0.54}$ & $1.09_{-0.21}^{+0.78}$ & $0.39_{-0.14}^{+0.08}$ & $115.3_{-86.3}^{+38.0}$ & $219_{-142}^{+76}$ & $185_{-24}^{+41}$  & $0.000_{-0.021}^{+0.021}$ & $-8.25_{-0.28}^{+0.11}$  & $0.882\pm 0.096$  & $0.88_{-0.03}^{+0.09}$ & $1.55_{-0.23}^{+0.11}$  \\ 
16 & HD106515A/HD106515B & $0.24_{-0.01}^{+0.02}$ & $0.43_{-0.07}^{+0.16}$ & $0.52_{-0.07}^{+0.10}$ & $159.7_{-10.8}^{+5.9}$ & $245_{-70}^{+18}$ & $304_{-8}^{+23}$  & $0.003_{-0.021}^{+0.021}$ & $-6.77_{-0.07}^{+0.04}$  & $2.954\pm 0.009$  & $2.95_{-0.01}^{+0.01}$ & $3.48_{-0.14}^{+0.11}$  \\ 
17 & HD133131B/HD133131A$^b$ & $0.51_{-0.09}^{+0.20}$ & $0.41_{-0.10}^{+0.36}$ & $0.29_{-0.19}^{+0.13}$ & $47.6_{-13.0}^{+13.7}$ & $216_{-145}^{+12}$ & $197_{-177}^{+5}$  & $-0.001_{-0.021}^{+0.021}$ & $-7.40_{-0.29}^{+0.16}$  & $1.522\pm 0.012$  & $1.52_{-0.02}^{+0.02}$ & $2.45_{-0.38}^{+0.24}$  \\ 
18 & HD135101A/HD135101B & $0.77_{-0.01}^{+0.09}$ & $1.12_{-0.13}^{+0.45}$ & $0.92_{-0.06}^{+0.02}$ & $35.0_{-15.4}^{+14.6}$ & $299_{-104}^{+26}$ & $214_{-5}^{+10}$  & $0.002_{-0.021}^{+0.021}$ & $-7.76_{-0.09}^{+0.03}$  & $1.624\pm 0.006$  & $1.62_{-0.01}^{+0.01}$ & $2.01_{-0.10}^{+0.06}$  \\ 
19 & HD143336/HD143336B & $2.88_{-0.79}^{+3.50}$ & $1.60_{-0.47}^{+2.48}$ & $0.91_{-0.05}^{+0.06}$ & $91.0_{-3.4}^{+4.1}$ & $173_{-67}^{+66}$ & $179_{-4}^{+9}$  & $-0.001_{-0.021}^{+0.021}$ & $-8.83_{-0.69}^{+0.27}$  & $0.352\pm 0.015$  & $0.35_{-0.04}^{+0.04}$ & $1.13_{-0.37}^{+0.19}$  \\ 
20 & HD147723/HD147722$^b$ & $0.14_{-0.01}^{+0.02}$ & $0.21_{-0.03}^{+0.11}$ & $0.81_{-0.07}^{+0.07}$ & $61.9_{-11.6}^{+8.9}$ & $144_{-16}^{+4}$ & $234_{-16}^{+32}$  & $0.002_{-0.021}^{+0.021}$ & $-6.16_{-0.10}^{+0.04}$  & $4.387\pm 0.008$  & $4.39_{-0.01}^{+0.01}$ & $5.34_{-0.30}^{+0.19}$  \\ 
21 & HD154363/GJ654 & $1.97_{-0.04}^{+0.13}$ & $2.82_{-0.48}^{+0.86}$ & $0.40_{-0.14}^{+0.17}$ & $109.0_{-1.5}^{+1.8}$ & $62_{-38}^{+37}$ & $-49_{-42}^{+67}$  & $0.001_{-0.021}^{+0.021}$ & $-8.78_{-0.05}^{+0.03}$  & $0.801\pm 0.008$  & $0.79_{-0.04}^{+0.04}$ & $0.99_{-0.03}^{+0.03}$  \\ 
22 & 36OphB/GJ663A$^b$ & $0.07_{-0.03}^{+0.20}$ & $0.03_{-0.02}^{+0.11}$ & $0.96_{-0.14}^{+0.02}$ & $116.6_{-20.5}^{+34.2}$ & $277_{-8}^{+2}$ & $178_{-2}^{+0}$  & $-0.001_{-0.021}^{+0.021}$ & $-5.69_{-1.21}^{+0.55}$  & $0.980\pm 0.004$  & $0.98_{-0.01}^{+0.01}$ & $6.27_{-3.14}^{+2.30}$  \\ 
23 & HD177730/BD-104948B & $1.35_{-0.23}^{+0.86}$ & $0.89_{-0.19}^{+0.93}$ & $0.65_{-0.11}^{+0.14}$ & $55.4_{-6.5}^{+11.7}$ & $177_{-84}^{+86}$ & $187_{-33}^{+13}$  & $-0.001_{-0.021}^{+0.021}$ & $-8.19_{-0.43}^{+0.16}$  & $0.797\pm 0.012$  & $0.80_{-0.03}^{+0.03}$ & $1.62_{-0.35}^{+0.16}$  \\ 
24 & HD189739/HD189760 & $5.81_{-0.03}^{+0.09}$ & $51.21_{-21.33}^{+52.12}$ & $0.89_{-0.08}^{+0.05}$ & $115.9_{-1.2}^{+1.4}$ & $25_{-6}^{+7}$ & $333_{-17}^{+15}$  & $0.035_{-0.019}^{+0.019}$ & $-9.39_{-0.02}^{+0.02}$  & $0.945\pm 0.014$  & $0.82_{-0.02}^{+0.02}$ & $0.85_{-0.02}^{+0.02}$  \\ 
25 & CPD-5211628/CPD-5211628A$^d$ & $1.98_{-0.29}^{+1.03}$ & $1.39_{-0.26}^{+1.27}$ & $0.87_{-0.28}^{+0.12}$ & $86.3_{-13.6}^{+15.5}$ & $191_{-22}^{+148}$ & $-141_{-29}^{+308}$  & $-0.001_{-0.021}^{+0.021}$ & $-8.46_{-0.36}^{+0.13}$  & $0.766\pm 0.029$  & $0.77_{-0.03}^{+0.03}$ & $1.43_{-0.27}^{+0.12}$  \\ 
26 & HD192344/HD192343$^b$ & $2.94_{-0.19}^{+0.61}$ & $3.38_{-0.49}^{+2.13}$ & $0.25_{-0.06}^{+0.17}$ & $157.9_{-17.5}^{+10.7}$ & $181_{-56}^{+39}$ & $69_{-21}^{+21}$  & $0.001_{-0.021}^{+0.021}$ & $-8.84_{-0.16}^{+0.06}$  & $0.845\pm 0.010$  & $0.84_{-0.01}^{+0.01}$ & $1.12_{-0.10}^{+0.05}$  \\ 
27 & HD196067/HD196068 & $0.68_{-0.01}^{+0.04}$ & $1.90_{-0.40}^{+0.94}$ & $0.76_{-0.05}^{+0.06}$ & $14.7_{-9.8}^{+12.8}$ & $254_{-120}^{+54}$ & $284_{-5}^{+8}$  & $0.007_{-0.020}^{+0.020}$ & $-7.54_{-0.05}^{+0.03}$  & $2.200\pm 0.005$  & $2.20_{-0.01}^{+0.01}$ & $2.43_{-0.08}^{+0.07}$  \\ 
28 & WASP-94A/WASP-94B & $3.95_{-0.71}^{+2.87}$ & $2.65_{-0.67}^{+2.90}$ & $0.76_{-0.33}^{+0.12}$ & $83.3_{-36.0}^{+49.5}$ & $185_{-138}^{+129}$ & $165_{-353}^{+27}$  & $-0.000_{-0.021}^{+0.021}$ & $-9.03_{-0.47}^{+0.17}$  & $0.473\pm 0.144$  & $0.46_{-0.05}^{+0.12}$ & $1.05_{-0.25}^{+0.11}$  \\ 
29 & HD201247A/HD201247B & $0.19_{-0.04}^{+0.16}$ & $0.12_{-0.03}^{+0.15}$ & $0.63_{-0.21}^{+0.16}$ & $75.6_{-7.2}^{+6.3}$ & $318_{-51}^{+29}$ & $176_{-4}^{+29}$  & $-0.001_{-0.021}^{+0.021}$ & $-6.51_{-0.52}^{+0.23}$  & $1.834\pm 0.020$  & $1.83_{-0.03}^{+0.03}$ & $4.20_{-1.08}^{+0.59}$  \\ 
30 & HD201796A/HD201796B$^e$ & $0.26_{-0.05}^{+0.12}$ &	$0.21_{-0.05}^{+0.20}$ & $0.84_{-0.15}^{+0.07}$ & $126.6_{-18.2}^{+30.0}$ & $150_{-25}^{+18}$ & $156_{-27}^{+9}$  & $0.000_{-0.021}^{+0.021}$ & $-6.75_{-0.32}^{+0.18}$  & $2.245\pm 0.009$ &	$2.24_{-0.01}^{+0.01}$ & $3.70_{-0.63}^{+0.40}$ \\
31 & HD206429/HD206428 & $12.04_{-4.64}^{+8.52}$ & $7.57_{-3.34}^{+8.22}$ & $0.62_{-0.23}^{+0.13}$ & $76.6_{-33.5}^{+59.3}$ & $93_{-15}^{+35}$ & $175_{-17}^{+5}$  & $-0.001_{-0.021}^{+0.021}$ & $-10.06_{-0.46}^{+0.42}$  & $0.251\pm 0.006$  & $0.25_{-0.01}^{+0.01}$ & $0.56_{-0.13}^{+0.15}$  \\ 
32 & HD221550A/HD221550B & $0.52_{-0.01}^{+0.08}$ & $0.50_{-0.04}^{+0.17}$ & $0.57_{-0.29}^{+0.24}$ & $98.4_{-2.3}^{+3.5}$ & $138_{-24}^{+6}$ & $229_{-17}^{+49}$  & $0.000_{-0.021}^{+0.021}$ & $-7.39_{-0.12}^{+0.04}$  & $1.731\pm 0.022$  & $1.73_{-0.04}^{+0.04}$ & $2.50_{-0.16}^{+0.09}$  \\ 
\hline
\end{tabular}
\end{center}
$^a$ Gaia DR3 identifications can be found in the data file given at Zenodo\citep{Chae:Zenodo2025}. $^b$ In these systems, the ordering of stars A and B is opposite to that of stars 1 and 2 from \cite{Saglia:2025} because this work follows the rule that star A is brighter (although this is dubious when the two stars are equally bright). $^c$ Corrected from the \cite{Saglia:2025} identifier ``G56.3B''. $^d$ Corrected from the \cite{Saglia:2025} identifier ``CPD-5211628B''. $^e$ There exists a degenerate parameter space localized around the periastron $\Delta\phi=360^\circ$ that is nearly indistinguishable from the solution given here. The prior probability distribution given by Equation~(\ref{eq:PrDelphi}) clearly prefers the latter. Also, the solution near the periastron predicts $a=2.88_{-0.99}^{+2.35}$~kau for the semimajor axis which is much larger than both the value given here and the sky-projected separation of $0.196$~kau. [Note: The argument of ascending node is not given here and in all the other following tables because it is related to the parameter $\theta$ that can be obtained by Equation~(\ref{eq:theta}) from the other parameters given here.]
\end{sidewaystable*}

\begin{sidewaystable*}
\caption{\small{Similar to Table~\ref{tab:prmt_newton} but in Generalized Gravity (with $\Gamma$ free)}}\label{tab:prmt_general}
\scriptsize
\begin{center}
  \begin{tabular}{lccccccccccccccccc}
  \hline
 \# & $r$ & $a$ &  $e$ & $i$ & $\phi_0$ & $\phi-\phi_0$ & $\log_{10} f_M$ & $\log_{10} g_{\rm N}$ & $v_{\rm obs}$ & $v_{\rm mod}$ & $v_{\rm esc}$ & $\Gamma$  \\
  & (kau) & (kau) &   & $(^\circ)$ & $(^\circ)$ & $(^\circ)$ &  & $(\text{m s}^{-2})$ & $(\text{km s}^{-1})$  & $(\text{km s}^{-1})$ & $(\text{km s}^{-1})$ &   \\
 \hline
1 & $15.65_{-4.53}^{+18.12}$ & $13.95_{-6.04}^{+20.75}$ & $0.81_{-0.07}^{+0.08}$  & $73.9_{-5.9}^{+8.9}$ & $70_{-25}^{+17}$ & $206_{-17}^{+25}$ & $-0.000_{-0.021}^{+0.021}$ & $-10.28_{-0.67}^{+0.30}$  & $0.569\pm 0.015$ & $0.56_{-0.04}^{+0.04}$ & $0.94_{-0.23}^{+0.53}$& $0.32_{-0.18}^{+0.24}$ \\ 
2 & $0.87_{-0.10}^{+0.39}$ & $0.75_{-0.24}^{+0.72}$ & $0.67_{-0.29}^{+0.20}$  & $109.6_{-1.7}^{+1.0}$ & $209_{-103}^{+72}$ & $174_{-42}^{+44}$ & $-0.000_{-0.021}^{+0.021}$ & $-7.98_{-0.32}^{+0.10}$  & $1.210\pm 0.010$ & $1.21_{-0.04}^{+0.04}$ & $2.05_{-0.50}^{+1.11}$& $0.12_{-0.14}^{+0.21}$ \\ 
3 & $5.19_{-0.23}^{+2.24}$ & $4.52_{-1.38}^{+3.71}$ & $0.50_{-0.25}^{+0.26}$  & $141.9_{-14.6}^{+4.1}$ & $115_{-74}^{+198}$ & $169_{-75}^{+14}$ & $-0.000_{-0.021}^{+0.021}$ & $-9.32_{-0.31}^{+0.05}$  & $0.799\pm 0.008$ & $0.80_{-0.02}^{+0.02}$ & $1.36_{-0.33}^{+0.79}$& $0.23_{-0.14}^{+0.21}$ \\ 
4 & $0.45_{-0.18}^{+0.31}$ & $0.38_{-0.16}^{+0.42}$ & $0.94_{-0.30}^{+0.05}$  & $89.3_{-5.2}^{+3.3}$ & $298_{-163}^{+13}$ & $170_{-282}^{+12}$ & $0.000_{-0.022}^{+0.021}$ & $-7.29_{-0.46}^{+0.44}$  & $1.026\pm 0.011$ & $1.02_{-0.03}^{+0.03}$ & $1.68_{-0.39}^{+0.89}$& $-0.18_{-0.16}^{+0.23}$ \\ 
5 & $3.90_{-0.16}^{+0.65}$ & $3.14_{-0.81}^{+2.01}$ & $0.60_{-0.28}^{+0.24}$  & $85.8_{-1.2}^{+0.9}$ & $161_{-59}^{+61}$ & $193_{-20}^{+57}$ & $-0.000_{-0.021}^{+0.021}$ & $-9.15_{-0.13}^{+0.04}$  & $0.635\pm 0.025$ & $0.63_{-0.04}^{+0.04}$ & $1.07_{-0.27}^{+0.62}$& $0.08_{-0.13}^{+0.20}$ \\ 
6 & $12.98_{-3.40}^{+6.53}$ & $10.62_{-3.75}^{+7.88}$ & $0.58_{-0.26}^{+0.27}$  & $89.8_{-0.5}^{+0.5}$ & $236_{-85}^{+65}$ & $175_{-48}^{+37}$ & $-0.000_{-0.021}^{+0.021}$ & $-10.19_{-0.35}^{+0.26}$  & $0.321\pm 0.030$ & $0.30_{-0.04}^{+0.05}$ & $0.52_{-0.14}^{+0.30}$& $0.03_{-0.15}^{+0.21}$ \\ 
7 & $0.31_{-0.03}^{+0.11}$ & $0.27_{-0.08}^{+0.22}$ & $0.57_{-0.26}^{+0.18}$  & $58.9_{-1.5}^{+4.0}$ & $247_{-85}^{+66}$ & $157_{-69}^{+29}$ & $0.000_{-0.021}^{+0.021}$ & $-7.16_{-0.27}^{+0.07}$  & $1.518\pm 0.009$ & $1.51_{-0.04}^{+0.04}$ & $2.47_{-0.54}^{+0.95}$& $0.01_{-0.11}^{+0.14}$ \\ 
8 & $1.98_{-0.13}^{+0.47}$ & $1.73_{-0.53}^{+1.30}$ & $0.48_{-0.22}^{+0.26}$  & $37.0_{-5.6}^{+10.3}$ & $244_{-112}^{+92}$ & $168_{-100}^{+23}$ & $-0.000_{-0.021}^{+0.021}$ & $-8.82_{-0.18}^{+0.06}$  & $0.604\pm 0.003$ & $0.60_{-0.02}^{+0.02}$ & $0.98_{-0.23}^{+0.54}$& $0.03_{-0.12}^{+0.19}$ \\ 
9 & $6.68_{-1.26}^{+3.45}$ & $5.74_{-2.00}^{+5.17}$ & $0.95_{-0.07}^{+0.02}$  & $61.5_{-19.0}^{+16.3}$ & $307_{-14}^{+17}$ & $165_{-14}^{+9}$ & $-0.000_{-0.021}^{+0.021}$ & $-9.52_{-0.36}^{+0.18}$  & $0.324\pm 0.010$ & $0.31_{-0.02}^{+0.03}$ & $0.53_{-0.13}^{+0.30}$& $-0.14_{-0.15}^{+0.21}$ \\ 
10 & $1.50_{-0.08}^{+0.49}$ & $1.30_{-0.39}^{+0.98}$ & $0.80_{-0.32}^{+0.14}$  & $74.1_{-15.1}^{+7.1}$ & $314_{-284}^{+28}$ & $199_{-14}^{+36}$ & $0.000_{-0.021}^{+0.021}$ & $-8.28_{-0.24}^{+0.05}$  & $1.133\pm 0.010$ & $1.13_{-0.02}^{+0.02}$ & $1.90_{-0.45}^{+1.04}$& $0.11_{-0.13}^{+0.20}$ \\ 
11 & $0.66_{-0.04}^{+0.34}$ & $0.59_{-0.20}^{+0.56}$ & $0.86_{-0.30}^{+0.12}$  & $78.1_{-23.0}^{+6.7}$ & $168_{-25}^{+37}$ & $194_{-11}^{+33}$ & $-0.000_{-0.021}^{+0.021}$ & $-7.57_{-0.35}^{+0.06}$  & $1.310\pm 0.012$ & $1.31_{-0.02}^{+0.02}$ & $2.21_{-0.53}^{+1.28}$& $0.01_{-0.14}^{+0.23}$ \\ 
12 & $28.70_{-7.18}^{+7.66}$ & $21.74_{-6.79}^{+12.87}$ & $0.64_{-0.28}^{+0.27}$  & $96.8_{-18.5}^{+3.4}$ & $266_{-34}^{+48}$ & $168_{-39}^{+17}$ & $0.000_{-0.021}^{+0.021}$ & $-10.84_{-0.21}^{+0.25}$  & $0.117\pm 0.004$ & $0.10_{-0.02}^{+0.03}$ & $0.18_{-0.05}^{+0.11}$& $-0.30_{-0.16}^{+0.22}$ \\ 
13 & $1.23_{-0.13}^{+0.77}$ & $1.16_{-0.41}^{+1.30}$ & $0.88_{-0.28}^{+0.09}$  & $95.9_{-31.7}^{+24.7}$ & $175_{-150}^{+158}$ & $168_{-114}^{+28}$ & $-0.000_{-0.021}^{+0.021}$ & $-8.12_{-0.42}^{+0.10}$  & $0.700\pm 0.098$ & $0.70_{-0.02}^{+0.08}$ & $1.17_{-0.28}^{+0.63}$& $-0.12_{-0.14}^{+0.22}$ \\ 
14 & $1.40_{-0.33}^{+1.10}$ & $1.02_{-0.38}^{+1.43}$ & $0.75_{-0.36}^{+0.22}$  & $95.9_{-33.9}^{+4.9}$ & $129_{-74}^{+41}$ & $183_{-14}^{+28}$ & $-0.000_{-0.021}^{+0.021}$ & $-8.23_{-0.50}^{+0.23}$  & $0.842\pm 0.013$ & $0.84_{-0.03}^{+0.03}$ & $1.65_{-0.48}^{+1.01}$& $0.05_{-0.13}^{+0.22}$ \\ 
15 & $1.48_{-0.24}^{+1.38}$ & $1.31_{-0.45}^{+1.49}$ & $0.46_{-0.24}^{+0.28}$  & $73.3_{-46.9}^{+74.1}$ & $237_{-144}^{+52}$ & $185_{-20}^{+86}$ & $0.000_{-0.021}^{+0.021}$ & $-8.29_{-0.57}^{+0.15}$  & $0.882\pm 0.096$ & $0.88_{-0.03}^{+0.09}$ & $1.53_{-0.34}^{+0.81}$& $0.05_{-0.16}^{+0.24}$ \\ 
16 & $0.25_{-0.01}^{+0.07}$ & $0.22_{-0.07}^{+0.16}$ & $0.44_{-0.17}^{+0.25}$  & $155.0_{-22.5}^{+9.9}$ & $239_{-196}^{+65}$ & $206_{-22}^{+90}$ & $-0.000_{-0.021}^{+0.021}$ & $-6.80_{-0.21}^{+0.06}$  & $2.954\pm 0.009$ & $2.96_{-0.01}^{+0.01}$ & $4.80_{-1.02}^{+2.13}$& $0.16_{-0.12}^{+0.17}$ \\ 
17 & $0.46_{-0.04}^{+0.10}$ & $0.63_{-0.17}^{+0.40}$ & $0.28_{-0.12}^{+0.26}$  & $40.9_{-8.9}^{+11.4}$ & $99_{-18}^{+50}$ & $317_{-51}^{+26}$ & $-0.000_{-0.021}^{+0.021}$ & $-7.31_{-0.17}^{+0.09}$  & $1.522\pm 0.012$ & $1.52_{-0.02}^{+0.02}$ & $1.95_{-0.20}^{+0.18}$& $-0.12_{-0.06}^{+0.04}$ \\ 
18 & $0.84_{-0.08}^{+0.43}$ & $0.75_{-0.23}^{+0.74}$ & $0.91_{-0.10}^{+0.04}$  & $44.3_{-22.0}^{+25.9}$ & $230_{-28}^{+103}$ & $201_{-12}^{+19}$ & $0.000_{-0.021}^{+0.021}$ & $-7.82_{-0.36}^{+0.09}$  & $1.624\pm 0.006$ & $1.62_{-0.01}^{+0.01}$ & $2.66_{-0.61}^{+1.36}$& $0.17_{-0.13}^{+0.20}$ \\ 
19 & $2.32_{-0.27}^{+1.33}$ & $2.07_{-0.70}^{+2.09}$ & $0.71_{-0.31}^{+0.21}$  & $91.0_{-3.7}^{+3.9}$ & $199_{-106}^{+65}$ & $175_{-42}^{+40}$ & $0.000_{-0.021}^{+0.021}$ & $-8.64_{-0.39}^{+0.11}$  & $0.352\pm 0.015$ & $0.34_{-0.04}^{+0.04}$ & $0.57_{-0.15}^{+0.34}$& $-0.31_{-0.16}^{+0.23}$ \\ 
20 & $0.15_{-0.02}^{+0.07}$ & $0.14_{-0.04}^{+0.12}$ & $0.77_{-0.28}^{+0.13}$  & $68.7_{-14.1}^{+10.3}$ & $174_{-30}^{+35}$ & $205_{-18}^{+37}$ & $-0.000_{-0.021}^{+0.021}$ & $-6.23_{-0.35}^{+0.10}$  & $4.387\pm 0.008$ & $4.39_{-0.01}^{+0.01}$ & $7.23_{-1.67}^{+3.57}$& $0.18_{-0.14}^{+0.19}$ \\ 
21 & $1.96_{-0.03}^{+0.14}$ & $1.82_{-0.54}^{+1.23}$ & $0.48_{-0.24}^{+0.23}$  & $109.3_{-1.6}^{+1.7}$ & $134_{-79}^{+43}$ & $-107_{-59}^{+179}$ & $-0.000_{-0.020}^{+0.021}$ & $-8.78_{-0.06}^{+0.03}$  & $0.801\pm 0.008$ & $0.80_{-0.04}^{+0.04}$ & $1.19_{-0.22}^{+0.50}$& $0.09_{-0.09}^{+0.15}$ \\ 
22 & $0.04_{-0.01}^{+0.02}$ & $0.03_{-0.01}^{+0.03}$ & $0.58_{-0.10}^{+0.21}$  & $144.6_{-24.8}^{+16.7}$ & $289_{-46}^{+39}$ & $145_{-54}^{+25}$ & $0.000_{-0.021}^{+0.021}$ & $-5.21_{-0.40}^{+0.14}$  & $0.980\pm 0.004$ & $0.98_{-0.01}^{+0.01}$ & $1.63_{-0.39}^{+0.93}$& $-0.68_{-0.14}^{+0.21}$ \\ 
23 & $1.28_{-0.17}^{+0.78}$ & $1.12_{-0.37}^{+1.08}$ & $0.65_{-0.28}^{+0.19}$  & $53.6_{-5.2}^{+12.2}$ & $147_{-78}^{+117}$ & $190_{-32}^{+51}$ & $-0.000_{-0.021}^{+0.021}$ & $-8.14_{-0.41}^{+0.12}$  & $0.797\pm 0.012$ & $0.79_{-0.03}^{+0.03}$ & $1.35_{-0.33}^{+0.80}$& $-0.06_{-0.14}^{+0.23}$ \\ 
24 & $6.54_{-0.73}^{+4.23}$ & $5.99_{-2.04}^{+6.20}$ & $0.74_{-0.31}^{+0.18}$  & $113.3_{-1.9}^{+1.5}$ & $103_{-35}^{+45}$ & $205_{-17}^{+49}$ & $-0.000_{-0.021}^{+0.021}$ & $-9.52_{-0.43}^{+0.10}$  & $0.945\pm 0.014$ & $0.94_{-0.04}^{+0.04}$ & $1.55_{-0.36}^{+0.84}$& $0.34_{-0.14}^{+0.22}$ \\ 
25 & $1.93_{-0.24}^{+1.15}$ & $1.71_{-0.58}^{+1.69}$ & $0.90_{-0.25}^{+0.08}$  & $86.0_{-14.5}^{+16.6}$ & $194_{-38}^{+144}$ & $135_{-312}^{+56}$ & $-0.000_{-0.021}^{+0.021}$ & $-8.44_{-0.41}^{+0.12}$  & $0.766\pm 0.029$ & $0.76_{-0.03}^{+0.03}$ & $1.28_{-0.31}^{+0.73}$& $-0.02_{-0.14}^{+0.22}$ \\ 
26 & $3.30_{-0.53}^{+2.81}$ & $2.99_{-1.10}^{+3.55}$ & $0.45_{-0.19}^{+0.29}$  & $145.3_{-30.3}^{+21.2}$ & $107_{-28}^{+91}$ & $156_{-98}^{+19}$ & $0.000_{-0.021}^{+0.021}$ & $-8.94_{-0.53}^{+0.15}$  & $0.845\pm 0.010$ & $0.84_{-0.01}^{+0.01}$ & $1.42_{-0.34}^{+0.79}$& $0.17_{-0.15}^{+0.23}$ \\ 
27 & $0.75_{-0.08}^{+0.36}$ & $0.69_{-0.23}^{+0.66}$ & $0.57_{-0.13}^{+0.18}$  & $32.5_{-21.9}^{+25.4}$ & $205_{-162}^{+92}$ & $216_{-25}^{+55}$ & $0.001_{-0.021}^{+0.021}$ & $-7.63_{-0.34}^{+0.10}$  & $2.200\pm 0.005$ & $2.20_{-0.01}^{+0.01}$ & $3.57_{-0.81}^{+1.58}$& $0.22_{-0.13}^{+0.18}$ \\ 
28 & $3.68_{-0.46}^{+2.17}$ & $3.24_{-1.09}^{+3.30}$ & $0.75_{-0.25}^{+0.13}$  & $98.1_{-58.3}^{+42.8}$ & $183_{-137}^{+128}$ & $98_{-292}^{+101}$ & $-0.000_{-0.021}^{+0.021}$ & $-8.97_{-0.40}^{+0.11}$  & $0.473\pm 0.144$ & $0.45_{-0.04}^{+0.10}$ & $0.79_{-0.22}^{+0.48}$& $-0.10_{-0.16}^{+0.23}$ \\ 
29 & $0.15_{-0.01}^{+0.06}$ & $0.14_{-0.04}^{+0.12}$ & $0.78_{-0.34}^{+0.15}$  & $63.7_{-19.3}^{+12.9}$ & $73_{-47}^{+238}$ & $160_{-33}^{+16}$ & $0.000_{-0.021}^{+0.021}$ & $-6.32_{-0.30}^{+0.08}$  & $1.834\pm 0.020$ & $1.83_{-0.03}^{+0.03}$ & $3.00_{-0.68}^{+1.41}$& $-0.17_{-0.13}^{+0.18}$ \\ 
30 & $0.25_{-0.04}^{+0.16}$ & $0.22_{-0.08}^{+0.24}$ & $0.88_{-0.15}^{+0.05}$  & $130.5_{-24.1}^{+28.9}$ & $140_{-23}^{+26}$ & $156_{-23}^{+14}$ & $-0.001_{-0.021}^{+0.021}$ & $-6.72_{-0.43}^{+0.15}$  & $2.245\pm 0.009$ & $2.24_{-0.01}^{+0.01}$ & $3.71_{-0.86}^{+1.81}$& $0.03_{-0.14}^{+0.19}$ \\ 
31 & $9.08_{-2.14}^{+7.54}$ & $7.46_{-2.70}^{+7.04}$ & $0.45_{-0.25}^{+0.31}$  & $76.1_{-46.2}^{+72.8}$ & $98_{-21}^{+142}$ & $22_{-211}^{+156}$ & $0.000_{-0.021}^{+0.021}$ & $-9.81_{-0.53}^{+0.23}$  & $0.251\pm 0.006$ & $0.25_{-0.01}^{+0.01}$ & $0.45_{-0.10}^{+0.26}$& $-0.12_{-0.16}^{+0.22}$ \\ 
32 & $0.53_{-0.02}^{+0.23}$ & $0.46_{-0.14}^{+0.38}$ & $0.69_{-0.30}^{+0.18}$  & $97.6_{-3.0}^{+3.9}$ & $165_{-41}^{+66}$ & $198_{-19}^{+43}$ & $-0.000_{-0.021}^{+0.021}$ & $-7.41_{-0.31}^{+0.05}$  & $1.731\pm 0.022$ & $1.73_{-0.04}^{+0.04}$ & $2.95_{-0.69}^{+1.62}$& $0.11_{-0.15}^{+0.21}$ \\ 
\hline
\end{tabular}
\end{center}
\end{sidewaystable*}

The right panels of Figure~\ref{fig:compare} exhibit individual values of $e$ and $\Gamma$ in generalized gravity. Here, the results with the general 3D model described in this work and the simple 3D model of \cite{Chae:2025} are compared. Two estimates of $e$ generally agree. However, there are some extreme values of $e\approx 1$ only from the simple 3D model for some systems, although the general 3D model also predicts high values $> 0.6$ for the same systems. This shows that the reduction of one of the two orientation angles in the simple 3D model can result in somewhat biased estimates of $e$ in some individual cases (despite overall agreement in a statistical sample). However, two estimates of $\Gamma$ generally agree well without any exceptional cases. This may not be surprising, because the simple 3D model of \cite{Chae:2025} was designed to mainly measure $\Gamma$ and tested with mock data, while the inclination and angle variables are merely nuisance parameters as stated in that paper. 

\subsection{Bayesian Inference of the Gravitational Anomaly Parameter $\Gamma$} \label{sec:inference}

While Section~\ref{sec:distribution} is concerned with statistical comparisons and tests of gravity models based on the Bayesian PDFs of the parameters, this subsection is concerned with direct measurements of $\Gamma$ through the consolidation of the individual PDFs. For this, the logical priors (arising from the randomness of the sample) must be imposed on inclination $i$ and phase $\Delta\phi$ because of the degeneracy between gravity and these parameters. Eccentricity $e$ can also have some degeneracy with gravity in general. However, it turns out that $e$ is reasonably constrained regardless of the prior when the 3D velocity $\mathbf{v}$ is very precise as in the present sample. Also, any prior on $e$ has to be empirical rather than a logical requirement. I consider two cases, i.e., the generic thermal prior ($f_{\rm pr}(e)=2e$) and the no/flat prior.

\begin{figure*}[!htb]
    \centering
    \includegraphics[width=1.\linewidth]{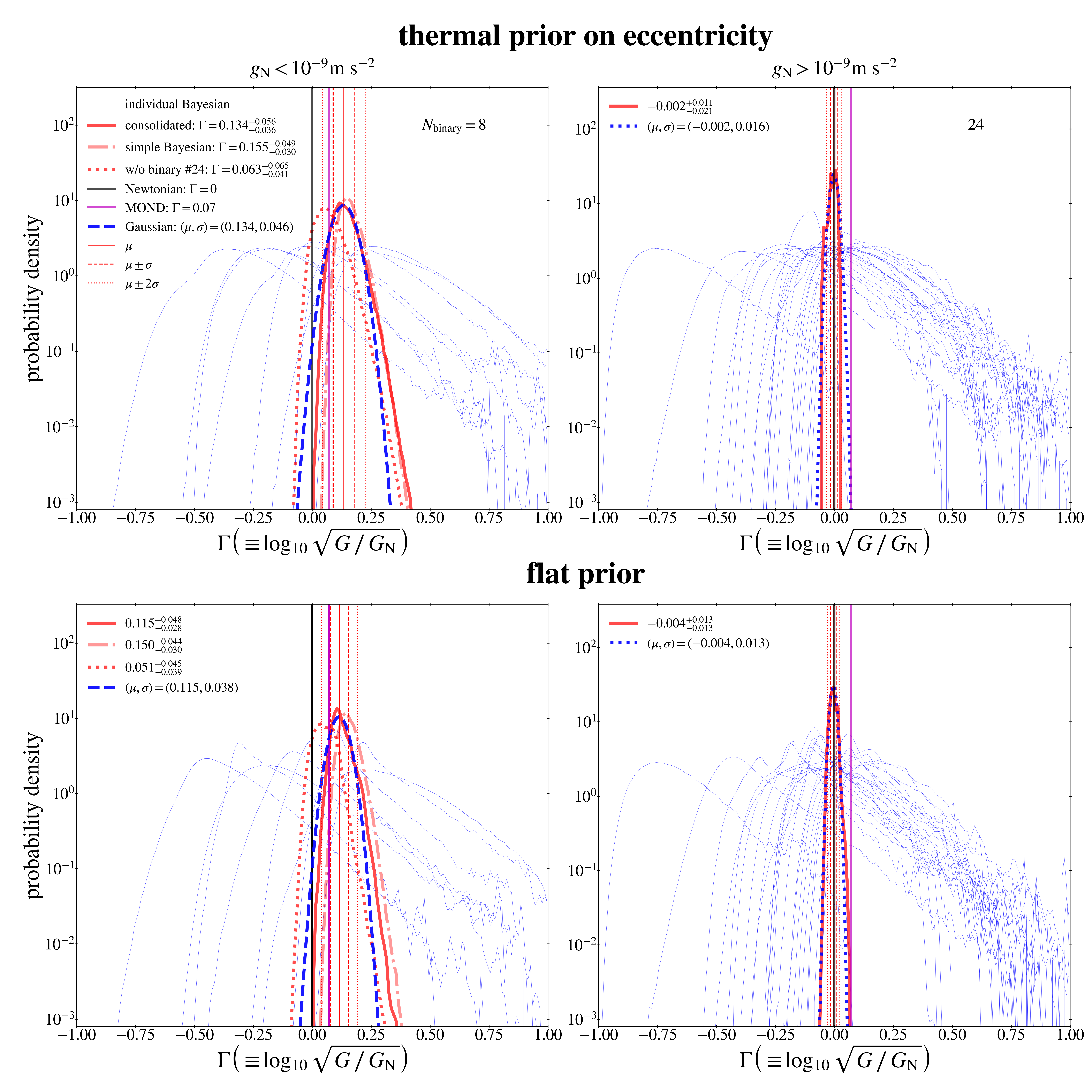}
    \caption{
    The individual Bayesian posterior PDFs of $\Gamma$ (Equation~(\ref{eq:Gamma})) and their consolidated distribution are exhibited for two subsamples of binaries with $g_{\rm N}<10^{-9}$ or $g_{\rm N}>10^{-9}$m\,s$^{-2}$. In the former, cases including or excluding Binary {\#}24 (Stars HD189739 and HD189760) (that is gravitationally unbound by the Newtonian criterion) are considered. The result based on the simple Bayesian algorithm of \cite{Chae:2025} is also shown for the case including {\#}24. The upper row shows the results with the thermal prior on eccentricity while the lower row is for the flat prior.
    }
    \label{fig:probdist}
\end{figure*}

Figure~\ref{fig:probdist} shows the consolidated PDFs of $\Gamma$ for two subsamples with $g_{\rm N}<10^{-9}$ or $g_{\rm N}>10^{-9}$m\,s$^{-2}$, where $g_{\rm N}$ is taken from the Bayesian results with generalized gravity (this is why the number in the low-acceleration regime is now 8, one less than the number with Newtonian gravity). Although the range $10^{-9}\lesssim g_{\rm N}\lesssim 10^{-8}$m\,s$^{-2}$ belongs to a smooth transition regime in MOND gravity, the range $g_{\rm N}>10^{-9}$m\,s$^{-2}$ is not split into finer ranges and referred to as the high-acceleration range because of the small sample size. The subsample of 24 binaries in the high-acceleration regime accurately reproduces Newtonian gravity with $\Gamma=-0.002_{-0.018}^{+0.012}$ (thermal prior) or $-0.004_{-0.013}^{+0.013}$ (flat prior). This agrees with the result of \cite{Chae:2025} for a much larger sample of 125 binaries (in a high-acceleration regime) that have, however, larger uncertainties of RVs. These results extend the validity of standard gravity to gravitational acceleration as low as $10^{-8}$ or $10^{-9}$~m\,s$^{-2}$.  

In contrast, the consolidated probability distribution of $\Gamma$ in the low-acceleration regime exhibits a discrepancy with Newton: $\Gamma=0.134_{-0.036}^{+0.056}$ (thermal prior) or $0.115_{-0.028}^{+0.048}$ (flat prior). These results represent $>3.7\sigma$ inconsistency with Newton $\Gamma=0$, where the quoted errors come from one half of the nominal $2\sigma$ width (rather than the nominal $1\sigma$ width) because the $2\sigma$ width better describes the overall behavior of the consolidated PDF that is asymmetrical and somewhat different from the Gaussian shape. If a stricter limit of $g_{\rm N} < 10^{-9.5}$~m\,s$^{-2}$ is considered\footnote{Although this limit is still not fully satisfactory, MONDian gravity is expected to dominate at accelerations lower than this (see Figure~1 of \cite{Chae:2024a} and also \cite{Pf-A:2025}). For the present sample, only 3 wide binaries satisfy $g_{\rm N} < a_0$.} for the internal acceleration to be more consistent with the MOND regime, only 6 wide binaries meet the limit. For these wide binaries, it is found that $\Gamma=0.143_{-0.041}^{+0.068}$ (thermal prior) or $0.149_{-0.044}^{+0.054}$ (flat prior), which are similar for the 8 wide binaries above.

The moderately high statistical significance for $\Gamma > 0$ is in large part due to one system, binary {\#}24 (Stars HD189739 and HD189760). \cite{Saglia:2025} suggest that Binary {\#}24 is gravitationally unbound. The Bayesian result with Newtonian gravity returns $v_{\rm obs}/v_{\rm escN}=1.116\pm 0.030$ confirming that the observed relative velocity between the pair exceeds the predicted Newtonian escape velocity. Moreover, Newtonian gravity predicts a much larger value of $a=51.2_{-21.3}^{+52.1}$~kau than the observed separation of $s=5.78$~kau, an increased mass with $\log_{10}f_M=0.035\pm 0.019$ (i.e., $\approx 8$\% increase), a phase close to the periastron ($\Delta\phi=333_{-17}^{+15}$ degrees), and  a significantly shifted posterior velocity of $v_{\rm mod}=0.82\pm 0.02$ km\,s$^{-1}$ (Equation~(\ref{eq:v})), which means $v_{\rm mod}/v_{\rm obs}=0.87\pm 0.02$ and $v_{\rm mod}/v_{\rm escN}=0.96\pm 0.03$. Thus, this system is clearly in tension with Newtonian gravity. 

While Newtonian gravity requires Binary {\#}24 to be a fly-by or chance-association in the 3D space, it is not so, based on MOND gravity theories, because $v/v_{\rm esc}<1$ (see Table~\ref{tab:prmt_general}) is feasible in MOND (existing MOND gravity theories \citep{Bekenstein:1984,Milgrom:2010} predict a velocity boost of about 20\%). Thus, if the scientific question is to test standard and nonstandard theories with wide binaries, {\#}24 could be taken as evidence for nonstandard theories. It is also important to remember that this system survived stringent observational selection criteria just like all the other systems carefully collected by \cite{Saglia:2025}. Moreover, the random chance-association probability for this system is extremely low or essentially zero as can be estimated through a Monte Carlo method. Specifically, the expected number of chance-association systems within 100~pc (like {\#}24) for the sky-projected separation $<6$~kau, the line-of-sight separation $<0.5$~pc (considering $2\sigma$ uncertainties in the distance measurements), and the 3D relative speed $<1$~km\,s$^{-1}$, is $\la 8\times 10^{-4}$ for the one-dimensional velocity dispersion $>30$~km~s$^{-1}$ of peculiar velocities which is a conservative limit for solar-type stars in the solar neighborhood. Moreover, if the scatter of metallicities is considered, the chance gets even lower because two stars of Binary {\#}24 have nearly identical metallicities \citep{Saglia:2025}. 

Another possibility to consider for Binary {\#}24 is a chance that it became gravitationally unbound just recently through the impulse due to a passing field star. \cite{BanikZhao:2018} considered this ``gravitational ionization'' for a binary with a 3D separation of 20~kau and found that the chance is negligibly small, because it will disperse too quickly to be observable during the age of the disk. Thus, it is extremely unlikely that Binary {\#}24 is a gravitationally unbound chance association in 3D space.

\begin{figure*}[!htb]
    \centering
    \includegraphics[width=1.\linewidth]{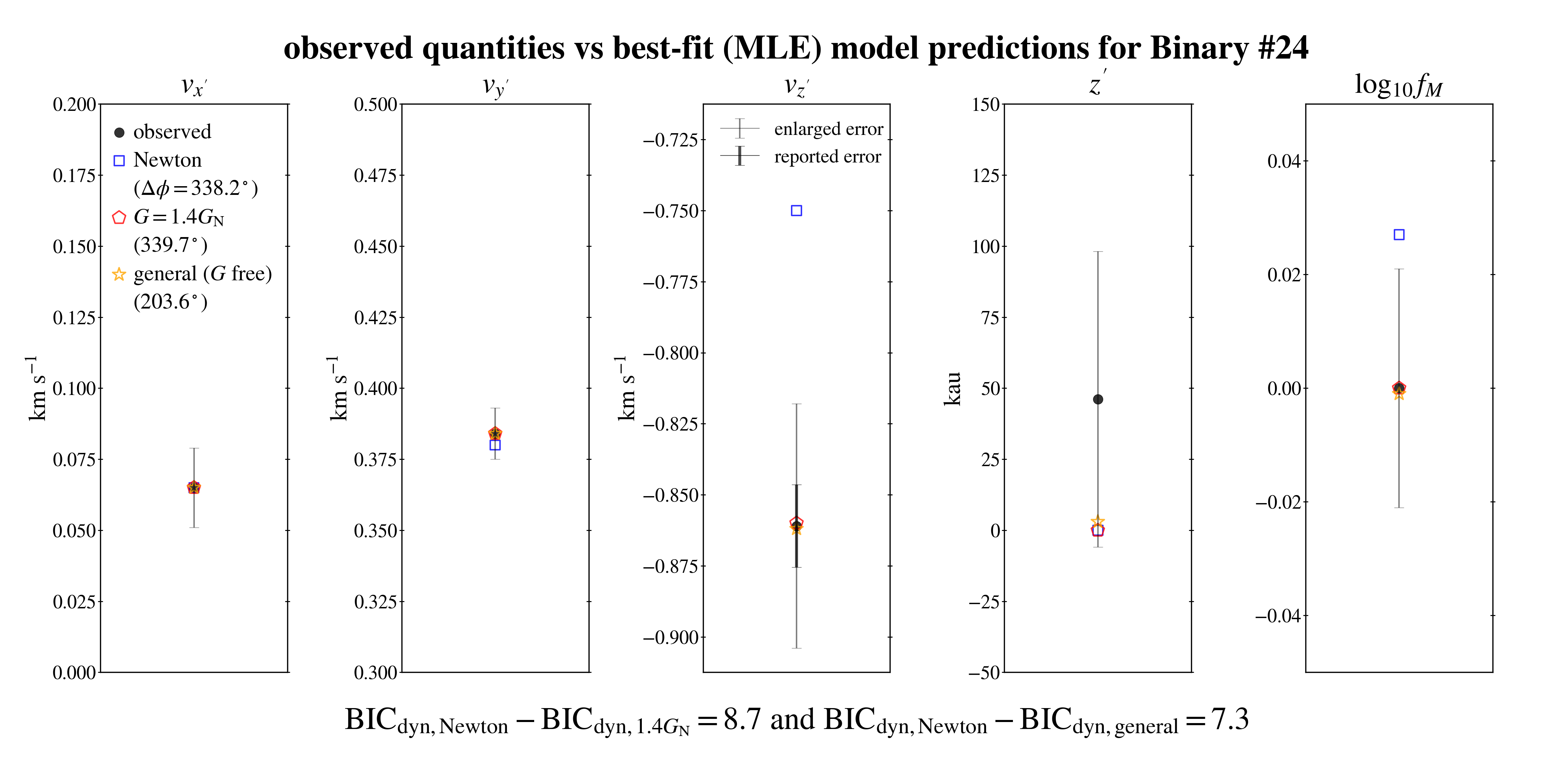}
    \caption{
    For Binary {\#}24 (Stars HD189739 and HD189760), the observed quantities are compared with the best-fit (MLE) predictions of three models: Newtonian, pseudo-Newtonian with $G=1.4G_{\rm N}$, and generalized with free $\Gamma$. The uncertain radial separation $z^\prime$ is not used in defining the best-fit models. The Newtonian model has difficulty in matching the observed value of the relative radial velocity $v_{z^\prime}$ between the stars. The discrepancy is $7.6$ times the reported uncertainty and $2.6$ times the enlarged uncertainty considering the effects of gravitational redshift and convective flow in stellar atmospheres. BIC for a dynamical model is defined in Appendix~\ref{sec:MLEtables}. The obtained BIC difference values indicate that Newtonian model is significantly disfavored even without regard to the issue of the plausibility of the orbit and orientation parameters (see the text).
    }
    \label{fig:binary24}
\end{figure*}

Figure~\ref{fig:binary24} explicitly compares the observed 3D velocity components and the total mass of Binary {\#}24 with the best-fit (MLE) values predicted by Newtonian, MOND-type ($G=1.4G_{\rm N}$), and generalized gravity ($\Gamma$ free) models. Newtonian model shows a large discrepancy in $v_{z^\prime}$ and a moderate discrepancy in $\log_{10}f_M$ while the other models reproduce the observed quantities well. The difference between the Newtonian prediction and the observed value of $v_{z^\prime}$ is $111$~m\,s$^{-1}$ which is $7.6$ times the nominal measurement uncertainty and $2.6$ times the enlarged uncertainty considering possible errors in correcting for the effects of gravitational redshift and convective flow (see \citealt{Saglia:2025}). Based on the enlarged error of $v_{z^\prime}$, $\Delta{\rm BIC}$ (see Appendix~\ref{sec:MLEtables}) between the Newtonian model and the other models is $\approx 8$ which is very significant. Moreover, the Newtonian model predicts an extremely unlikely (or impossible) combination of $(e,\Delta\phi)=(1.00,338.2^\circ)$. It is interesting to note that the pseudo-Newtonian model with $G=1.4G_{\rm N}$ predicts $(e,\Delta\phi)=(0.91,339.7^\circ)$, which seems somewhat rare but not impossible, while the generalized gravity model predicts $(e,\Delta\phi)=(0.67,203.6^\circ)$. Because $z^\prime$ is not used in obtaining the MLE (see Appendix~\ref{sec:MLEtables}), no models match the observed $z^\prime$ (nearly) exactly. Interestingly, all models predict $z^\prime \approx 0$, which is within its (large) measurement uncertainty.

Although it is extremely unlikely that Binary {\#}24 is a chance association, I also consider results without it just in case the system may be kinematically contaminated (see Section~\ref{sec:systematic} below.) The thick dotted red curves in the left panels of Figure~\ref{fig:probdist} show the consolidated PDFs of $\Gamma$: $\Gamma=0.063_{-0.041}^{+0.065}$ (thermal prior) or $0.051_{-0.039}^{+0.045}$ (flat prior). These results are only weakly discrepant with Newton and cannot distinguish well between Newton and MOND. This is not surprising given the small number of wide binaries. What is surprising is the presence of Binary {\#}24 whose inclusion results in a discrepancy with Newton. It will be interesting to see whether future observational studies can discover more wide binaries like Binary {\#}24 that are extremely unlikely to be chance associations but have significantly boosted 3D velocity ($v_{\rm obs}/v_{\rm escN}>1$). Finally, it is worth noting that the median value of $\Gamma=0.063$ without Binary {\#}24 means a gravity boost factor of $\gamma\approx 1.3$ that agrees with MOND numerical solutions by \cite{Pf-A:2025}.

\subsection{Systematic errors?} \label{sec:systematic}

As described in Sections~\ref{sec:distribution} and \ref{sec:inference}, the reasonable interpretation of the \cite{Saglia:2025} sample indicates tension with Newton in favor of MOND-based modified gravity. Considering the profound implication of this (though moderate) conclusion, it is warranted to investigate any possible systematic effects that can bias the gravity inference. 

Can this tension be attributed to an unrecognized bias in the \cite{Saglia:2025} sample, or known issues in the Gaia data? Considering the tension in the distribution of $\Delta\phi$ (Section~\ref{sec:distribution}), one may wonder whether the \cite{Saglia:2025} sample selection preferred certain phase (e.g.\ apastron or periastron). This seems unlikely because the \cite{Saglia:2025} selection is primarily based on the availability of the HARPS radial velocities from binaries of diverse separations, and it is impossible to know the phase in selecting each binary without an accurate orbit reconstruction (in other words, the sample selection is blind to the phase). Known biases in Gaia DR3 parallaxes are potential concerns \citep{Lindegren:2021}. For the \cite{Saglia:2025} sample, the maximum R.A. and decl.\ separations are $0.111^\circ$ and $0.086^\circ$. On the scale of $0.1^\circ$, the RMS amplitude of the smoothed variations in derived bias patterns (Figure~14 of \cite{Lindegren:2021}) is $7.7\mu\rm{as}$, which is smaller than the reported minimal nominal error of $12\mu\rm{as}$ in the \cite{Saglia:2025} sample. Since the known parallax bias is smaller than the actually used errors of parallaxes, it would have not significantly affected the Bayesian results. Nevertheless, the parallax bias highlights the fact that the current data have significant limitations with regard to the radial separation.

The input stellar masses can potentially have the largest systematic effect. \cite{Saglia:2025} estimated stellar masses through state-of-the-art stellar evolution modeling based on a wealth of spectroscopic, photometric, and parallax data, and quoted a systematic error of at most 10\% at the 95\% confidence level. The good agreement with the Newtonian expectation for the 24 binaries in the high-acceleration regime (the right panel of Figure~\ref{fig:probdist}) may strengthen the reliability of their stellar masses. Noting this, it seems contrived to suspect that only the eight wide binaries in the low-acceleration regime have systematically biased stellar masses. Nevertheless, if a 10\% increase in stellar mass is uniformly applied to all stars in the eight wide binaries, the consolidated value of $\Gamma$ through the MCMC procedure is adjusted only by $\Delta\Gamma \approx -0.02$, meaning that the statistical significance of the anomaly is only weakly reduced. With the 10\% increased mass, the Bayesian outputs in Newtonian gravity for binary {\#}24 would give $v_{\rm obs}/v_{\rm escN}=1.073\pm 0.029$, $\log_{10}f_M = 0.028\pm 0.019$, and $v_{\rm mod}/v_{\rm obs}=0.90\pm 0.03$.

Given that the statistical significance of $\Gamma >0$ from this pilot study is largely driven by one system (Binary \#24), some further consideration of this system is warranted. Since a chance association is extremely unlikely for this system as discussed above, the remaining possibility for systematic error is kinematic contamination (e.g., \citealt{Clarke:2020}) by an undetected companion within the Gaia angular resolution limit of $\approx 0.5$ - $1.0$~arcsec \citep{Gaia:2018}, which corresponds to $\approx 50$ - 100 kau for the distance of $\approx 100$~pc. There are several observational tests of kinematic contamination by a hidden close companion, including Gaia's \texttt{ruwe} value \citep{Belokurov:2020}, location in the color-magnitude(CM) diagram, variability of RVs in time \citep{ElBadry:2024}, and direct high-resolution imaging by Speckle interferometry (e.g., \citealt{Manchanda:2023,Tokovinin:2023}).

The two stars HD189739 and HD189760 have fully acceptable \texttt{ruwe} values of $1.103$ and $0.969$ and normal luminosities at their measured Gaia BP$-$RP colors. The HARPS value of the relative RV $v_r=-0.861\pm 0.015$ is well consistent with the Gaia DR3 value of $-1.092\pm 0.239$ (although the precision of the latter is relatively poor), indicating stability over $\ga 3$~years. Thus, nominal interpretation would be that both are single stars. However, a companion of 0.1 - 0.25 $M_\odot$ and orbit radius of 15 - 30 au, can still pass the \texttt{ruwe} test due to long periods, and being $\la1$\% of the solar luminosity and a fraction of an arcsec separation, will not be obvious in the Gaia DR3 or currently available ground-based imaging or 3-year baseline measurements of RVs. Such an undetected companion can induce a perturbation required for the observed boosted velocity. Thus, an RV monitoring over a longer time baseline and direct Speckle high-resolution imaging are needed to completely resolve the issue. One caveat of this consideration is that from an observational point of view, the same concern can be applicable to all the other 7 systems in the low-acceleration regime, because all satisfy the same selection criteria. For example, if the chance of a hidden companion is 1/8, one has to consider removing \emph{any} of the 8 systems with an equal probability from the observational point of view. 

\section{Summary and Outlook} \label{sec:summary}

This work presents a fully realistic and general Bayesian algorithm for modeling the 3D geometry of the wide binary orbit and the gravity anomaly parameter $\Gamma$. The algorithm is designed for a binary whose relative positions on the sky are (nearly) exactly known as in the Gaia DR3 database. The full potential of the algorithm can be achieved only when the relative RV and radial separation between the pair are measured with high precision (recall that high-precision sky-projected velocities are already available for large numbers of wide binaries). While high-precision RVs are being gathered in increasingly large numbers, high-precision radial separations are rare, because the present precision of parallaxes is usually not sufficient to measure the relatively tiny radial separation compared with their distances from the Sun. In principle, if all six components of $\mathbf{r}$ and $\mathbf{v}$ are precisely measured, the algorithm can be used to obtain the \emph{unique} solution (that is, a narrow range of parameters) of the orbit and 3D geometry for a \emph{fixed} gravitational constant.

The new algorithm is different from the algorithm by \cite{Chae:2025} in that all parameters of the former have correct physical meanings, while the latter is mainly designed to measure $\Gamma$ (e.g., the inclination parameter $i$ is not a true angle between the sky plane and the orbital plane in the latter). 

A pilot study with the general algorithm is carried out for the \cite{Saglia:2025} 32 binaries with precise HARPS RVs. Two approaches are considered to test or infer gravity in wide binaries: statistical testing or direct gravity inference. In the statistical testing approach, distributions of the inferred orbit and orientation parameters are statistically tested for the assumed gravity. In this approach, three models (Newtonian, MOND-type toy, and general) of gravity are considered with or without priors. In the gravity inference approach, the logical priors on inclination $i$ and phase $\Delta\phi$ (the true anomaly parameter) are imposed and the strength of gravity is measured through the statistical consolidation of the individual PDFs of $\Gamma$.

From the statistical testing approach the following are found:
\begin{itemize}
    \item With priors imposed on all parameters, the posterior distribution of $\Delta\phi$ in Newtonian gravity is strongly disfavored compared with that in the generalized gravity with $\Delta\rm{BIC}_{\Delta\phi}\approx 65.5$. This value is largely contributed by the low-acceleration subsample. For $e$ and $i$, there is no significant distinction between Newton and the generalized gravity. 
    \item With no/flat priors on $e$, $i$, and $\Delta\phi$, the posterior distribution of $\Delta\phi$ in Newtonian gravity has the A-D statistic $A_n^2=5.4$ with $p=1.9\times 10^{-3}$ while that in the MOND-type toy model has $A_n^2=4.9$ with $p=3.3\times 10^{-3}$. In addition, the BIC values of the two models show $\Delta\rm{BIC}_{\Delta\phi}\approx 33.8$ strongly disfavoring Newton.
    \item The results with Newtonian gravity agree overall with the orbit solutions derived by \cite{Saglia:2025} except for system {\#}22 (Stars 36OphB and GJ663A) for which \cite{Saglia:2025} obtain an unusually large value of $a/s\approx 40$ (with $e=0.25$) while the Bayesian result with priors predicts $a/s\approx 1$ (with $e=0.96_{-0.14}^{+0.02}$) which is much more natural.
    \item For Binary {\#}24 (Stars HD189739 and HD189760),\footnote{Gaia DR3 identifiers: 4235732073427592704 and 4235731867269159680.} the Bayesian modeling in Newtonian gravity returns $\log_{10}f_M=0.035\pm 0.019$ (i.e., $\approx 8$\% increase in mass), a semi-major axis $a=51.2_{-21.3}^{+52.1}$~kau (much larger than the sky-projected separation $s=5.78$~kau), and a phase close to the periastron ($\Delta\phi=333_{-17}^{+15}$ degrees). Consequently, the Newtonian model predicts a 3D relative velocity $v_{\rm mod}$ inconsistent with the observed velocity $v_{\rm obs}$ with $v_{\rm mod}/v_{\rm obs}=0.87\pm 0.02$ and $v_{\rm obs}/v_{\rm escN}=1.116\pm 0.030$ confirming that $v_{\rm obs}$ exceeds the predicted Newtonian escape velocity as was noticed by \cite{Saglia:2025}. However, this binary system can be consistent with MOND-type modified gravity. 
\end{itemize}

From the gravity inference approach the following are found:
\begin{itemize}
    \item For 24 wide binaries with $g_{\rm N}>10^{-9}$m\,s$^{-2}$, the inferred gravity is consistent with Newton: $\Gamma=-0.002_{-0.018}^{+0.012}$ (thermal prior) or $-0.004_{-0.013}^{+0.013}$ (flat prior).
    \item For 8 wide binaries with $g_{\rm N}<10^{-9}$m\,s$^{-2}$, the inferred gravity is boosted: $\Gamma=0.134_{-0.036}^{+0.056}$ (thermal prior) or $0.115_{-0.028}^{+0.048}$ (flat prior). For 6 wide binaries with $g_{\rm N}<10^{-9.5}$m\,s$^{-2}$, it is found that $\Gamma=0.143_{-0.041}^{+0.068}$ (thermal prior) or $0.149_{-0.044}^{+0.054}$ (flat prior). These results are $\ga 3.5\sigma$ discrepant with Newton, but can be consistent with MOND-type modified gravity within $2\sigma$. 
    \item If Binary {\#}24 is excluded from the the sample with $g_{\rm N}<10^{-9}$m\,s$^{-2}$, the inferred gravity is $\Gamma=0.063_{-0.041}^{+0.065}$ (thermal prior) or $0.051_{-0.039}^{+0.045}$ (flat prior), which would be well consistent with MOND but also consistent with Newton within $1.5\sigma$. For Binary {\#}24, a chance association is extremely unlikely. Also, there is no obvious observational indication for kinematical contamination by a hidden companion. However, possibilities of kinematic contamination cannot be completely ruled out based on the currently available observational information, and thus further observational studies are required.
\end{itemize}

The above results with all 32 wide binaries from the \cite{Saglia:2025} sample are consistent with the results from various recent statistical analyses (mostly based on the sky-projected velocities) by two groups (e.g., \citealt{Chae:2023,Chae:2024a,Chae:2024b,Hernandez:2023,Hernandez:2024a,Hernandez:2024review,HernandezKroupa:2025,Chae:2025,Yoon:2025}). However, if Binary {\#}24 is excluded (e.g., assuming that it will turn out to be kinematically contaminated from future observations), the distinction between Newton and MOND is largely removed because of the small sample size. Considering the importance of this system and the recent wide binary results in the literature supporting Newton (e.g., \citealt{Pittordis:2023,Banik:2024,Pittordis:2025,Makarov:2026}),\footnote{Since this work is mainly intended to be a presentation of the 3D methodology, a review of recent works is not considered here.} it will be interesting to see whether the system happens to be a rare exception/fluke or more such systems are discovered from larger samples of similar data qualities in the future.

The pilot study of this work based on the 32 binaries with precise RVs demonstrates the power of the general Bayesian 3D modeling algorithm to measure gravity at low acceleration. While this work was limited by the small sample size, it is expected that $\Gamma$ will be measured with greater and greater precision in the low-acceleration regime based on larger and larger samples of wide binaries with precise RVs. 
Also, distributions of the inferred orbit parameters (in particular, the phase true anomaly parameter $\Delta\phi$) in a fixed gravity model may provide a decisive test of the assumed gravity with much larger samples of wide binaries in the future. Thus, standard and nonstandard gravities may be decisively distinguished through 3D velocity data in the future. Larger samples of 3D velocity data than the \cite{Saglia:2025} sample are currently being gathered by multiple groups around the world (see, e.g., \citealt{Saad:2025,Chae:2026}). The methodology presented here may provide a useful avenue in interpreting future data for experimental studies of low-acceleration gravity. 

\begin{acknowledgments}
It is the author's great pleasure to thank two anonymous referees and the Statistics Editor of the AAS journals for useful suggestions including valuable statistical methods that prompted additional analyses and led to much more improved presentations. The author also thanks a Data Editor for the suggestion of checking and giving the identifications for all the stars of the \cite{Saglia:2025} sample using the SIMBAD Astronomical Database (see Table~\ref{tab:prmt_newton} for the correction of two typos in their tables). The author gratefully acknowledges interesting discussions with Xavier Hernandez, Arthur Kosowsky, Luca Pasquini, and David Turnshek. The author also thanks all collaborators of follow-up works. Finally, the author would like to thank the Scientific Editor Fred Rasio for the patient and thoughtful handling of this review process. The initial results from this work were presented at a PITT PACC Astrophysics Seminar (28 August 2025, University of Pittsburgh), at the workshop ``MOND: alternative paths in the dark matter problem'' (1 - 5 September 2025) held at the Lorentz center of Leiden University,\footnote{https://www.lorentzcenter.nl/mond-alternative-paths-in-the-dark-matter-problem.html} and by an invited talk given online at 2025 China-Japan-Korea International Workshop on the Milky Way and Exoplanets (22 - 26 September 2025) held at Hangzhou, China. This work was supported by the National Research Foundation of Korea (NRF-2022R1A2C1092306).
\end{acknowledgments}

\bibliographystyle{aasjournalv7.bst}
\bibliography{ms.bib}{}

@ARTICLE{Chae:2026,
       author = {{Chae}, K.-H. and {Lee}, B.-C. and {Hernandez}, X. and {Orlov}, V.~G. and {Lim}, D. and {Turnshek}, D.~A. and {Lee}, Y.-W.},
        title = "{Detection of Gravitational Anomaly at Low Acceleration from a Highest-quality Sample of 36 Wide Binaries with Accurate 3D Velocities}",
      journal = {arXiv e-prints},
         year = 2026,
        month = jan,
          eid = {arXiv:2601.21728},
        pages = {arXiv:2601.21728},
          doi = {10.48550/arXiv.2601.21728},
archivePrefix = {arXiv},
       eprint = {2601.21728},
 primaryClass = {astro-ph.GA},
       adsurl = {https://ui.adsabs.harvard.edu/abs/2026arXiv260121728C}
}

@ARTICLE{Banik:2019,
       author = {{Banik}, Indranil},
        title = "{A new line on the wide binary test of gravity}",
      journal = {\mnras},
         year = 2019,
        month = aug,
       volume = {487},
       number = {4},
        pages = {5291-5303},
          doi = {10.1093/mnras/stz1551},
archivePrefix = {arXiv},
       eprint = {1902.01857},
 primaryClass = {astro-ph.GA},
       adsurl = {https://ui.adsabs.harvard.edu/abs/2019MNRAS.487.5291B}
}

@ARTICLE{Acedo:2020,
       author = {{Acedo}, Luis},
        title = "{Modified Newtonian Gravity, Wide Binaries and the Tully-Fisher Relation}",
      journal = {Universe},
         year = 2020,
        month = nov,
       volume = {6},
       number = {11},
          eid = {209},
        pages = {209},
          doi = {10.3390/universe6110209},
       adsurl = {https://ui.adsabs.harvard.edu/abs/2020Univ....6..209A}
}

@ARTICLE{BanikZhao:2022,
       author = {{Banik}, Indranil and {Zhao}, Hongsheng},
        title = "{From Galactic Bars to the Hubble Tension: Weighing Up the Astrophysical Evidence for Milgromian Gravity}",
      journal = {Symmetry},
         year = 2022,
        month = jun,
       volume = {14},
       number = {7},
          eid = {1331},
        pages = {1331},
          doi = {10.3390/sym14071331},
archivePrefix = {arXiv},
       eprint = {2110.06936},
 primaryClass = {astro-ph.CO},
       adsurl = {https://ui.adsabs.harvard.edu/abs/2022Symm...14.1331B}
}

@ARTICLE{Makarov:2026,
       author = {{Makarov}, Valeri V.},
        title = "{Distributions of Wide Binary Stars in Theory and in Gaia Data. III. Orbital Momenta, Masses, and Manifestations of MOND}",
      journal = {\aj},
         year = 2026,
        month = feb,
       volume = {171},
       number = {2},
          eid = {79},
        pages = {79},
          doi = {10.3847/1538-3881/ae2757},
archivePrefix = {arXiv},
       eprint = {2512.25002},
 primaryClass = {astro-ph.SR},
       adsurl = {https://ui.adsabs.harvard.edu/abs/2026AJ....171...79M}
}

@ARTICLE{Saad:2025,
       author = {{Saad}, Serat Mahmud and {Ting}, Yuan-Sen},
        title = "{High-Precision Differential Radial Velocities of C3PO Wide Binaries: A Test of Modified Newtonian Dynamics (MOND)}",
      journal = {arXiv e-prints},
         year = 2025,
        month = dec,
          eid = {arXiv:2512.19652},
        pages = {arXiv:2512.19652},
          doi = {10.48550/arXiv.2512.19652},
archivePrefix = {arXiv},
       eprint = {2512.19652},
 primaryClass = {astro-ph.SR},
       adsurl = {https://ui.adsabs.harvard.edu/abs/2025arXiv251219652M}
}

@ARTICLE{Pittordis:2025,
       author = {{Pittordis}, Charalambos and {Sutherland}, Will and {Shepherd}, Paul},
        title = "{Wide Binaries from GAIA DR3 : testing GR vs MOND with realistic triple modelling}",
      journal = {The Open Journal of Astrophysics},
         year = 2025,
        month = aug,
       volume = {8},
          eid = {109},
        pages = {109},
          doi = {10.33232/001c.142887},
archivePrefix = {arXiv},
       eprint = {2504.07569},
 primaryClass = {astro-ph.GA},
       adsurl = {https://ui.adsabs.harvard.edu/abs/2025OJAp....8E.109P}
}

@ARTICLE{Pittordis:2023,
       author = {{Pittordis}, Charalambos and {Sutherland}, Will},
        title = "{Wide Binaries from GAIA EDR3: preference for GR over MOND?}",
      journal = {The Open Journal of Astrophysics},
         year = 2023,
        month = feb,
       volume = {6},
          eid = {4},
        pages = {4},
          doi = {10.21105/astro.2205.02846},
archivePrefix = {arXiv},
       eprint = {2205.02846},
 primaryClass = {astro-ph.GA},
       adsurl = {https://ui.adsabs.harvard.edu/abs/2023OJAp....6E...4P}
}

@ARTICLE{Banik:2024,
       author = {{Banik}, Indranil and {Pittordis}, Charalambos and {Sutherland}, Will and {Famaey}, Benoit and {Ibata}, Rodrigo and {Mieske}, Steffen and {Zhao}, Hongsheng},
        title = "{Strong constraints on the gravitational law from Gaia DR3 wide binaries}",
      journal = {\mnras},
         year = 2024,
        month = jan,
       volume = {527},
       number = {3},
        pages = {4573-4615},
          doi = {10.1093/mnras/stad3393},
archivePrefix = {arXiv},
       eprint = {2311.03436},
 primaryClass = {astro-ph.SR},
       adsurl = {https://ui.adsabs.harvard.edu/abs/2024MNRAS.527.4573B}
}

@ARTICLE{Manchanda:2023,
       author = {{Manchanda}, Dhruv and {Sutherland}, Will and {Pittordis}, Charalambos},
        title = "{Wide Binaries as a Modified Gravity test: prospects for detecting triple-system contamination}",
      journal = {The Open Journal of Astrophysics},
         year = 2023,
        month = jan,
       volume = {6},
        pages = {E2},
          doi = {10.21105/astro.2210.07781},
archivePrefix = {arXiv},
       eprint = {2210.07781},
 primaryClass = {astro-ph.GA},
       adsurl = {https://ui.adsabs.harvard.edu/abs/2023OJAp....6E...2M}
}

@ARTICLE{Tokovinin:2023,
       author = {{Tokovinin}, Andrei},
        title = "{Exploring Thousands of Nearby Hierarchical Systems with Gaia and Speckle Interferometry}",
      journal = {\aj},
         year = 2023,
        month = apr,
       volume = {165},
       number = {4},
          eid = {180},
        pages = {180},
          doi = {10.3847/1538-3881/acc464},
       adsurl = {https://ui.adsabs.harvard.edu/abs/2023AJ....165..180T}
}

@ARTICLE{Clarke:2020,
       author = {{Clarke}, C.~J.},
        title = "{The distribution of relative proper motions of wide binaries in Gaia DR2: MOND or multiplicity?}",
      journal = {\mnras},
         year = 2020,
        month = jan,
       volume = {491},
       number = {1},
        pages = {L72-L75},
          doi = {10.1093/mnrasl/slz161},
archivePrefix = {arXiv},
       eprint = {1910.10256},
 primaryClass = {astro-ph.SR},
       adsurl = {https://ui.adsabs.harvard.edu/abs/2020MNRAS.491L..72C}
}

@ARTICLE{Belokurov:2020,
       author = {{Belokurov}, Vasily and {Penoyre}, Zephyr and {Oh}, Semyeong and {Iorio}, Giuliano and {Hodgkin}, Simon and {Evans}, N. Wyn and {Everall}, Andrew and {Koposov}, Sergey E. and {Tout}, Christopher A. and {Izzard}, Robert and {Clarke}, Cathie J. and {Brown}, Anthony G.~A.},
        title = "{Unresolved stellar companions with Gaia DR2 astrometry}",
      journal = {\mnras},
         year = 2020,
        month = aug,
       volume = {496},
       number = {2},
        pages = {1922-1940},
          doi = {10.1093/mnras/staa1522},
archivePrefix = {arXiv},
       eprint = {2003.05467},
 primaryClass = {astro-ph.SR},
       adsurl = {https://ui.adsabs.harvard.edu/abs/2020MNRAS.496.1922B}
}

@ARTICLE{Gaia:2018,
       author = {{Gaia Collaboration} and {Brown}, A.~G.~A. and {Vallenari}, A. and {Prusti}, T. and {de Bruijne}, J.~H.~J. and {Babusiaux}, C. and {Bailer-Jones}, C.~A.~L. and {Biermann}, M. and {Evans}, D.~W. and {Eyer}, L. and {Jansen}, F. and {Jordi}, C. and {Klioner}, S.~A. and {Lammers}, U. and {Lindegren}, L. and {Luri}, X. and {Mignard}, F. and {Panem}, C. and {Pourbaix}, D. and {Randich}, S. and {Sartoretti}, P. and {Siddiqui}, H.~I. and {Soubiran}, C. and {van Leeuwen}, F. and {Walton}, N.~A. and {Arenou}, F. and {Bastian}, U. and {Cropper}, M. and {Drimmel}, R. and {Katz}, D. and {Lattanzi}, M.~G. and {Bakker}, J. and {Cacciari}, C. and {Casta{\~n}eda}, J. and {Chaoul}, L. and {Cheek}, N. and {De Angeli}, F. and {Fabricius}, C. and {Guerra}, R. and {Holl}, B. and {Masana}, E. and {Messineo}, R. and {Mowlavi}, N. and {Nienartowicz}, K. and {Panuzzo}, P. and {Portell}, J. and {Riello}, M. and {Seabroke}, G.~M. and {Tanga}, P. and {Th{\'e}venin}, F. and {Gracia-Abril}, G. and {Comoretto}, G. and {Garcia-Reinaldos}, M. and {Teyssier}, D. and {Altmann}, M. and {Andrae}, R. and {Audard}, M. and {Bellas-Velidis}, I. and {Benson}, K. and {Berthier}, J. and {Blomme}, R. and {Burgess}, P. and {Busso}, G. and {Carry}, B. and {Cellino}, A. and {Clementini}, G. and {Clotet}, M. and {Creevey}, O. and {Davidson}, M. and {De Ridder}, J. and {Delchambre}, L. and {Dell'Oro}, A. and {Ducourant}, C. and {Fern{\'a}ndez-Hern{\'a}ndez}, J. and {Fouesneau}, M. and {Fr{\'e}mat}, Y. and {Galluccio}, L. and {Garc{\'\i}a-Torres}, M. and {Gonz{\'a}lez-N{\'u}{\~n}ez}, J. and {Gonz{\'a}lez-Vidal}, J.~J. and {Gosset}, E. and {Guy}, L.~P. and {Halbwachs}, J.-L. and {Hambly}, N.~C. and {Harrison}, D.~L. and {Hern{\'a}ndez}, J. and {Hestroffer}, D. and {Hodgkin}, S.~T. and {Hutton}, A. and {Jasniewicz}, G. and {Jean-Antoine-Piccolo}, A. and {Jordan}, S. and {Korn}, A.~J. and {Krone-Martins}, A. and {Lanzafame}, A.~C. and {Lebzelter}, T. and {L{\"o}ffler}, W. and {Manteiga}, M. and {Marrese}, P.~M. and {Mart{\'\i}n-Fleitas}, J.~M. and {Moitinho}, A. and {Mora}, A. and {Muinonen}, K. and {Osinde}, J. and {Pancino}, E. and {Pauwels}, T. and {Petit}, J.-M. and {Recio-Blanco}, A. and {Richards}, P.~J. and {Rimoldini}, L. and {Robin}, A.~C. and {Sarro}, L.~M. and {Siopis}, C. and {Smith}, M. and {Sozzetti}, A. and {S{\"u}veges}, M. and {Torra}, J. and {van Reeven}, W. and {Abbas}, U. and {Abreu Aramburu}, A. and {Accart}, S. and {Aerts}, C. and {Altavilla}, G. and {{\'A}lvarez}, M.~A. and {Alvarez}, R. and {Alves}, J. and {Anderson}, R.~I. and {Andrei}, A.~H. and {Anglada Varela}, E. and {Antiche}, E. and {Antoja}, T. and {Arcay}, B. and {Astraatmadja}, T.~L. and {Bach}, N. and {Baker}, S.~G. and {Balaguer-N{\'u}{\~n}ez}, L. and {Balm}, P. and {Barache}, C. and {Barata}, C. and {Barbato}, D. and {Barblan}, F. and {Barklem}, P.~S. and {Barrado}, D. and {Barros}, M. and {Barstow}, M.~A. and {Bartholom{\'e} Mu{\~n}oz}, S. and {Bassilana}, J.-L. and {Becciani}, U. and {Bellazzini}, M. and {Berihuete}, A. and {Bertone}, S. and {Bianchi}, L. and {Bienaym{\'e}}, O. and {Blanco-Cuaresma}, S. and {Boch}, T. and {Boeche}, C. and {Bombrun}, A. and {Borrachero}, R. and {Bossini}, D. and {Bouquillon}, S. and {Bourda}, G. and {Bragaglia}, A. and {Bramante}, L. and {Breddels}, M.~A. and {Bressan}, A. and {Brouillet}, N. and {Br{\"u}semeister}, T. and {Brugaletta}, E. and {Bucciarelli}, B. and {Burlacu}, A. and {Busonero}, D. and {Butkevich}, A.~G. and {Buzzi}, R. and {Caffau}, E. and {Cancelliere}, R. and {Cannizzaro}, G. and {Cantat-Gaudin}, T. and {Carballo}, R. and {Carlucci}, T. and {Carrasco}, J.~M. and {Casamiquela}, L. and {Castellani}, M. and {Castro-Ginard}, A. and {Charlot}, P. and {Chemin}, L. and {Chiavassa}, A. and {Cocozza}, G. and {Costigan}, G. and {Cowell}, S. and {Crifo}, F. and {Crosta}, M. and {Crowley}, C. and {Cuypers}, J. and {Dafonte}, C. and {Damerdji}, Y. and {Dapergolas}, A. and {David}, P. and {David}, M. and {de Laverny}, P. and {De Luise}, F.},
        title = "{Gaia Data Release 2. Summary of the contents and survey properties}",
      journal = {\aap},
         year = 2018,
        month = aug,
       volume = {616},
          eid = {A1},
        pages = {A1},
          doi = {10.1051/0004-6361/201833051},
archivePrefix = {arXiv},
       eprint = {1804.09365},
 primaryClass = {astro-ph.GA},
       adsurl = {https://ui.adsabs.harvard.edu/abs/2018A&A...616A...1G}
}

@ARTICLE{Pf-A:2025,
       author = {{Pflamm-Altenburg}, J.},
        title = "{Numerical solutions of the complete two-body system in QUMOND}",
      journal = {\aap},
         year = 2025,
        month = nov,
       volume = {703},
          eid = {A68},
        pages = {A68},
          doi = {10.1051/0004-6361/202555656},
archivePrefix = {arXiv},
       eprint = {2509.01493},
 primaryClass = {astro-ph.IM},
       adsurl = {https://ui.adsabs.harvard.edu/abs/2025A&A...703A..68P}
}

@article{Wasserstein:2016,
author = {Ronald L. Wasserstein and Nicole A. Lazar},
title = {The ASA Statement on p-Values: Context, Process, and Purpose},
journal = {The American Statistician},
volume = {70},
number = {2},
pages = {129--133},
year = {2016},
publisher = {ASA Website},
doi = {10.1080/00031305.2016.1154108},
URL = {https://doi.org/10.1080/00031305.2016.1154108},
eprint = {https://doi.org/10.1080/00031305.2016.1154108}
}

@article{Wasserstein:2019,
author = {Ronald L. Wasserstein and Allen L. Schirm and Nicole A. Lazar},
title = {Moving to a World Beyond “p<0.05”},
journal = {The American Statistician},
volume = {73},
number = {sup1},
pages = {1--19},
year = {2019},
publisher = {ASA Website},
doi = {10.1080/00031305.2019.1583913},
URL = {https://doi.org/10.1080/00031305.2019.1583913},
eprint = {https://doi.org/10.1080/00031305.2019.1583913}
}

@article{AndersonDarling:1954,
author = {T. W. Anderson and D. A. Darling},
title = {A Test of Goodness of Fit},
journal = {Journal of the American Statistical Association},
volume = {49},
number = {268},
pages = {765--769},
year = {1954},
publisher = {ASA Website},
doi = {10.1080/01621459.1954.10501232},
URL = {https://www.tandfonline.com/doi/abs/10.1080/01621459.1954.10501232},
eprint = {https://www.tandfonline.com/doi/pdf/10.1080/01621459.1954.10501232}
}

@article{Marsaglia:2004,
 title={Evaluating the Anderson-Darling Distribution},
 volume={9},
 url={https://www.jstatsoft.org/index.php/jss/article/view/v009i02},
 doi={10.18637/jss.v009.i02},
 number={2},
 journal={Journal of Statistical Software},
 author={Marsaglia, George and Marsaglia, John},
 year={2004},
 pages={1–5}
}

@article{Kass:1995,
    author = "Kass, Robert E. and Raftery, Adrian E.",
    title = "{Bayes Factors}",
    doi = "10.1080/01621459.1995.10476572",
    journal = "J. Am. Statist. Assoc.",
    volume = "90",
    number = "430",
    pages = "773--795",
    year = "1995"
}

@BOOK{Jeffreys:1939,
       author = {{Jeffreys}, Harold},
        title = "{Theory of Probability}",
         year = 1939,
       adsurl = {https://ui.adsabs.harvard.edu/abs/1939thpr.book.....J}
}

@software{Chae:Zenodo2025,
       author = {{Chae}, Kyu-Hyun},
        title = "{Bayesian 3D modeling of the orbital dynamics of wide binary stars: General Python algorithms to infer gravity v2}",
         year = 2025,
        month = sep,
          eid = {10.5281/zenodo.17113129},
          doi = {10.5281/zenodo.17113129},
    publisher = {Zenodo},
          url = {https://doi.org/10.5281/zenodo.17113129},
}

@ARTICLE{PittordisSutherland:2018,
       author = {{Pittordis}, Charalambos and {Sutherland}, Will},
        title = "{Testing modified-gravity theories via wide binaries and GAIA}",
      journal = {\mnras},
         year = 2018,
        month = oct,
       volume = {480},
       number = {2},
        pages = {1778-1795},
          doi = {10.1093/mnras/sty1578},
archivePrefix = {arXiv},
       eprint = {1711.10867},
 primaryClass = {astro-ph.CO},
       adsurl = {https://ui.adsabs.harvard.edu/abs/2018MNRAS.480.1778P}
}

@ARTICLE{BanikKroupa:2019,
       author = {{Banik}, Indranil and {Kroupa}, Pavel},
        title = "{Directly testing gravity with Proxima Centauri}",
      journal = {\mnras},
         year = 2019,
        month = aug,
       volume = {487},
       number = {2},
        pages = {1653-1661},
          doi = {10.1093/mnras/stz1379},
archivePrefix = {arXiv},
       eprint = {1906.08264},
 primaryClass = {astro-ph.SR},
       adsurl = {https://ui.adsabs.harvard.edu/abs/2019MNRAS.487.1653B}
}

@ARTICLE{BanikZhao:2018,
       author = {{Banik}, Indranil and {Zhao}, Hongsheng},
        title = "{Testing gravity with wide binary stars like {\ensuremath{\alpha}} Centauri}",
      journal = {\mnras},
         year = 2018,
        month = oct,
       volume = {480},
       number = {2},
        pages = {2660-2688},
          doi = {10.1093/mnras/sty2007},
archivePrefix = {arXiv},
       eprint = {1805.12273},
 primaryClass = {astro-ph.GA},
       adsurl = {https://ui.adsabs.harvard.edu/abs/2018MNRAS.480.2660B}
}

@ARTICLE{Scarpa:2017,
       author = {{Scarpa}, Riccardo and {Ottolina}, Riccardo and {Falomo}, Renato and {Treves}, Aldo},
        title = "{Dynamics of wide binary stars: A case study for testing Newtonian dynamics in the low acceleration regime}",
      journal = {International Journal of Modern Physics D},
         year = 2017,
        month = jan,
       volume = {26},
       number = {7},
          eid = {1750067},
        pages = {1750067},
          doi = {10.1142/S0218271817500675},
archivePrefix = {arXiv},
       eprint = {1611.08635},
 primaryClass = {astro-ph.GA},
       adsurl = {https://ui.adsabs.harvard.edu/abs/2017IJMPD..2650067S}
}

@ARTICLE{Lindegren:2021,
       author = {{Lindegren}, L. and {Klioner}, S.~A. and {Hern{\'a}ndez}, J. and {Bombrun}, A. and {Ramos-Lerate}, M. and {Steidelm{\"u}ller}, H. and {Bastian}, U. and {Biermann}, M. and {de Torres}, A. and {Gerlach}, E. and {Geyer}, R. and {Hilger}, T. and {Hobbs}, D. and {Lammers}, U. and {McMillan}, P.~J. and {Stephenson}, C.~A. and {Casta{\~n}eda}, J. and {Davidson}, M. and {Fabricius}, C. and {Gracia-Abril}, G. and {Portell}, J. and {Rowell}, N. and {Teyssier}, D. and {Torra}, F. and {Bartolom{\'e}}, S. and {Clotet}, M. and {Garralda}, N. and {Gonz{\'a}lez-Vidal}, J.~J. and {Torra}, J. and {Abbas}, U. and {Altmann}, M. and {Anglada Varela}, E. and {Balaguer-N{\'u}{\~n}ez}, L. and {Balog}, Z. and {Barache}, C. and {Becciani}, U. and {Bernet}, M. and {Bertone}, S. and {Bianchi}, L. and {Bouquillon}, S. and {Brown}, A.~G.~A. and {Bucciarelli}, B. and {Busonero}, D. and {Butkevich}, A.~G. and {Buzzi}, R. and {Cancelliere}, R. and {Carlucci}, T. and {Charlot}, P. and {Cioni}, M. -R.~L. and {Crosta}, M. and {Crowley}, C. and {del Peloso}, E.~F. and {del Pozo}, E. and {Drimmel}, R. and {Esquej}, P. and {Fienga}, A. and {Fraile}, E. and {Gai}, M. and {Garcia-Reinaldos}, M. and {Guerra}, R. and {Hambly}, N.~C. and {Hauser}, M. and {Jan{\ss}en}, K. and {Jordan}, S. and {Kostrzewa-Rutkowska}, Z. and {Lattanzi}, M.~G. and {Liao}, S. and {Licata}, E. and {Lister}, T.~A. and {L{\"o}ffler}, W. and {Marchant}, J.~M. and {Masip}, A. and {Mignard}, F. and {Mints}, A. and {Molina}, D. and {Mora}, A. and {Morbidelli}, R. and {Murphy}, C.~P. and {Pagani}, C. and {Panuzzo}, P. and {Pe{\~n}alosa Esteller}, X. and {Poggio}, E. and {Re Fiorentin}, P. and {Riva}, A. and {Sagrist{\`a} Sell{\'e}s}, A. and {Sanchez Gimenez}, V. and {Sarasso}, M. and {Sciacca}, E. and {Siddiqui}, H.~I. and {Smart}, R.~L. and {Souami}, D. and {Spagna}, A. and {Steele}, I.~A. and {Taris}, F. and {Utrilla}, E. and {van Reeven}, W. and {Vecchiato}, A.},
        title = "{Gaia Early Data Release 3. The astrometric solution}",
      journal = {\aap},
         year = 2021,
        month = may,
       volume = {649},
          eid = {A2},
        pages = {A2},
          doi = {10.1051/0004-6361/202039709},
archivePrefix = {arXiv},
       eprint = {2012.03380},
 primaryClass = {astro-ph.IM},
       adsurl = {https://ui.adsabs.harvard.edu/abs/2021A&A...649A...2L}
}

@ARTICLE{emcee,
       author = {{Foreman-Mackey}, Daniel and {Hogg}, David W. and {Lang}, Dustin and {Goodman}, Jonathan},
        title = "{emcee: The MCMC Hammer}",
      journal = {\pasp},
         year = 2013,
        month = mar,
       volume = {125},
       number = {925},
        pages = {306},
          doi = {10.1086/670067},
archivePrefix = {arXiv},
       eprint = {1202.3665},
 primaryClass = {astro-ph.IM},
       adsurl = {https://ui.adsabs.harvard.edu/abs/2013PASP..125..306F}
}

@ARTICLE{Hill:2011,
       author = {{Hill}, Theodore P. and {Miller}, Jack},
        title = "{How to combine independent data sets for the same quantity}",
      journal = {Chaos},
         year = 2011,
        month = sep,
       volume = {21},
       number = {3},
          eid = {033102},
        pages = {033102},
          doi = {10.1063/1.3593373},
archivePrefix = {arXiv},
       eprint = {1005.4978},
 primaryClass = {physics.data-an},
       adsurl = {https://ui.adsabs.harvard.edu/abs/2011Chaos..21c3102H}
}

@ARTICLE{Akeson:2021,
       author = {{Akeson}, Rachel and {Beichman}, Charles and {Kervella}, Pierre and {Fomalont}, Edward and {Benedict}, G. Fritz},
        title = "{Precision Millimeter Astrometry of the {\ensuremath{\alpha}} Centauri AB System}",
      journal = {\aj},
         year = 2021,
        month = jul,
       volume = {162},
       number = {1},
          eid = {14},
        pages = {14},
          doi = {10.3847/1538-3881/abfaff},
archivePrefix = {arXiv},
       eprint = {2104.10086},
 primaryClass = {astro-ph.SR},
       adsurl = {https://ui.adsabs.harvard.edu/abs/2021AJ....162...14A}
}

@ARTICLE{Kervella:2017,
       author = {{Kervella}, P. and {Th{\'e}venin}, F. and {Lovis}, C.},
        title = "{Proxima's orbit around {\ensuremath{\alpha}} Centauri}",
      journal = {\aap},
         year = 2017,
        month = feb,
       volume = {598},
          eid = {L7},
        pages = {L7},
          doi = {10.1051/0004-6361/201629930},
archivePrefix = {arXiv},
       eprint = {1611.03495},
 primaryClass = {astro-ph.SR},
       adsurl = {https://ui.adsabs.harvard.edu/abs/2017A&A...598L...7K}
}

@ARTICLE{Hernandez:2024review,
       author = {{Hernandez}, X. and {Chae}, Kyu-Hyun and {Aguayo-Ortiz}, A.},
        title = "{A critical review of recent Gaia wide binary gravity tests}",
      journal = {\mnras},
         year = 2024,
        month = sep,
       volume = {533},
       number = {1},
        pages = {729-742},
          doi = {10.1093/mnras/stae1823},
archivePrefix = {arXiv},
       eprint = {2312.03162},
 primaryClass = {astro-ph.GA},
       adsurl = {https://ui.adsabs.harvard.edu/abs/2024MNRAS.533..729H}
}

@ARTICLE{Yoon:2025,
       author = {{Yoon}, Youngsub and {Tian}, Yong and {Chae}, Kyu-Hyun},
        title = "{Probing the Nature of Gravity in the Low-acceleration Limit: Wide Binaries of Extreme Separations with Perspective Effects}",
      journal = {\apj},
         year = 2025,
        month = oct,
       volume = {992},
       number = {1},
          eid = {102},
        pages = {102},
          doi = {10.3847/1538-4357/ae0190},
archivePrefix = {arXiv},
       eprint = {2507.13177},
 primaryClass = {astro-ph.GA},
       adsurl = {https://ui.adsabs.harvard.edu/abs/2025arXiv250713177Y}
}

@ARTICLE{Duchene:2013,
       author = {{Duch{\^e}ne}, Gaspard and {Kraus}, Adam},
        title = "{Stellar Multiplicity}",
      journal = {\araa},
         year = 2013,
        month = aug,
       volume = {51},
       number = {1},
        pages = {269-310},
          doi = {10.1146/annurev-astro-081710-102602},
archivePrefix = {arXiv},
       eprint = {1303.3028},
 primaryClass = {astro-ph.SR},
       adsurl = {https://ui.adsabs.harvard.edu/abs/2013ARA&A..51..269D}
}

@ARTICLE{ElBadry:2024,
       author = {{El-Badry}, Kareem},
        title = "{Gaia's binary star renaissance}",
      journal = {\nar},
         year = 2024,
        month = jun,
       volume = {98},
          eid = {101694},
        pages = {101694},
          doi = {10.1016/j.newar.2024.101694},
archivePrefix = {arXiv},
       eprint = {2403.12146},
 primaryClass = {astro-ph.SR},
       adsurl = {https://ui.adsabs.harvard.edu/abs/2024NewAR..9801694E}
}

@ARTICLE{Gaia:2023,
       author = {{Vallenari}, A. and {Brown}, A.~G.~A. and {Prusti}, T. and {de Bruijne}, J.~H.~J. and {Arenou}, F. and {Babusiaux}, C. and {Biermann}, M. and {Creevey}, O.~L. and {Ducourant}, C. and {Evans}, D.~W. and {Eyer}, L. and {Guerra}, R. and {Hutton}, A. and {Jordi}, C. and {Klioner}, S.~A. and {Lammers}, U.~L. and {Lindegren}, L. and {Luri}, X. and {Mignard}, F. and {Panem}, C. and {Pourbaix}, D. and {Randich}, S. and {Sartoretti}, P. and {Soubiran}, C. and {Tanga}, P. and {Walton}, N.~A. and {Bailer-Jones}, C.~A.~L. and {Bastian}, U. and {Drimmel}, R. and {Jansen}, F. and {Katz}, D. and {Lattanzi}, M.~G. and {van Leeuwen}, F. and {Bakker}, J. and {Cacciari}, C. and {Casta{\~n}eda}, J. and {De Angeli}, F. and {Fabricius}, C. and {Fouesneau}, M. and {Fr{\'e}mat}, Y. and {Galluccio}, L. and {Guerrier}, A. and {Heiter}, U. and {Masana}, E. and {Messineo}, R. and {Mowlavi}, N. and {Nicolas}, C. and {Nienartowicz}, K. and {Pailler}, F. and {Panuzzo}, P. and {Riclet}, F. and {Roux}, W. and {Seabroke}, G.~M. and {Sordo}, R. and {Th{\'e}venin}, F. and {Gracia-Abril}, G. and {Portell}, J. and {Teyssier}, D. and {Altmann}, M. and {Andrae}, R. and {Audard}, M. and {Bellas-Velidis}, I. and {Benson}, K. and {Berthier}, J. and {Blomme}, R. and {Burgess}, P.~W. and {Busonero}, D. and {Busso}, G. and {C{\'a}novas}, H. and {Carry}, B. and {Cellino}, A. and {Cheek}, N. and {Clementini}, G. and {Damerdji}, Y. and {Davidson}, M. and {de Teodoro}, P. and {Nu{\~n}ez Campos}, M. and {Delchambre}, L. and {Dell'Oro}, A. and {Esquej}, P. and {Fern{\'a}ndez-Hern{\'a}ndez}, J. and {Fraile}, E. and {Garabato}, D. and {Garc{\'\i}a-Lario}, P. and {Gosset}, E. and {Haigron}, R. and {Halbwachs}, J. -L. and {Hambly}, N.~C. and {Harrison}, D.~L. and {Hern{\'a}ndez}, J. and {Hestroffer}, D. and {Hodgkin}, S.~T. and {Holl}, B. and {Jan{\ss}en}, K. and {Jevardat de Fombelle}, G. and {Jordan}, S. and {Krone-Martins}, A. and {Lanzafame}, A.~C. and {L{\"o}ffler}, W. and {Marchal}, O. and {Marrese}, P.~M. and {Moitinho}, A. and {Muinonen}, K. and {Osborne}, P. and {Pancino}, E. and {Pauwels}, T. and {Recio-Blanco}, A. and {Reyl{\'e}}, C. and {Riello}, M. and {Rimoldini}, L. and {Roegiers}, T. and {Rybizki}, J. and {Sarro}, L.~M. and {Siopis}, C. and {Smith}, M. and {Sozzetti}, A. and {Utrilla}, E. and {van Leeuwen}, M. and {Abbas}, U. and {{\'A}brah{\'a}m}, P. and {Abreu Aramburu}, A. and {Aerts}, C. and {Aguado}, J.~J. and {Ajaj}, M. and {Aldea-Montero}, F. and {Altavilla}, G. and {{\'A}lvarez}, M.~A. and {Alves}, J. and {Anders}, F. and {Anderson}, R.~I. and {Anglada Varela}, E. and {Antoja}, T. and {Baines}, D. and {Baker}, S.~G. and {Balaguer-N{\'u}{\~n}ez}, L. and {Balbinot}, E. and {Balog}, Z. and {Barache}, C. and {Barbato}, D. and {Barros}, M. and {Barstow}, M.~A. and {Bartolom{\'e}}, S. and {Bassilana}, J. -L. and {Bauchet}, N. and {Becciani}, U. and {Bellazzini}, M. and {Berihuete}, A. and {Bernet}, M. and {Bertone}, S. and {Bianchi}, L. and {Binnenfeld}, A. and {Blanco-Cuaresma}, S. and {Blazere}, A. and {Boch}, T. and {Bombrun}, A. and {Bossini}, D. and {Bouquillon}, S. and {Bragaglia}, A. and {Bramante}, L. and {Breedt}, E. and {Bressan}, A. and {Brouillet}, N. and {Brugaletta}, E. and {Bucciarelli}, B. and {Burlacu}, A. and {Butkevich}, A.~G. and {Buzzi}, R. and {Caffau}, E. and {Cancelliere}, R. and {Cantat-Gaudin}, T. and {Carballo}, R. and {Carlucci}, T. and {Carnerero}, M.~I. and {Carrasco}, J.~M. and {Casamiquela}, L. and {Castellani}, M. and {Castro-Ginard}, A. and {Chaoul}, L. and {Charlot}, P. and {Chemin}, L. and {Chiaramida}, V. and {Chiavassa}, A. and {Chornay}, N. and {Comoretto}, G. and {Contursi}, G. and {Cooper}, W.~J. and {Cornez}, T. and {Cowell}, S. and {Crifo}, F. and {Cropper}, M. and {Crosta}, M. and {Crowley}, C. and {Dafonte}, C. and {Dapergolas}, A. and {David}, M. and {David}, P. and {de Laverny}, P. and {De Luise}, F. and {De March}, R.},
        title = "{Gaia Data Release 3. Summary of the content and survey properties}",
      journal = {\aap},
         year = 2023,
        month = jun,
       volume = {674},
          eid = {A1},
        pages = {A1},
          doi = {10.1051/0004-6361/202243940},
archivePrefix = {arXiv},
       eprint = {2208.00211},
 primaryClass = {astro-ph.GA},
       adsurl = {https://ui.adsabs.harvard.edu/abs/2023A&A...674A...1G}
}

@ARTICLE{Sanders:2002,
       author = {{Sanders}, Robert H. and {McGaugh}, Stacy S.},
        title = "{Modified Newtonian Dynamics as an Alternative to Dark Matter}",
      journal = {\araa},
         year = 2002,
        month = jan,
       volume = {40},
        pages = {263-317},
          doi = {10.1146/annurev.astro.40.060401.093923},
archivePrefix = {arXiv},
       eprint = {astro-ph/0204521},
 primaryClass = {astro-ph},
       adsurl = {https://ui.adsabs.harvard.edu/abs/2002ARA&A..40..263S}
}

@ARTICLE{Famaey:2012,
       author = {{Famaey}, Beno{\^\i}t and {McGaugh}, Stacy S.},
        title = "{Modified Newtonian Dynamics (MOND): Observational Phenomenology and Relativistic Extensions}",
      journal = {Living Reviews in Relativity},
         year = 2012,
        month = dec,
       volume = {15},
       number = {1},
          eid = {10},
        pages = {10},
          doi = {10.12942/lrr-2012-10},
archivePrefix = {arXiv},
       eprint = {1112.3960},
 primaryClass = {astro-ph.CO},
       adsurl = {https://ui.adsabs.harvard.edu/abs/2012LRR....15...10F}
}

@ARTICLE{Milgrom:1983,
       author = {{Milgrom}, M.},
        title = "{A modification of the Newtonian dynamics as a possible alternative to the hidden mass hypothesis.}",
      journal = {\apj},
         year = 1983,
        month = jul,
       volume = {270},
        pages = {365-370},
          doi = {10.1086/161130},
       adsurl = {https://ui.adsabs.harvard.edu/abs/1983ApJ...270..365M}
}

@ARTICLE{Hernandez:2012,
       author = {{Hernandez}, X. and {Jim{\'e}nez}, M.~A. and {Allen}, C.},
        title = "{Wide binaries as a critical test of classical gravity}",
      journal = {European Physical Journal C},
         year = 2012,
        month = feb,
       volume = {72},
          eid = {1884},
        pages = {1884},
          doi = {10.1140/epjc/s10052-012-1884-6},
archivePrefix = {arXiv},
       eprint = {1105.1873},
 primaryClass = {astro-ph.GA},
       adsurl = {https://ui.adsabs.harvard.edu/abs/2012EPJC...72.1884H}
}

@ARTICLE{Hernandez:2023,
       author = {{Hernandez}, X.},
        title = "{Internal kinematics of Gaia DR3 wide binaries: anomalous behaviour in the low acceleration regime}",
      journal = {\mnras},
         year = 2023,
        month = oct,
       volume = {525},
       number = {1},
        pages = {1401-1415},
          doi = {10.1093/mnras/stad2306},
archivePrefix = {arXiv},
       eprint = {2304.07322},
 primaryClass = {astro-ph.GA},
       adsurl = {https://ui.adsabs.harvard.edu/abs/2023MNRAS.525.1401H}
}

@ARTICLE{Hernandez:2024a,
       author = {{Hernandez}, X. and {Verteletskyi}, V. and {Nasser}, L. and {Aguayo-Ortiz}, A.},
        title = "{Statistical analysis of the gravitational anomaly in Gaia wide binaries}",
      journal = {\mnras},
         year = 2024,
        month = mar,
       volume = {528},
       number = {3},
        pages = {4720-4732},
          doi = {10.1093/mnras/stad3446},
archivePrefix = {arXiv},
       eprint = {2309.10995},
 primaryClass = {astro-ph.GA},
       adsurl = {https://ui.adsabs.harvard.edu/abs/2024MNRAS.528.4720H}
}

@ARTICLE{Bekenstein:1984,
       author = {{Bekenstein}, J. and {Milgrom}, M.},
        title = "{Does the missing mass problem signal the breakdown of Newtonian gravity?}",
      journal = {\apj},
         year = 1984,
        month = nov,
       volume = {286},
        pages = {7-14},
          doi = {10.1086/162570},
       adsurl = {https://ui.adsabs.harvard.edu/abs/1984ApJ...286....7B}
}

@ARTICLE{Milgrom:2010,
       author = {{Milgrom}, Mordehai},
        title = "{Quasi-linear formulation of MOND}",
      journal = {\mnras},
         year = 2010,
        month = apr,
       volume = {403},
       number = {2},
        pages = {886-895},
          doi = {10.1111/j.1365-2966.2009.16184.x},
archivePrefix = {arXiv},
       eprint = {0911.5464},
 primaryClass = {astro-ph.CO},
       adsurl = {https://ui.adsabs.harvard.edu/abs/2010MNRAS.403..886M}
}

@ARTICLE{Chae:2023,
       author = {{Chae}, Kyu-Hyun},
        title = "{Breakdown of the Newton-Einstein Standard Gravity at Low Acceleration in Internal Dynamics of Wide Binary Stars}",
      journal = {\apj},
         year = 2023,
        month = aug,
       volume = {952},
       number = {2},
          eid = {128},
        pages = {128},
          doi = {10.3847/1538-4357/ace101},
archivePrefix = {arXiv},
       eprint = {2305.04613},
 primaryClass = {astro-ph.GA},
       adsurl = {https://ui.adsabs.harvard.edu/abs/2023ApJ...952..128C}
}

@ARTICLE{Chae:2024a,
       author = {{Chae}, Kyu-Hyun},
        title = "{Robust Evidence for the Breakdown of Standard Gravity at Low Acceleration from Statistically Pure Binaries Free of Hidden Companions}",
      journal = {\apj},
         year = 2024,
        month = jan,
       volume = {960},
       number = {2},
          eid = {114},
        pages = {114},
          doi = {10.3847/1538-4357/ad0ed5},
archivePrefix = {arXiv},
       eprint = {2309.10404},
 primaryClass = {astro-ph.GA},
       adsurl = {https://ui.adsabs.harvard.edu/abs/2024ApJ...960..114C}
}

@ARTICLE{Chae:2024b,
       author = {{Chae}, Kyu-Hyun},
        title = "{Measurements of the Low-acceleration Gravitational Anomaly from the Normalized Velocity Profile of Gaia Wide Binary Stars and Statistical Testing of Newtonian and Milgromian Theories}",
      journal = {\apj},
         year = 2024,
        month = sep,
       volume = {972},
       number = {2},
          eid = {186},
        pages = {186},
          doi = {10.3847/1538-4357/ad61e9},
archivePrefix = {arXiv},
       eprint = {2402.05720},
 primaryClass = {astro-ph.GA},
       adsurl = {https://ui.adsabs.harvard.edu/abs/2024ApJ...972..186C}
}

@ARTICLE{Chae:2025,
       author = {{Chae}, Kyu-Hyun},
        title = "{Low-acceleration Gravitational Anomaly from Bayesian 3D Modeling of Wide Binary Orbits: Methodology and Results with Gaia Data Release 3}",
      journal = {\apj},
         year = 2025,
        month = jun,
       volume = {985},
       number = {2},
          eid = {210},
        pages = {210},
          doi = {10.3847/1538-4357/adce09},
archivePrefix = {arXiv},
       eprint = {2502.09373},
 primaryClass = {astro-ph.GA},
       adsurl = {https://ui.adsabs.harvard.edu/abs/2025ApJ...985..210C}
}

@ARTICLE{Saglia:2025,
       author = {{Saglia}, R. and {Pasquini}, L. and {Patat}, F. and {Ludwig}, H. -G. and {Giribaldi}, R. and {Leao}, I. and {de Medeiros}, J.~R. and {Murphy}, Michael T.},
        title = "{Testing gravity with wide binaries: 3D velocities and distances of wide binaries from Gaia and HARPS}",
      journal = {\aap},
         year = 2025,
        month = jul,
       volume = {699},
          eid = {A151},
        pages = {A151},
          doi = {10.1051/0004-6361/202555115},
archivePrefix = {arXiv},
       eprint = {2506.05049},
 primaryClass = {astro-ph.GA},
       adsurl = {https://ui.adsabs.harvard.edu/abs/2025A&A...699A.151S}
}

@ARTICLE{Shaya:2011,
       author = {{Shaya}, Ed J. and {Olling}, Rob P.},
        title = "{Very Wide Binaries and Other Comoving Stellar Companions: A Bayesian Analysis of the Hipparcos Catalogue}",
      journal = {\apjs},
         year = 2011,
        month = jan,
       volume = {192},
       number = {1},
          eid = {2},
        pages = {2},
          doi = {10.1088/0067-0049/192/1/2},
archivePrefix = {arXiv},
       eprint = {1007.0425},
 primaryClass = {astro-ph.GA},
       adsurl = {https://ui.adsabs.harvard.edu/abs/2011ApJS..192....2S}
}

@ARTICLE{HernandezKroupa:2025,
       author = {{Hernandez}, X. and {Kroupa}, Pavel},
        title = "{A recent confirmation of the wide binary gravitational anomaly}",
      journal = {\mnras},
         year = 2025,
        month = mar,
       volume = {537},
       number = {3},
        pages = {2925-2930},
          doi = {10.1093/mnras/staf210},
archivePrefix = {arXiv},
       eprint = {2410.17178},
 primaryClass = {astro-ph.GA},
       adsurl = {https://ui.adsabs.harvard.edu/abs/2025MNRAS.537.2925H}
}

%\vspace{5mm}

\appendix

\section{Maximum Likelihood Estimates and Additional Tables}\label{sec:MLEtables}

Additional tables of the fitted and derived parameters are given here. Table~\ref{tab:prmt_noprior} gives the Bayesian results in Newtonian ($G=G_{\rm N}$) or pseudo-Newtonian ($G=1.4G_{\rm N}$) gravity with no/flat priors on the orientation/orbit parameters $e$, $i$, and $\Delta\phi$. The Newtonian results can be compared with the results given in Table~\ref{tab:prmt_newton}.

The other tables (Tables~\ref{tab:MLE_newton}, \ref{tab:MLE_noprior}, and \ref{tab:MLE_general}) give the MLE values of the parameters. The MLE or ``best-fit'' values of the parameters for each binary are defined as follows, from the MCMC output of $2\times 10^6$ sets of the parameters. The best-fit dynamical model is defined to be the one that minimizes the function given by
\begin{equation}
    \chi^2_{\rm dyn} = \sum_{j=x^\prime,y^\prime,z^\prime} \left( \frac{v_{j,{\rm mod}}-v_{j,{\rm obs}}}{\sigma_{v_j}} \right)^2 + \left(\frac{\log_{10}f_M}{0.021}\right)^2,
    \label{eq:chi2}
\end{equation}
where $v_{j,{\rm obs}}$ and $\sigma_{v_j}$ are the observed velocity components (Equation~(\ref{eq:relvel_obs})) and their uncertainties. Equation~(\ref{eq:chi2}) does not include radial displacement $z^\prime$ because its uncertain value can pick a biased model when the other constraints are not sufficient as in the present case. 

With Equation~(\ref{eq:chi2}) for $\chi^2_{\rm dyn}$, the BIC for a dynamical model can be defined by ${\rm BIC}_{\rm dyn}=\chi^2_{\rm dyn,min}+N_{\rm prmt}\ln N_{\rm obs}$ where $\chi^2_{\rm dyn,min}$ is the minimum value of $\chi^2_{\rm dyn}$, the number of observational constraints $N_{\rm obs}=4$, and the number of adjustable parameters $N_{\rm prmt}$ is $5$ for Newtonian and pseudo-Newtonian models and $6$ for the generalized gravity model. This quantity can be used to compare models for an individual system.

Tables~\ref{tab:MLE_newton} and \ref{tab:MLE_noprior} present the results with or without priors on $e$, $i$, and $\Delta\phi$ for models with fixed gravity for each binary. For 9 binaries with $g_{\rm N}<10^{-9}$~m\,s$^{-2}$ (based on the Newtonian modeling criterion), both Newtonian and pseudo-Newtonian gravity with $G=1.4G_{\rm N}$ are presented. Table~\ref{tab:MLE_general} presents the results with priors for the generalized gravity with $\Gamma$ free for each binary.

%\begin{sidewaystable*}
\begin{table*}
\caption{\small{Similar to Table~\ref{tab:prmt_newton} but in Newtonian or pseudo-Newtonian Gravity with No/Flat Priors}}\label{tab:prmt_noprior}
\scriptsize
\begin{center}
  \begin{tabular}{lccccccccccc}
  \hline
 \# & G & $r$ & $a$ &  $e$ & $i$ & $\phi_0$ & $\phi-\phi_0$ & $\log_{10} f_M$ & $\log_{10} g_{\rm N}$ & $v_{\rm mod}$ & $v_{\rm esc}$  \\
  &  & (kau) & (kau) &   & $(^\circ)$ & $(^\circ)$ & $(^\circ)$ &  & $(\text{m s}^{-2})$  & $(\text{km s}^{-1})$ & $(\text{km s}^{-1})$   \\
 \hline
1 & $G_{\rm N}$ & $11.17_{-1.13}^{+1.46}$ & $40.98_{-22.20}^{+102.75}$ & $0.87_{-0.12}^{+0.09}$ & $67.7_{-2.5}^{+2.2}$ & $22_{-6}^{+9}$ & $260_{-7}^{+5}$  & $0.003_{-0.021}^{+0.021}$ & $-9.98_{-0.10}^{+0.09}$  & $0.54_{-0.03}^{+0.03}$ & $0.59_{-0.03}^{+0.03}$  \\ 
2 & $G_{\rm N}$ & $0.78_{-0.01}^{+0.15}$ & $0.78_{-0.06}^{+0.33}$ & $0.18_{-0.13}^{+0.37}$ & $109.8_{-0.8}^{+0.8}$ & $203_{-124}^{+89}$ & $160_{-58}^{+94}$  & $-0.004_{-0.020}^{+0.020}$ & $-7.89_{-0.15}^{+0.03}$  & $1.22_{-0.04}^{+0.04}$ & $1.73_{-0.13}^{+0.05}$  \\ 
3 & $G_{\rm N}$ & $5.02_{-0.07}^{+0.13}$ & $13.41_{-3.39}^{+7.82}$ & $0.63_{-0.12}^{+0.14}$ & $144.5_{-2.6}^{+2.7}$ & $175_{-3}^{+2}$ & $346_{-8}^{+8}$  & $0.000_{-0.021}^{+0.021}$ & $-9.28_{-0.03}^{+0.02}$  & $0.80_{-0.02}^{+0.02}$ & $0.88_{-0.02}^{+0.02}$  \\ 
4 & $G_{\rm N}$ & $0.55_{-0.27}^{+0.95}$ & $0.34_{-0.18}^{+1.22}$ & $0.71_{-0.09}^{+0.16}$ & $89.2_{-1.0}^{+0.5}$ & $107_{-61}^{+61}$ & $190_{-14}^{+43}$  & $-0.002_{-0.021}^{+0.021}$ & $-7.47_{-0.87}^{+0.59}$  & $1.03_{-0.03}^{+0.03}$ & $2.36_{-0.93}^{+0.94}$  \\ 
5 & $G_{\rm N}$ & $4.06_{-0.26}^{+0.39}$ & $4.09_{-0.51}^{+0.86}$ & $0.16_{-0.11}^{+0.25}$ & $86.3_{-0.7}^{+0.6}$ & $162_{-65}^{+165}$ & $213_{-143}^{+63}$  & $-0.001_{-0.020}^{+0.020}$ & $-9.18_{-0.08}^{+0.06}$  & $0.63_{-0.03}^{+0.04}$ & $0.90_{-0.04}^{+0.03}$  \\ 
6 & $G_{\rm N}$ & $14.16_{-3.68}^{+6.45}$ & $13.18_{-4.41}^{+15.46}$ & $0.39_{-0.16}^{+0.23}$ & $89.8_{-0.4}^{+0.4}$ & $289_{-76}^{+40}$ & $126_{-38}^{+53}$  & $-0.002_{-0.021}^{+0.021}$ & $-10.27_{-0.33}^{+0.26}$  & $0.33_{-0.04}^{+0.05}$ & $0.48_{-0.08}^{+0.08}$  \\ 
7 & $G_{\rm N}$ & $0.84_{-0.03}^{+0.05}$ & $7.30_{-3.91}^{+17.28}$ & $0.96_{-0.04}^{+0.03}$ & $72.9_{-0.6}^{+0.7}$ & $333_{-3}^{+2}$ & $106_{-2}^{+2}$  & $0.009_{-0.019}^{+0.020}$ & $-8.03_{-0.04}^{+0.03}$  & $1.49_{-0.03}^{+0.03}$ & $1.54_{-0.04}^{+0.04}$  \\ 
8 & $G_{\rm N}$ & $1.96_{-0.11}^{+0.45}$ & $1.70_{-0.20}^{+0.72}$ & $0.28_{-0.09}^{+0.12}$ & $37.4_{-5.5}^{+9.4}$ & $194_{-106}^{+95}$ & $164_{-38}^{+65}$  & $-0.004_{-0.021}^{+0.021}$ & $-8.82_{-0.18}^{+0.06}$  & $0.61_{-0.02}^{+0.02}$ & $0.94_{-0.09}^{+0.04}$  \\ 
9 & $G_{\rm N}$ & $11.69_{-5.00}^{+8.21}$ & $8.43_{-4.37}^{+12.43}$ & $0.86_{-0.12}^{+0.10}$ & $80.8_{-17.4}^{+4.9}$ & $307_{-6}^{+16}$ & $161_{-24}^{+12}$  & $-0.002_{-0.021}^{+0.021}$ & $-10.01_{-0.46}^{+0.48}$  & $0.32_{-0.02}^{+0.03}$ & $0.58_{-0.14}^{+0.19}$  \\ 
10 & $G_{\rm N}$ & $1.72_{-0.27}^{+0.46}$ & $2.25_{-0.71}^{+2.66}$ & $0.55_{-0.11}^{+0.25}$ & $79.3_{-7.2}^{+2.8}$ & $298_{-45}^{+24}$ & $276_{-52}^{+60}$  & $-0.001_{-0.021}^{+0.021}$ & $-8.39_{-0.20}^{+0.14}$  & $1.13_{-0.02}^{+0.02}$ & $1.44_{-0.16}^{+0.12}$  \\ 
11 & $G_{\rm N}$ & $1.08_{-0.35}^{+0.38}$ & $1.12_{-0.56}^{+1.29}$ & $0.41_{-0.13}^{+0.33}$ & $85.3_{-4.0}^{+1.4}$ & $152_{-64}^{+26}$ & $260_{-49}^{+76}$  & $-0.001_{-0.021}^{+0.021}$ & $-7.99_{-0.26}^{+0.34}$  & $1.31_{-0.02}^{+0.02}$ & $1.82_{-0.26}^{+0.40}$  \\ 
12 & $G_{\rm N}$ & $32.12_{-7.15}^{+7.45}$ & $18.79_{-4.74}^{+5.77}$ & $0.81_{-0.04}^{+0.05}$ & $96.0_{-1.4}^{+2.0}$ & $256_{-10}^{+11}$ & $171_{-8}^{+5}$  & $-0.002_{-0.021}^{+0.021}$ & $-10.93_{-0.18}^{+0.22}$  & $0.12_{-0.03}^{+0.03}$ & $0.33_{-0.03}^{+0.05}$  \\ 
13 & $G_{\rm N}$ & $2.80_{-1.41}^{+1.33}$ & $2.62_{-1.61}^{+2.83}$ & $0.35_{-0.25}^{+0.43}$ & $86.5_{-3.4}^{+8.0}$ & $155_{-119}^{+154}$ & $163_{-136}^{+55}$  & $-0.002_{-0.021}^{+0.021}$ & $-8.84_{-0.33}^{+0.61}$  & $0.72_{-0.04}^{+0.13}$ & $1.10_{-0.19}^{+0.46}$  \\ 
14 & $G_{\rm N}$ & $2.00_{-0.71}^{+1.02}$ & $1.72_{-0.84}^{+2.23}$ & $0.34_{-0.11}^{+0.27}$ & $95.5_{-1.7}^{+3.1}$ & $116_{-79}^{+61}$ & $213_{-44}^{+73}$  & $-0.002_{-0.021}^{+0.021}$ & $-8.54_{-0.35}^{+0.38}$  & $0.84_{-0.03}^{+0.03}$ & $1.31_{-0.24}^{+0.33}$  \\ 
15 & $G_{\rm N}$ & $1.53_{-0.29}^{+0.79}$ & $1.26_{-0.39}^{+1.14}$ & $0.36_{-0.20}^{+0.12}$ & $112.1_{-79.3}^{+39.1}$ & $207_{-121}^{+92}$ & $187_{-32}^{+76}$  & $-0.002_{-0.021}^{+0.021}$ & $-8.31_{-0.36}^{+0.18}$  & $0.87_{-0.02}^{+0.10}$ & $1.49_{-0.28}^{+0.16}$  \\ 
16 & $G_{\rm N}$ & $0.24_{-0.01}^{+0.02}$ & $0.43_{-0.07}^{+0.19}$ & $0.52_{-0.07}^{+0.11}$ & $160.9_{-12.4}^{+5.0}$ & $242_{-63}^{+20}$ & $304_{-7}^{+24}$  & $0.000_{-0.021}^{+0.021}$ & $-6.77_{-0.08}^{+0.03}$  & $2.95_{-0.01}^{+0.01}$ & $3.48_{-0.16}^{+0.11}$  \\ 
17 & $G_{\rm N}$ & $0.65_{-0.12}^{+0.07}$ & $0.65_{-0.21}^{+0.12}$ & $0.06_{-0.05}^{+0.24}$ & $58.2_{-9.0}^{+3.4}$ & $180_{-116}^{+54}$ & $190_{-183}^{+40}$  & $-0.003_{-0.020}^{+0.020}$ & $-7.62_{-0.08}^{+0.18}$  & $1.52_{-0.02}^{+0.02}$ & $2.16_{-0.09}^{+0.24}$  \\ 
18 & $G_{\rm N}$ & $0.79_{-0.03}^{+0.20}$ & $1.21_{-0.19}^{+1.67}$ & $0.93_{-0.05}^{+0.01}$ & $36.9_{-18.0}^{+19.6}$ & $247_{-69}^{+75}$ & $215_{-6}^{+18}$  & $-0.001_{-0.021}^{+0.021}$ & $-7.78_{-0.19}^{+0.04}$  & $1.62_{-0.01}^{+0.01}$ & $1.98_{-0.19}^{+0.08}$  \\ 
19 & $G_{\rm N}$ & $2.80_{-0.71}^{+4.94}$ & $1.55_{-0.42}^{+3.77}$ & $0.90_{-0.05}^{+0.06}$ & $90.7_{-3.0}^{+3.5}$ & $157_{-59}^{+69}$ & $181_{-7}^{+9}$  & $-0.002_{-0.021}^{+0.021}$ & $-8.80_{-0.88}^{+0.25}$  & $0.36_{-0.04}^{+0.04}$ & $1.14_{-0.46}^{+0.18}$  \\ 
20 & $G_{\rm N}$ & $0.15_{-0.02}^{+0.03}$ & $0.30_{-0.10}^{+0.55}$ & $0.80_{-0.06}^{+0.09}$ & $70.0_{-12.3}^{+5.6}$ & $130_{-25}^{+16}$ & $262_{-35}^{+38}$  & $0.000_{-0.021}^{+0.021}$ & $-6.25_{-0.16}^{+0.11}$  & $4.39_{-0.01}^{+0.01}$ & $5.08_{-0.43}^{+0.33}$  \\ 
21 & $G_{\rm N}$ & $1.97_{-0.04}^{+0.12}$ & $2.81_{-0.51}^{+1.02}$ & $0.37_{-0.16}^{+0.19}$ & $108.8_{-1.5}^{+1.5}$ & $57_{-37}^{+280}$ & $-31_{-49}^{+58}$  & $0.002_{-0.021}^{+0.021}$ & $-8.78_{-0.05}^{+0.03}$  & $0.79_{-0.04}^{+0.04}$ & $0.99_{-0.04}^{+0.03}$  \\ 
22 & $G_{\rm N}$ & $0.54_{-0.50}^{+0.85}$ & $0.34_{-0.32}^{+1.12}$ & $0.66_{-0.36}^{+0.31}$ & $92.9_{-1.8}^{+42.3}$ & $278_{-52}^{+27}$ & $170_{-76}^{+8}$  & $-0.002_{-0.021}^{+0.021}$ & $-7.53_{-0.82}^{+2.21}$  & $0.98_{-0.01}^{+0.01}$ & $2.17_{-0.82}^{+5.58}$  \\ 
23 & $G_{\rm N}$ & $1.59_{-0.45}^{+1.66}$ & $1.12_{-0.40}^{+2.76}$ & $0.62_{-0.09}^{+0.17}$ & $61.0_{-10.6}^{+13.1}$ & $213_{-118}^{+92}$ & $177_{-50}^{+23}$  & $-0.002_{-0.021}^{+0.021}$ & $-8.33_{-0.62}^{+0.29}$  & $0.80_{-0.03}^{+0.03}$ & $1.49_{-0.45}^{+0.27}$  \\ 
24 & $G_{\rm N}$ & $5.81_{-0.03}^{+0.09}$ & $132.97_{-79.25}^{+354.22}$ & $0.96_{-0.06}^{+0.03}$ & $115.5_{-1.1}^{+1.3}$ & $24_{-6}^{+7}$ & $334_{-17}^{+15}$  & $0.032_{-0.019}^{+0.019}$ & $-9.39_{-0.02}^{+0.02}$  & $0.83_{-0.02}^{+0.02}$ & $0.84_{-0.02}^{+0.02}$  \\ 
25 & $G_{\rm N}$ & $3.54_{-1.31}^{+1.54}$ & $3.65_{-1.99}^{+5.83}$ & $0.57_{-0.28}^{+0.26}$ & $87.8_{-2.6}^{+5.3}$ & $169_{-90}^{+163}$ & $-52_{-93}^{+172}$  & $-0.001_{-0.021}^{+0.021}$ & $-8.97_{-0.31}^{+0.40}$  & $0.77_{-0.03}^{+0.03}$ & $1.07_{-0.18}^{+0.28}$  \\ 
26 & $G_{\rm N}$ & $2.79_{-0.05}^{+0.44}$ & $3.06_{-0.27}^{+1.18}$ & $0.21_{-0.03}^{+0.12}$ & $165.7_{-18.0}^{+5.2}$ & $153_{-48}^{+53}$ & $76_{-21}^{+19}$  & $-0.000_{-0.021}^{+0.021}$ & $-8.80_{-0.12}^{+0.03}$  & $0.84_{-0.01}^{+0.01}$ & $1.15_{-0.07}^{+0.04}$  \\ 
27 & $G_{\rm N}$ & $0.67_{-0.00}^{+0.04}$ & $1.84_{-0.36}^{+0.93}$ & $0.76_{-0.05}^{+0.07}$ & $5.2_{-4.0}^{+17.4}$ & $221_{-87}^{+86}$ & $282_{-4}^{+7}$  & $-0.000_{-0.021}^{+0.021}$ & $-7.53_{-0.04}^{+0.03}$  & $2.20_{-0.01}^{+0.01}$ & $2.43_{-0.08}^{+0.07}$  \\ 
28 & $G_{\rm N}$ & $5.34_{-1.99}^{+6.78}$ & $4.62_{-2.29}^{+9.64}$ & $0.45_{-0.28}^{+0.40}$ & $83.8_{-16.9}^{+29.3}$ & $182_{-127}^{+125}$ & $18_{-215}^{+180}$  & $-0.002_{-0.021}^{+0.021}$ & $-9.29_{-0.71}^{+0.40}$  & $0.48_{-0.06}^{+0.22}$ & $0.90_{-0.30}^{+0.23}$  \\ 
29 & $G_{\rm N}$ & $0.34_{-0.16}^{+0.27}$ & $0.25_{-0.15}^{+0.51}$ & $0.49_{-0.12}^{+0.23}$ & $81.6_{-7.6}^{+3.5}$ & $270_{-78}^{+55}$ & $202_{-28}^{+67}$  & $-0.002_{-0.021}^{+0.021}$ & $-7.00_{-0.51}^{+0.56}$  & $1.83_{-0.03}^{+0.03}$ & $3.18_{-0.81}^{+1.21}$  \\ 
30$^a$ & $G_{\rm N}$ & $0.61_{-0.35}^{+0.08}$ & $2.13_{-1.92}^{+6.96}$ & $0.84_{-0.15}^{+0.09}$ & $98.5_{-1.2}^{+27.6}$ & $110_{-3}^{+72}$ & $257_{-148}^{+100}$  & $0.003_{-0.021}^{+0.021}$ & $-7.49_{-0.10}^{+0.73}$  & $2.25_{-0.01}^{+0.01}$ & $2.44_{-0.15}^{+1.25}$  \\ 
31 & $G_{\rm N}$ & $18.56_{-11.05}^{+9.48}$ & $13.60_{-9.29}^{+12.95}$ & $0.44_{-0.26}^{+0.30}$ & $78.0_{-20.3}^{+44.8}$ & $104_{-22}^{+58}$ & $171_{-36}^{+9}$  & $-0.002_{-0.021}^{+0.021}$ & $-10.43_{-0.36}^{+0.78}$  & $0.25_{-0.01}^{+0.01}$ & $0.45_{-0.09}^{+0.26}$  \\ 
32 & $G_{\rm N}$ & $0.59_{-0.08}^{+0.08}$ & $0.64_{-0.15}^{+0.22}$ & $0.30_{-0.09}^{+0.29}$ & $96.2_{-0.9}^{+2.5}$ & $120_{-53}^{+20}$ & $270_{-42}^{+63}$  & $-0.000_{-0.021}^{+0.021}$ & $-7.50_{-0.11}^{+0.12}$  & $1.73_{-0.04}^{+0.04}$ & $2.36_{-0.15}^{+0.16}$  \\ 
\hline
1 & $1.4G_{\rm N}$ & $11.50_{-1.42}^{+2.12}$ & $15.89_{-5.29}^{+12.76}$ & $0.74_{-0.06}^{+0.09}$ & $68.3_{-3.0}^{+2.9}$ & $42_{-11}^{+12}$ & $239_{-10}^{+9}$  & $0.002_{-0.021}^{+0.021}$ & $-10.01_{-0.15}^{+0.12}$  & $0.54_{-0.03}^{+0.03}$ & $0.69_{-0.06}^{+0.05}$  \\ 
3 & $1.4G_{\rm N}$ & $5.59_{-0.29}^{+0.73}$ & $7.72_{-1.44}^{+3.68}$ & $0.36_{-0.12}^{+0.18}$ & $139.0_{-5.6}^{+3.9}$ & $191_{-12}^{+14}$ & $300_{-9}^{+7}$  & $0.004_{-0.021}^{+0.021}$ & $-9.37_{-0.11}^{+0.05}$  & $0.79_{-0.02}^{+0.02}$ & $0.99_{-0.06}^{+0.04}$  \\ 
5 & $1.4G_{\rm N}$ & $4.00_{-0.25}^{+0.59}$ & $3.19_{-0.41}^{+0.79}$ & $0.36_{-0.14}^{+0.23}$ & $86.2_{-1.0}^{+0.7}$ & $187_{-43}^{+66}$ & $191_{-49}^{+17}$  & $-0.002_{-0.021}^{+0.021}$ & $-9.17_{-0.12}^{+0.06}$  & $0.64_{-0.04}^{+0.05}$ & $1.06_{-0.07}^{+0.04}$  \\ 
6 & $1.4G_{\rm N}$ & $14.44_{-4.03}^{+6.14}$ & $10.67_{-3.73}^{+7.49}$ & $0.52_{-0.09}^{+0.18}$ & $89.9_{-0.4}^{+0.5}$ & $254_{-47}^{+45}$ & $163_{-31}^{+20}$  & $-0.001_{-0.021}^{+0.021}$ & $-10.28_{-0.31}^{+0.28}$  & $0.31_{-0.04}^{+0.04}$ & $0.56_{-0.09}^{+0.10}$  \\ 
9 & $1.4G_{\rm N}$ & $5.59_{-1.23}^{+4.30}$ & $3.13_{-0.75}^{+2.96}$ & $0.79_{-0.14}^{+0.10}$ & $78.6_{-6.2}^{+4.8}$ & $139_{-32}^{+27}$ & $177_{-2}^{+10}$  & $-0.001_{-0.021}^{+0.021}$ & $-9.37_{-0.49}^{+0.21}$  & $0.32_{-0.03}^{+0.03}$ & $1.00_{-0.25}^{+0.13}$  \\ 
12 & $1.4G_{\rm N}$ & $31.35_{-7.12}^{+7.46}$ & $17.27_{-4.23}^{+4.78}$ & $0.88_{-0.04}^{+0.05}$ & $96.2_{-1.8}^{+2.3}$ & $253_{-9}^{+10}$ & $175_{-5}^{+3}$  & $-0.002_{-0.021}^{+0.021}$ & $-10.91_{-0.19}^{+0.22}$  & $0.12_{-0.03}^{+0.03}$ & $0.40_{-0.04}^{+0.06}$  \\ 
24 & $1.4G_{\rm N}$ & $5.83_{-0.04}^{+0.18}$ & $20.79_{-6.71}^{+15.46}$ & $0.75_{-0.13}^{+0.11}$ & $113.7_{-1.3}^{+1.5}$ & $29_{-11}^{+11}$ & $326_{-24}^{+22}$  & $0.013_{-0.020}^{+0.020}$ & $-9.41_{-0.03}^{+0.02}$  & $0.90_{-0.03}^{+0.03}$ & $0.97_{-0.03}^{+0.02}$  \\ 
28 & $1.4G_{\rm N}$ & $4.59_{-1.29}^{+6.22}$ & $3.07_{-1.10}^{+5.96}$ & $0.69_{-0.33}^{+0.20}$ & $96.9_{-38.4}^{+27.9}$ & $167_{-117}^{+141}$ & $167_{-352}^{+23}$  & $-0.001_{-0.021}^{+0.021}$ & $-9.16_{-0.74}^{+0.28}$  & $0.47_{-0.05}^{+0.16}$ & $1.15_{-0.40}^{+0.20}$  \\ 
31 & $1.4G_{\rm N}$ & $15.79_{-7.65}^{+11.19}$ & $9.77_{-5.25}^{+10.27}$ & $0.65_{-0.23}^{+0.16}$ & $80.3_{-26.0}^{+45.8}$ & $93_{-11}^{+21}$ & $175_{-13}^{+4}$  & $-0.002_{-0.021}^{+0.021}$ & $-10.29_{-0.47}^{+0.57}$  & $0.25_{-0.01}^{+0.01}$ & $0.58_{-0.14}^{+0.23}$  \\ 
\hline
\end{tabular}
\end{center}
$^a$ Unlike Table~\ref{tab:prmt_newton}, the parameter ranges for this system are based on two degenerate solutions including one localized around the periastron $\Delta\phi=360^\circ$ because the prior on $\Delta\phi$ is not used by design.
%\end{sidewaystable*}
\end{table*}

\begin{table*}
\caption{MLE values of the Fitted and Derived Parameters in Newtonian or pseudo-Newtonian Gravity with Priors}\label{tab:MLE_newton}
\scriptsize
\begin{center}
  \begin{tabular}{lccccccccccc}
  \hline
 \# & G & $r$ & $a$ &  $e$ & $i$ & $\phi_0$ & $\phi-\phi_0$ & $\log_{10} f_M$ & $\log_{10} g_{\rm N}$ & $v_{\rm mod}$ & $v_{\rm esc}$  \\
  &  & (kau) & (kau) &   & $(^\circ)$ & $(^\circ)$ & $(^\circ)$ &  & $(\text{m s}^{-2})$  & $(\text{km s}^{-1})$ & $(\text{km s}^{-1})$   \\
  \hline
1 & $G_{\rm N}$ & $11.58$ & $221.18$ & $0.98$ & $68.7$ & $264.5$ & $18.3$ & $-0.001$ & $-10.02$ & $0.570$ & $0.577$ \\ 
2 & $G_{\rm N}$ & $1.31$ & $3.27$ & $0.85$ & $107.7$ & $110.8$ & $307.6$ & $0.001$ & $-8.33$ & $1.209$ & $1.353$ \\ 
3 & $G_{\rm N}$ & $5.01$ & $13.16$ & $0.62$ & $144.6$ & $347.5$ & $175.6$ & $0.001$ & $-9.28$ & $0.799$ & $0.888$ \\ 
4 & $G_{\rm N}$ & $0.27$ & $0.15$ & $0.84$ & $88.3$ & $175.5$ & $165.3$ & $-0.001$ & $-6.86$ & $1.025$ & $3.347$ \\ 
5 & $G_{\rm N}$ & $3.73$ & $3.43$ & $0.35$ & $85.7$ & $236.5$ & $128.0$ & $0.000$ & $-9.10$ & $0.633$ & $0.938$ \\ 
6 & $G_{\rm N}$ & $10.02$ & $7.29$ & $0.87$ & $89.6$ & $199.1$ & $135.9$ & $0.001$ & $-9.96$ & $0.320$ & $0.571$ \\ 
7 & $G_{\rm N}$ & $0.83$ & $11.07$ & $0.97$ & $72.8$ & $105.9$ & $334.0$ & $0.001$ & $-8.02$ & $1.516$ & $1.545$ \\ 
8 & $G_{\rm N}$ & $2.21$ & $2.03$ & $0.21$ & $44.3$ & $234.5$ & $72.8$ & $-0.001$ & $-8.92$ & $0.604$ & $0.895$ \\ 
9 & $G_{\rm N}$ & $5.45$ & $3.19$ & $0.98$ & $40.1$ & $175.8$ & $276.1$ & $-0.000$ & $-9.35$ & $0.325$ & $0.857$ \\ 
10 & $G_{\rm N}$ & $1.46$ & $1.53$ & $0.93$ & $57.0$ & $202.8$ & $320.7$ & $-0.000$ & $-8.25$ & $1.135$ & $1.571$ \\ 
11 & $G_{\rm N}$ & $1.13$ & $1.23$ & $0.27$ & $85.5$ & $272.2$ & $144.9$ & $-0.000$ & $-8.03$ & $1.310$ & $1.783$ \\ 
12 & $G_{\rm N}$ & $41.99$ & $25.01$ & $0.80$ & $95.1$ & $169.1$ & $264.6$ & $0.000$ & $-11.17$ & $0.117$ & $0.293$ \\ 
13 & $G_{\rm N}$ & $1.12$ & $0.67$ & $0.92$ & $68.8$ & $189.3$ & $6.8$ & $0.000$ & $-8.04$ & $0.702$ & $1.743$ \\ 
14 & $G_{\rm N}$ & $4.21$ & $15.02$ & $0.77$ & $92.8$ & $304.4$ & $338.5$ & $-0.000$ & $-9.18$ & $0.841$ & $0.907$ \\ 
15 & $G_{\rm N}$ & $1.25$ & $0.87$ & $0.47$ & $162.5$ & $168.3$ & $50.9$ & $0.000$ & $-8.13$ & $0.877$ & $1.656$ \\ 
16 & $G_{\rm N}$ & $0.24$ & $0.39$ & $0.51$ & $165.8$ & $297.9$ & $203.7$ & $-0.000$ & $-6.75$ & $2.956$ & $3.532$ \\ 
17 & $G_{\rm N}$ & $0.84$ & $1.14$ & $0.27$ & $65.8$ & $15.4$ & $62.3$ & $-0.000$ & $-7.84$ & $1.523$ & $1.914$ \\ 
18 & $G_{\rm N}$ & $0.87$ & $1.58$ & $0.93$ & $39.1$ & $216.6$ & $194.8$ & $-0.000$ & $-7.85$ & $1.624$ & $1.907$ \\ 
19 & $G_{\rm N}$ & $2.59$ & $1.42$ & $0.87$ & $90.8$ & $184.8$ & $136.5$ & $-0.000$ & $-8.74$ & $0.350$ & $1.193$ \\ 
20 & $G_{\rm N}$ & $0.16$ & $0.32$ & $0.73$ & $70.5$ & $265.3$ & $128.5$ & $-0.000$ & $-6.26$ & $4.388$ & $5.061$ \\ 
21 & $G_{\rm N}$ & $2.11$ & $3.51$ & $0.42$ & $107.1$ & $31.8$ & $353.4$ & $0.001$ & $-8.84$ & $0.801$ & $0.958$ \\ 
22 & $G_{\rm N}$ & $0.37$ & $0.21$ & $0.75$ & $94.3$ & $174.3$ & $277.6$ & $-0.000$ & $-7.20$ & $0.981$ & $2.635$ \\ 
23 & $G_{\rm N}$ & $1.16$ & $0.73$ & $0.59$ & $54.7$ & $181.6$ & $200.7$ & $0.000$ & $-8.05$ & $0.797$ & $1.761$ \\ 
24 & $G_{\rm N}$ & $5.78$ & $1857.69$ & $1.00$ & $114.9$ & $338.2$ & $22.3$ & $0.027$ & $-9.39$ & $0.843$ & $0.844$ \\ 
25 & $G_{\rm N}$ & $2.41$ & $1.85$ & $0.84$ & $97.1$ & $157.1$ & $336.0$ & $-0.000$ & $-8.63$ & $0.765$ & $1.297$ \\ 
26 & $G_{\rm N}$ & $3.01$ & $3.55$ & $0.27$ & $154.8$ & $69.9$ & $189.1$ & $-0.000$ & $-8.86$ & $0.846$ & $1.115$ \\ 
27 & $G_{\rm N}$ & $0.68$ & $1.90$ & $0.76$ & $13.9$ & $283.4$ & $140.0$ & $0.000$ & $-7.54$ & $2.201$ & $2.429$ \\ 
28 & $G_{\rm N}$ & $3.48$ & $2.12$ & $0.89$ & $150.3$ & $191.2$ & $294.2$ & $-0.000$ & $-8.92$ & $0.475$ & $1.120$ \\ 
29 & $G_{\rm N}$ & $0.37$ & $0.30$ & $0.38$ & $82.4$ & $211.2$ & $259.3$ & $-0.000$ & $-7.09$ & $1.834$ & $3.027$ \\ 
30 & $G_{\rm N}$ & $0.27$ & $0.22$ & $0.82$ & $124.0$ & $154.1$ & $150.8$ & $-0.001$ & $-6.78$ & $2.244$ & $3.644$ \\ 
31 & $G_{\rm N}$ & $22.75$ & $18.25$ & $0.33$ & $107.2$ & $149.9$ & $119.7$ & $-0.001$ & $-10.61$ & $0.251$ & $0.409$ \\ 
32 & $G_{\rm N}$ & $0.56$ & $0.57$ & $0.33$ & $96.7$ & $254.1$ & $133.1$ & $0.001$ & $-7.46$ & $1.730$ & $2.421$ \\ 
\hline
1 & $1.4G_{\rm N}$ & $11.33$ & $17.75$ & $0.77$ & $68.4$ & $242.1$ & $41.2$ & $-0.001$ & $-10.00$ & $0.570$ & $0.691$ \\ 
3 & $1.4G_{\rm N}$ & $7.04$ & $18.98$ & $0.70$ & $129.0$ & $302.3$ & $171.1$ & $0.001$ & $-9.58$ & $0.799$ & $0.886$ \\ 
5 & $1.4G_{\rm N}$ & $4.77$ & $4.10$ & $0.29$ & $86.7$ & $136.7$ & $262.0$ & $0.000$ & $-9.32$ & $0.636$ & $0.982$ \\ 
6 & $1.4G_{\rm N}$ & $18.27$ & $15.52$ & $0.49$ & $89.8$ & $136.3$ & $283.9$ & $-0.000$ & $-10.49$ & $0.321$ & $0.500$ \\ 
9 & $1.4G_{\rm N}$ & $5.50$ & $3.07$ & $0.80$ & $78.2$ & $177.6$ & $140.7$ & $-0.000$ & $-9.35$ & $0.324$ & $1.010$ \\ 
12 & $1.4G_{\rm N}$ & $31.59$ & $17.28$ & $0.87$ & $96.3$ & $175.2$ & $252.7$ & $0.000$ & $-10.92$ & $0.117$ & $0.400$ \\ 
24 & $1.4G_{\rm N}$ & $5.78$ & $59.66$ & $0.91$ & $112.2$ & $339.6$ & $20.9$ & $-0.000$ & $-9.41$ & $0.944$ & $0.968$ \\ 
28 & $1.4G_{\rm N}$ & $6.70$ & $4.45$ & $0.52$ & $74.2$ & $186.8$ & $287.1$ & $-0.000$ & $-9.49$ & $0.475$ & $0.955$ \\ 
31 & $1.4G_{\rm N}$ & $11.67$ & $6.77$ & $0.73$ & $55.2$ & $176.8$ & $86.6$ & $0.000$ & $-10.03$ & $0.251$ & $0.676$ \\ 
 \hline

\end{tabular}
\end{center}
\end{table*}

\begin{table*}
\caption{Similar to Table~\ref{tab:MLE_newton} but with No/Flat Priors on $e$, $i$, and $\Delta\phi$}\label{tab:MLE_noprior}
\scriptsize
\begin{center}
  \begin{tabular}{lccccccccccc}
  \hline
 \# & G & $r$ & $a$ &  $e$ & $i$ & $\phi_0$ & $\phi-\phi_0$ & $\log_{10} f_M$ & $\log_{10} g_{\rm N}$ & $v_{\rm mod}$ & $v_{\rm esc}$  \\
  &  & (kau) & (kau) &   & $(^\circ)$ & $(^\circ)$ & $(^\circ)$ &  & $(\text{m s}^{-2})$  & $(\text{km s}^{-1})$ & $(\text{km s}^{-1})$   \\
  \hline
1 & $G_{\rm N}$ & $11.44$ & $121.79$ & $0.96$ & $68.5$ & $263.8$ & $19.2$ & $0.000$ & $-10.01$ & $0.568$ & $0.581$ \\ 
2 & $G_{\rm N}$ & $1.16$ & $1.98$ & $0.77$ & $109.1$ & $246.1$ & $61.4$ & $0.000$ & $-8.22$ & $1.209$ & $1.438$ \\ 
3 & $G_{\rm N}$ & $5.24$ & $17.08$ & $0.70$ & $142.1$ & $335.1$ & $171.6$ & $0.000$ & $-9.32$ & $0.798$ & $0.867$ \\ 
4 & $G_{\rm N}$ & $0.28$ & $0.16$ & $0.82$ & $88.4$ & $176.2$ & $159.4$ & $-0.000$ & $-6.89$ & $1.026$ & $3.290$ \\ 
5 & $G_{\rm N}$ & $4.36$ & $4.69$ & $0.14$ & $86.5$ & $66.4$ & $324.9$ & $-0.000$ & $-9.24$ & $0.635$ & $0.868$ \\ 
6 & $G_{\rm N}$ & $13.41$ & $11.53$ & $0.30$ & $89.8$ & $136.6$ & $270.7$ & $-0.001$ & $-10.22$ & $0.319$ & $0.493$ \\ 
7 & $G_{\rm N}$ & $0.85$ & $45.68$ & $0.99$ & $73.2$ & $104.8$ & $335.5$ & $0.000$ & $-8.04$ & $1.518$ & $1.525$ \\ 
8 & $G_{\rm N}$ & $1.89$ & $1.54$ & $0.23$ & $34.8$ & $181.8$ & $154.2$ & $0.001$ & $-8.78$ & $0.604$ & $0.971$ \\ 
9 & $G_{\rm N}$ & $5.40$ & $3.15$ & $0.98$ & $39.4$ & $175.8$ & $273.1$ & $-0.000$ & $-9.34$ & $0.325$ & $0.861$ \\ 
10 & $G_{\rm N}$ & $1.42$ & $1.43$ & $0.82$ & $68.4$ & $214.8$ & $325.6$ & $0.000$ & $-8.22$ & $1.133$ & $1.594$ \\ 
11 & $G_{\rm N}$ & $1.41$ & $2.17$ & $0.37$ & $86.5$ & $331.1$ & $93.1$ & $-0.000$ & $-8.22$ & $1.311$ & $1.596$ \\ 
12 & $G_{\rm N}$ & $35.44$ & $20.47$ & $0.81$ & $95.8$ & $171.5$ & $258.9$ & $0.001$ & $-11.02$ & $0.117$ & $0.319$ \\ 
13 & $G_{\rm N}$ & $3.22$ & $3.00$ & $0.56$ & $94.8$ & $130.1$ & $158.9$ & $-0.000$ & $-8.96$ & $0.700$ & $1.028$ \\ 
14 & $G_{\rm N}$ & $1.69$ & $1.30$ & $0.31$ & $96.4$ & $187.0$ & $117.0$ & $0.000$ & $-8.39$ & $0.843$ & $1.430$ \\ 
15 & $G_{\rm N}$ & $2.06$ & $1.94$ & $0.17$ & $124.9$ & $240.5$ & $220.8$ & $-0.000$ & $-8.57$ & $0.884$ & $1.289$ \\ 
16 & $G_{\rm N}$ & $0.24$ & $0.42$ & $0.48$ & $156.7$ & $316.4$ & $263.7$ & $-0.000$ & $-6.77$ & $2.956$ & $3.494$ \\ 
17 & $G_{\rm N}$ & $0.55$ & $0.48$ & $0.18$ & $51.7$ & $198.2$ & $229.9$ & $-0.001$ & $-7.48$ & $1.522$ & $2.353$ \\ 
18 & $G_{\rm N}$ & $0.77$ & $1.07$ & $0.93$ & $20.2$ & $210.0$ & $298.4$ & $-0.000$ & $-7.75$ & $1.623$ & $2.028$ \\ 
19 & $G_{\rm N}$ & $2.06$ & $1.11$ & $0.88$ & $91.0$ & $177.0$ & $193.6$ & $0.000$ & $-8.54$ & $0.353$ & $1.339$ \\ 
20 & $G_{\rm N}$ & $0.15$ & $0.25$ & $0.75$ & $67.1$ & $249.2$ & $137.9$ & $-0.000$ & $-6.21$ & $4.388$ & $5.225$ \\ 
21 & $G_{\rm N}$ & $1.94$ & $2.73$ & $0.32$ & $108.6$ & $326.6$ & $39.6$ & $-0.001$ & $-8.77$ & $0.801$ & $0.997$ \\ 
22 & $G_{\rm N}$ & $0.75$ & $0.52$ & $0.50$ & $92.1$ & $163.4$ & $287.5$ & $-0.000$ & $-7.81$ & $0.981$ & $1.851$ \\ 
23 & $G_{\rm N}$ & $1.75$ & $1.27$ & $0.80$ & $54.4$ & $202.7$ & $84.1$ & $0.001$ & $-8.41$ & $0.797$ & $1.432$ \\ 
24 & $G_{\rm N}$ & $5.78$ & $5564.19$ & $1.00$ & $115.0$ & $337.7$ & $22.2$ & $0.030$ & $-9.38$ & $0.847$ & $0.847$ \\ 
25 & $G_{\rm N}$ & $2.70$ & $2.22$ & $0.48$ & $86.0$ & $220.0$ & $192.2$ & $-0.000$ & $-8.73$ & $0.767$ & $1.226$ \\ 
26 & $G_{\rm N}$ & $3.00$ & $3.50$ & $0.26$ & $155.3$ & $71.2$ & $187.4$ & $0.001$ & $-8.86$ & $0.845$ & $1.118$ \\ 
27 & $G_{\rm N}$ & $0.73$ & $2.96$ & $0.82$ & $28.3$ & $294.6$ & $303.6$ & $-0.000$ & $-7.59$ & $2.202$ & $2.351$ \\ 
28 & $G_{\rm N}$ & $3.25$ & $1.94$ & $0.87$ & $148.2$ & $190.8$ & $326.8$ & $-0.001$ & $-8.86$ & $0.469$ & $1.157$ \\ 
29 & $G_{\rm N}$ & $0.39$ & $0.31$ & $0.37$ & $82.6$ & $214.7$ & $255.0$ & $-0.000$ & $-7.12$ & $1.834$ & $2.974$ \\ 
30 & $G_{\rm N}$ & $0.38$ & $0.42$ & $0.67$ & $107.9$ & $126.7$ & $168.6$ & $0.000$ & $-7.09$ & $2.244$ & $3.046$ \\ 
31 & $G_{\rm N}$ & $7.38$ & $4.21$ & $0.76$ & $26.1$ & $182.4$ & $290.0$ & $-0.001$ & $-9.63$ & $0.251$ & $0.718$ \\ 
32 & $G_{\rm N}$ & $0.51$ & $0.47$ & $0.61$ & $98.9$ & $226.5$ & $142.0$ & $0.000$ & $-7.37$ & $1.730$ & $2.549$ \\ 
\hline
1 & $1.4G_{\rm N}$ & $11.39$ & $17.77$ & $0.77$ & $68.4$ & $242.1$ & $41.1$ & $-0.001$ & $-10.00$ & $0.568$ & $0.689$ \\ 
3 & $1.4G_{\rm N}$ & $6.38$ & $12.16$ & $0.57$ & $132.9$ & $301.5$ & $178.7$ & $0.000$ & $-9.49$ & $0.799$ & $0.930$ \\ 
5 & $1.4G_{\rm N}$ & $3.79$ & $2.84$ & $0.64$ & $85.1$ & $209.2$ & $139.7$ & $-0.000$ & $-9.12$ & $0.634$ & $1.101$ \\ 
6 & $1.4G_{\rm N}$ & $18.99$ & $16.61$ & $0.50$ & $89.9$ & $133.4$ & $288.0$ & $-0.000$ & $-10.52$ & $0.321$ & $0.491$ \\ 
9 & $1.4G_{\rm N}$ & $6.08$ & $3.43$ & $0.77$ & $79.5$ & $178.9$ & $133.4$ & $-0.001$ & $-9.44$ & $0.323$ & $0.959$ \\ 
12 & $1.4G_{\rm N}$ & $20.64$ & $10.94$ & $0.90$ & $98.8$ & $178.1$ & $236.2$ & $0.000$ & $-10.55$ & $0.117$ & $0.494$ \\ 
24 & $1.4G_{\rm N}$ & $5.85$ & $71.26$ & $0.92$ & $111.7$ & $356.0$ & $13.6$ & $0.000$ & $-9.42$ & $0.942$ & $0.962$ \\ 
28 & $1.4G_{\rm N}$ & $5.27$ & $3.28$ & $0.88$ & $55.4$ & $167.5$ & $117.0$ & $0.000$ & $-9.28$ & $0.479$ & $1.077$ \\ 
31 & $1.4G_{\rm N}$ & $12.20$ & $7.13$ & $0.72$ & $123.4$ & $175.4$ & $91.3$ & $-0.001$ & $-10.07$ & $0.251$ & $0.661$ \\ 
 \hline

\end{tabular}
\end{center}
\end{table*}

\begin{table*}
\caption{MLE values of the Fitted and Derived Parameters in Generalized (Free $\Gamma$) Gravity with Priors}\label{tab:MLE_general}
\scriptsize
\begin{center}
  \begin{tabular}{lccccccccccc}
  \hline
 \# & $\Gamma$ & $r$ & $a$ &  $e$ & $i$ & $\phi_0$ & $\phi-\phi_0$ & $\log_{10} f_M$ & $\log_{10} g_{\rm N}$ & $v_{\rm mod}$ & $v_{\rm esc}$  \\
  &  & (kau) & (kau) &   & $(^\circ)$ & $(^\circ)$ & $(^\circ)$ &  & $(\text{m s}^{-2})$  & $(\text{km s}^{-1})$ & $(\text{km s}^{-1})$   \\
  \hline
1 & $0.199$ & $23.77$ & $58.29$ & $0.87$ & $78.7$ & $242.9$ & $30.3$ & $-0.000$ & $-10.64$ & $0.569$ & $0.638$ \\ 
2 & $0.089$ & $0.82$ & $0.61$ & $0.44$ & $109.8$ & $154.6$ & $226.6$ & $0.000$ & $-7.92$ & $1.209$ & $2.103$ \\ 
3 & $0.224$ & $5.65$ & $4.20$ & $0.44$ & $138.7$ & $155.9$ & $71.8$ & $0.001$ & $-9.38$ & $0.800$ & $1.400$ \\ 
4 & $-0.178$ & $0.26$ & $0.16$ & $0.73$ & $91.9$ & $194.6$ & $175.0$ & $-0.001$ & $-6.83$ & $1.027$ & $2.270$ \\ 
5 & $0.101$ & $3.86$ & $2.75$ & $0.43$ & $86.1$ & $192.7$ & $182.4$ & $0.001$ & $-9.13$ & $0.637$ & $1.165$ \\ 
6 & $-0.157$ & $10.00$ & $13.74$ & $0.30$ & $89.7$ & $324.0$ & $60.7$ & $-0.000$ & $-9.96$ & $0.318$ & $0.398$ \\ 
7 & $-0.079$ & $0.38$ & $0.53$ & $0.60$ & $61.6$ & $101.3$ & $309.1$ & $-0.000$ & $-7.35$ & $1.515$ & $1.894$ \\ 
8 & $-0.171$ & $1.86$ & $5.85$ & $0.70$ & $31.8$ & $27.3$ & $349.6$ & $0.000$ & $-8.76$ & $0.605$ & $0.659$ \\ 
9 & $-0.041$ & $6.83$ & $4.37$ & $0.96$ & $61.4$ & $171.8$ & $303.9$ & $-0.001$ & $-9.54$ & $0.325$ & $0.696$ \\ 
10 & $0.231$ & $1.58$ & $0.98$ & $0.73$ & $77.5$ & $193.5$ & $13.7$ & $0.000$ & $-8.32$ & $1.134$ & $2.568$ \\ 
11 & $0.128$ & $0.90$ & $0.59$ & $0.64$ & $84.0$ & $196.7$ & $210.0$ & $-0.000$ & $-7.83$ & $1.311$ & $2.688$ \\ 
12 & $-0.175$ & $28.42$ & $18.73$ & $0.64$ & $96.9$ & $162.6$ & $262.6$ & $-0.000$ & $-10.83$ & $0.117$ & $0.238$ \\ 
13 & $-0.225$ & $1.33$ & $1.45$ & $0.62$ & $102.2$ & $122.0$ & $21.3$ & $-0.001$ & $-8.20$ & $0.699$ & $0.951$ \\ 
14 & $0.088$ & $1.90$ & $1.28$ & $0.49$ & $95.8$ & $187.5$ & $112.4$ & $0.001$ & $-8.49$ & $0.843$ & $1.656$ \\ 
15 & $-0.127$ & $1.25$ & $1.28$ & $0.18$ & $161.1$ & $93.1$ & $127.0$ & $0.001$ & $-8.14$ & $0.883$ & $1.235$ \\ 
16 & $0.166$ & $0.24$ & $0.18$ & $0.42$ & $160.0$ & $203.1$ & $8.6$ & $-0.001$ & $-6.76$ & $2.955$ & $5.159$ \\ 
17 & $-0.097$ & $0.51$ & $0.65$ & $0.23$ & $47.9$ & $329.3$ & $95.8$ & $0.000$ & $-7.41$ & $1.522$ & $1.957$ \\ 
18 & $-0.025$ & $0.76$ & $1.30$ & $0.92$ & $27.9$ & $215.9$ & $312.1$ & $0.001$ & $-7.74$ & $1.621$ & $1.928$ \\ 
19 & $-0.089$ & $7.12$ & $5.53$ & $0.91$ & $90.6$ & $198.3$ & $88.2$ & $-0.000$ & $-9.61$ & $0.350$ & $0.587$ \\ 
20 & $0.123$ & $0.14$ & $0.11$ & $0.84$ & $57.3$ & $204.1$ & $169.2$ & $0.000$ & $-6.14$ & $4.383$ & $7.210$ \\ 
21 & $0.064$ & $1.94$ & $1.86$ & $0.15$ & $108.6$ & $246.2$ & $119.5$ & $-0.000$ & $-8.76$ & $0.802$ & $1.157$ \\ 
22 & $-0.701$ & $0.04$ & $0.03$ & $0.56$ & $147.6$ & $148.9$ & $302.7$ & $-0.000$ & $-5.18$ & $0.982$ & $1.683$ \\ 
23 & $-0.158$ & $1.20$ & $1.07$ & $0.56$ & $47.9$ & $226.1$ & $101.7$ & $-0.001$ & $-8.08$ & $0.796$ & $1.201$ \\ 
24 & $0.361$ & $6.51$ & $4.56$ & $0.67$ & $113.9$ & $203.6$ & $126.2$ & $-0.001$ & $-9.52$ & $0.945$ & $1.768$ \\ 
25 & $0.063$ & $2.88$ & $2.09$ & $0.78$ & $95.0$ & $156.4$ & $328.5$ & $-0.000$ & $-8.79$ & $0.765$ & $1.373$ \\ 
26 & $0.313$ & $4.05$ & $2.48$ & $0.66$ & $132.5$ & $172.5$ & $97.8$ & $-0.001$ & $-9.12$ & $0.846$ & $1.975$ \\ 
27 & $0.115$ & $0.69$ & $0.68$ & $0.52$ & $18.9$ & $237.2$ & $185.2$ & $-0.001$ & $-7.56$ & $2.199$ & $3.131$ \\ 
28 & $0.276$ & $15.47$ & $9.95$ & $0.76$ & $81.4$ & $163.4$ & $114.8$ & $0.001$ & $-10.21$ & $0.473$ & $1.004$ \\ 
29 & $-0.229$ & $0.15$ & $0.13$ & $0.91$ & $42.2$ & $158.7$ & $57.6$ & $-0.000$ & $-6.30$ & $1.833$ & $2.805$ \\ 
30 & $-0.046$ & $0.22$ & $0.18$ & $0.89$ & $145.2$ & $158.3$ & $144.5$ & $0.000$ & $-6.62$ & $2.243$ & $3.591$ \\ 
31 & $-0.100$ & $24.04$ & $32.69$ & $0.32$ & $73.8$ & $47.9$ & $220.4$ & $-0.001$ & $-10.66$ & $0.251$ & $0.316$ \\ 
32 & $-0.014$ & $0.53$ & $0.54$ & $0.47$ & $97.6$ & $244.9$ & $133.6$ & $0.000$ & $-7.40$ & $1.732$ & $2.418$ \\ 
 \hline

\end{tabular}
\end{center}
\end{table*}

\end{document}